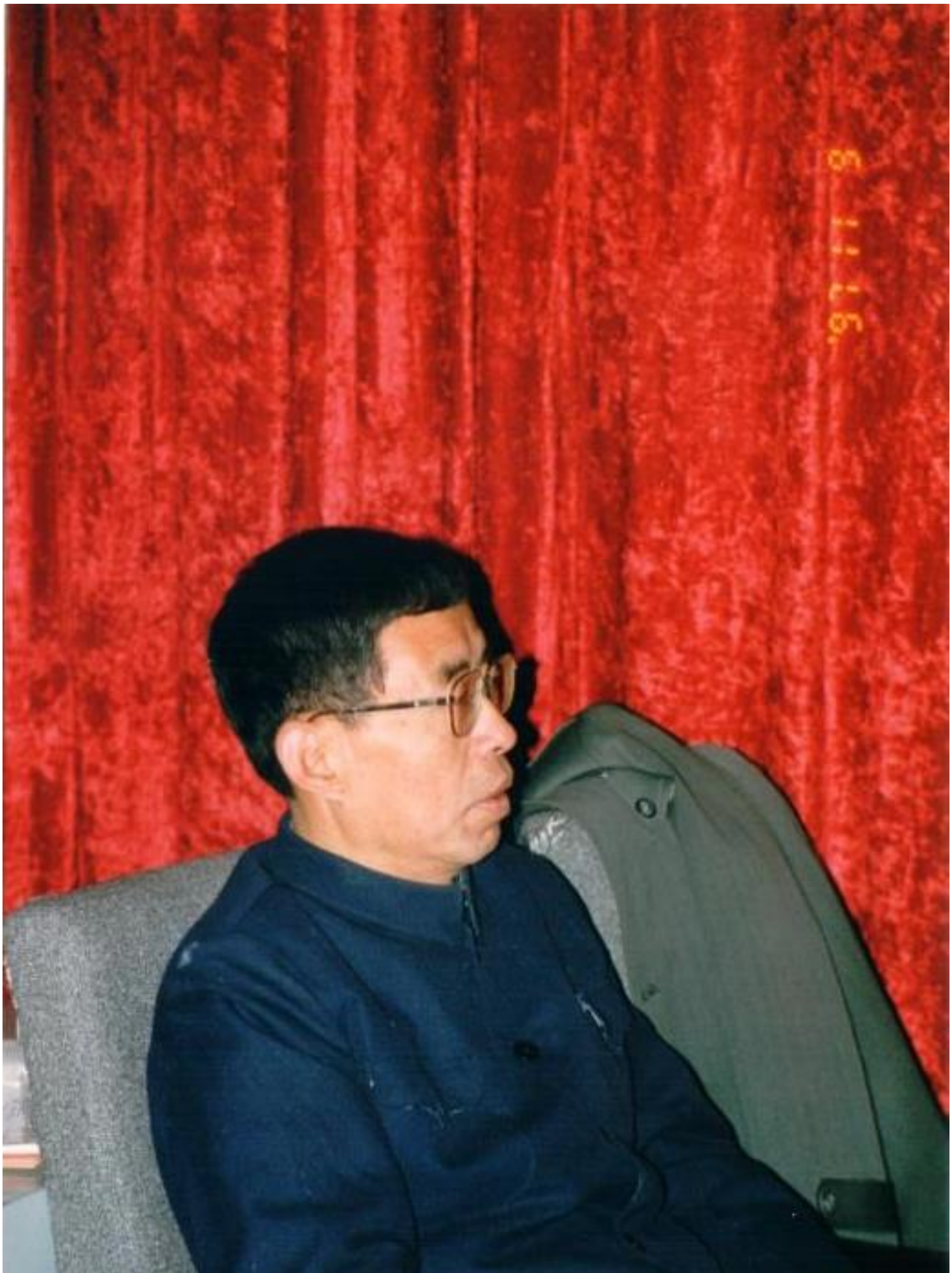

Qian [Tsien] Jian at the State Key Laboratory of Turbulence, Peking University, China, 09 November 1997.

## Qian Jian (1939–2018) and His Contribution to Small-Scale Turbulence Studies

**John Z. Shi***

State Key Laboratory of Ocean Engineering, School of Naval Architecture, Ocean and Civil Engineering, Shanghai Jiao Tong University, 1954 Hua Shan Road, Shanghai 200030, China *zshi@sjtu.edu.cn

## Abstract

Qian [Tsien] Jian (1939–2018), a Chinese theoretical physicist and fluid dynamicist, devoted the second part of his scientific life to the physical understanding of small-scale turbulence to the exclusion of all else. To place Qian's contribution in an appropriate position in the field of small-scale turbulence, a historical overview and a state-of-the art review are attempted. Qian developed his own statistical theory of small-scale turbulence, based on the Liouville (1853) equation and a perturbation variational approach to non-equilibrium statistical mechanics, which is compatible with the Kolmogorov–Oboukhov energy spectrum. Qian's statistical theory of small-scale turbulence, which appears mathematically and physically valid, successfully led to his contributions to (i) the closure problem of turbulence; (ii) one-dimensional turbulence; (iii) two-dimensional turbulence; (iv) the turbulent passive scalar field; (v) the cascade model of turbulence; (vi) the universal equilibrium range of turbulence; (vii) a simple model of the bump phenomenon; (viii) universal constants of turbulence; (ix) the intermittency of turbulence; and perhaps most importantly, (x) the effect of the Taylor microscale Reynolds number ($R_\lambda$) on both the width of the inertial range of finite $R_\lambda$ turbulence and the scaling exponents of velocity structure functions. In particular, Qian found that the inertial range cannot exist when $R_\lambda \ll 2000$. In contrast to the prevailing intermittency models, he discovered that normal scaling is valid in the real Kolmogorov inertial range when $R_\lambda$ approaches infinity while the anomalous scaling observed in experiments reflects the finite $R_\lambda$ effect ($Q_e$). He then made a correction to the famous Kolmogorov (1941c) equation and obtained the finite $R_\lambda$ effect equation or the Kolmogorov–Novikov–Qian equation. He also independently derived the decay law of the finite $R_\lambda$ effect. Following up Kraichnan, Qian steered all of us along the right path to an improved understanding of small-scale turbulence and solutions to its problems. Qian is credited with his contribution to enhanced knowledge about the finite $R_\lambda$ effect of turbulence, which has profoundly shaped and stimulated thinking about the K41 turbulence, the K62 turbulence and the finite $R_\lambda$ turbulence.

## Introduction

> Some mathematicians are birds, others are frogs……Frogs live in the mud below and see only the flowers that grow nearby. They delight in the details of particular objects, and they solve problems one at a time.
>
> Dyson (2009, page 212, lines 1-2)

Qian [Tsien] Jian (1939–2018) was a Chinese theoretical physicist and fluid dynamicist. In light of Qian's personality and work, he was indeed a quiet and lonely frog, who ever lived in the mud below and saw only the beautiful flowers that grow nearby, and ever delighted in the details of a particular object, i.e. small-scale turbulence, and solved problems one at a time. A brief account of his life and work can be found in an English obituary by Shi (2018) and a Chinese obituary by Shi (2019). Qian's significant contribution is briefly highlighted in Shi (2020). However, consider the fundamental contributions which Qian made to small-scale turbulence, he has not been given enough credit; and a detailed account of his life and work is still required.

Studies of small-scale turbulence have been undertaken in mainland China since the 1950s (e.g. Chou and Cai 1957; Huang and Chou 1981; Fong 1982; He, Chen, Kraichnan, Zhang, and Zhou, 1998; He and Zhang, 2006; Ran 2006, 2008, 2009a, b; He, Jin, and Zhao 2009; Huang, Schmitt, Lu, Fougairolles, Gagne and Liu 2010; He 2011; Ran 2011; Ran and Yuan 2013; Liao and Su 2015; He, Jin, and Yang 2017; Wang, Wan, Chen and Chen 2018; Wu, Fang, Shao and Lu 2018; and Yang, Pumir and Xu 2018). In the present document Qian's work will be introduced in two separate sections in detail, and a list of his publications are left for those sections. Zhen-Su She's work on this topic was mainly undertaken in the U.S.A. and will be briefly reviewed in other sections wherever applicable. Chou P. and Chou R. (1995) gave a brief overview of 50 years of turbulence research in China, which highlighted the work done by Chou Peiyuan and his associates from 1940 to 1988. A short historical overview of small-scale turbulence research in China can also be found in Shi (2020).

This article takes for its scope the whole of Qian's work on small-scale turbulence, with particular emphasis on his contribution to the finite $R_\lambda$ effect of turbulence and the finite $R_\lambda$ turbulence. However, the author can not do more than scratch the surface of what Qian has done, partly because Qian's mind was extremely mathematical. It is also not such an easy task to place his contribution in an appropriate position in the history of small-scale turbulence studies, partly because of a slow march to the great ocean of truth of turbulence and partly because so many people have been working in this field all over the world. This historical overview and a state-of-the-art review may be helpful for doing so. Although small-scale turbulence has already been reviewed in some detail (e.g. Hunt and Vassilicos 1991; Nelkin 1994; Frisch 1995; Sreenivasan and Antonia 1997; Barenblatt and Chorin 1998; Sreenivasan 1999; Lumley and Yaglom 2001; Moffatt 2012; and McComb 2014, Chapter 6, pages 143-187), the author will attempt to add updated material together with some refined and unpublished historical notes. It is hoped that the present account of Qian's work on small-scale turbulence can be placed within the general context of small-scale turbulence research conducted since Kolmogorov (1941a, b, c) and Oboukhov (1941).

The overall structure of the main text is organized as follows: Section II Qian's biographical sketch; Section III Qian's academic pedigree; Section IV Some physical backgrounds; Section V Qian's major work on small-scale turbulence between 1983 and 1996; Section VI Qian's work on the finite $R_\lambda$ effect of turbulence between 1997 and 2006; Section VII Significance of Qian's work on the finite $R_\lambda$ effect of turbulence; Section VIII The continuing legacy of Qian; and Section IX Concluding Remarks.

## Qian's biographical sketch

......a well-written Life is almost as rare as a well-spent one.
Thomas Carlyle (1795–1881) in his Jean Paul Friedrich Richter (1827)(Carlyle 1872, page 1)

Qian Jian (Figure 1, right) was born on 20 November 1939 in Wuxi City, Jiangsu Province, China. In September 1957, he entered the Department of Physics, Peking University, China. He graduated with the B.Sc. degree from Peking University in August 1963.

During the first part of his scientific life (ca. 17 years from 1963 to 1979) after graduating, Qian held posts at several institutes and mainly worked on Chinese military defense projects. From 1963 to 1965, Qian worked on high temperature gas transport at the No. 11 Laboratory, the Institute of Mechanics, the Chinese Academy of Sciences, Beijing, China. From August 1965 to June 1966, he participated in the Four Cleans Movement in the Yongji County, Shanxi Province. In July 1966, he returned to the Institute of Mechanics, working on the physics of gas dynamics, together with other members of the No. 11 Laboratory and other participants from other institutes of the Chinese Academy of Sciences. In February 1967, the project group was established with Qian as a member. In May 1970, the project group was reorganized as the 207 Institute, the 2nd Academy of China Aerospace, i.e. the Beijing Institute of Environmental Features. Qian remained there from August 1974 until July 1988.

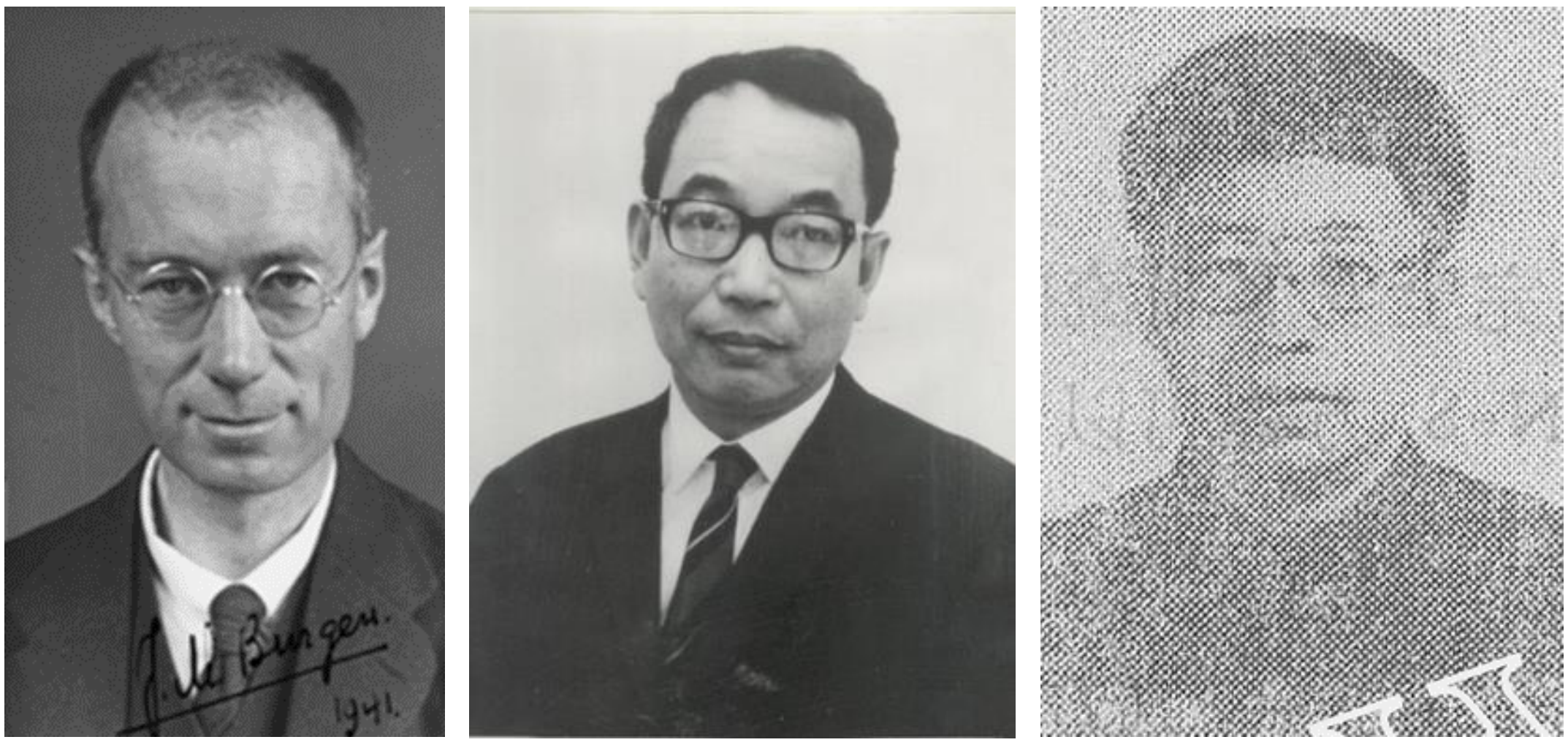

Figure 1 Johannes Martinus Burgers in 1941 (left)[Photo Credit: https://www.dwc.knaw.nl/]; Chan-Mou Tchen (middle)[Photo Credit: The City College of The City University of New York, U.S.A.]; and Qian Jian in circa 1988 (right).

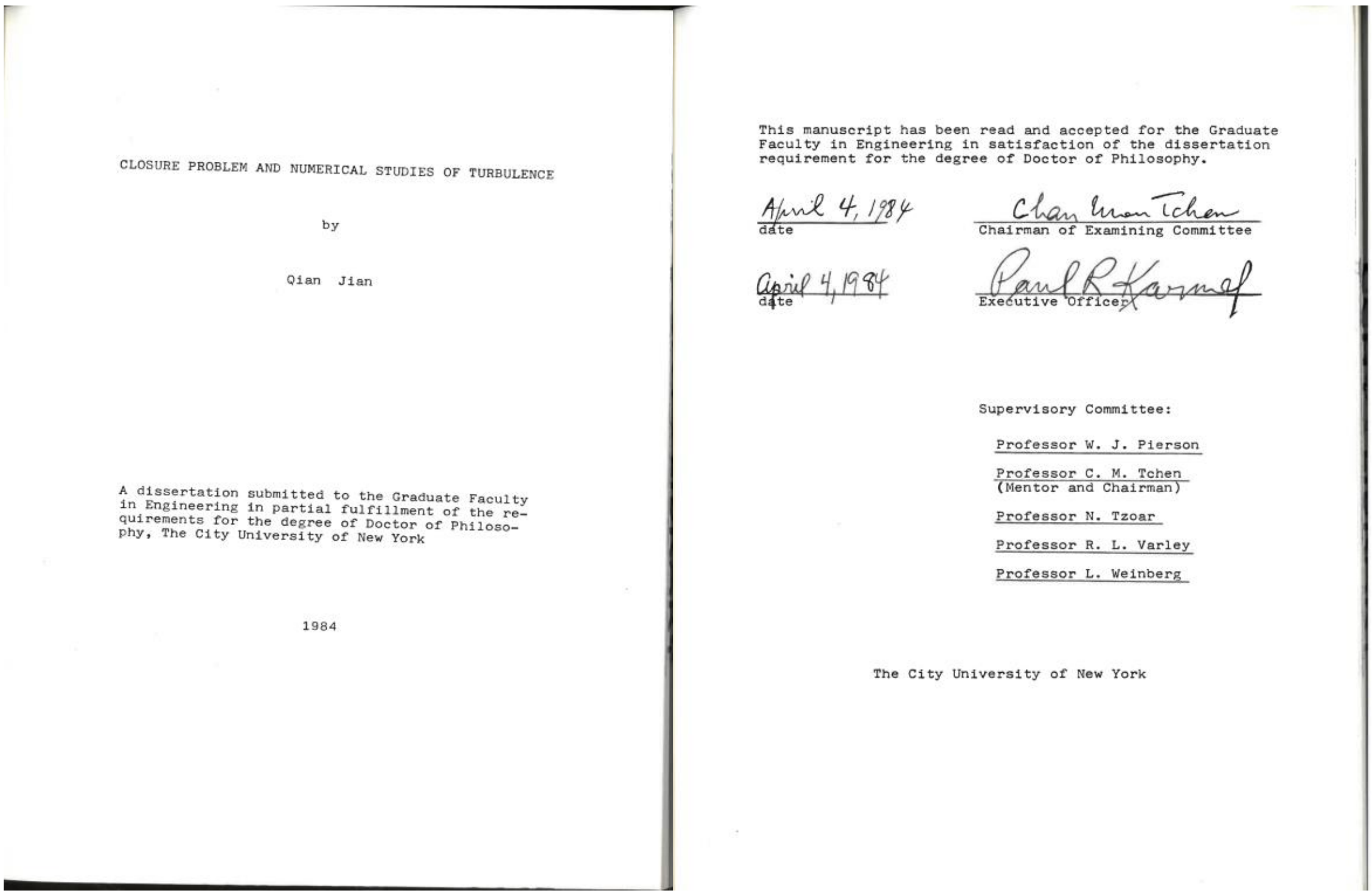

CLOSURE PROBLEM AND NUMERICAL STUDIES OF TURBULENCE

by

Qian Jian

A dissertation submitted to the Graduate Faculty in Engineering in partial fulfillment of the requirements for the degree of Doctor of Philosophy, The City University of New York

1984

This manuscript has been read and accepted for the Graduate Faculty in Engineering in satisfaction of the dissertation requirement for the degree of Doctor of Philosophy.

date — Chairman of Examining Committee

date — Executive Officer

Supervisory Committee:

Professor W. J. Pierson

Professor C. M. Tchen (Mentor and Chairman)

Professor N. Tzoar

Professor R. L. Varley

Professor L. Weinberg

The City University of New York

Figure 2 Qian Jian's PhD Thesis front page and his Supervisory Committee page.

In 1979, when China had just opened its doors to the West, Qian went on to the Department of Mechanical Engineering, The City College of The City University of New York, U.S.A., and did his Ph.D. research under the supervision of Herbert G. Kayser Professor Chan-Mou Tchen (1914–2004)(Figure 1, middle). Qian's Thesis was read and accepted for the Graduate Faculty in Engineering in satisfaction of the dissertation requirement for the degree of Doctor of Philosophy on 4 April 1984. His Ph.D. thesis is entitled "Closure Problems and Numerical Studies of Turbulence" (Figure 2). Chairman of Examing Committee of Qian's PhD Thesis was Chan-Mou Tchen, the Executive Officer was Paul R. Karmel, and the Supervisory Committee included Willard J. Pierson, Chan Mou Tchen, N. Tzoar, R.L. Varley, and L. Weinberg. David H. Cheng and Paul R. Karmel were Deans of the School. Peter Ganatos gave Qian much help in using computers.

Qian returned to China from the U.S.A. after obtaining his PhD in 1984. In August 1988, Qian was transferred from the 2nd Academy of China Aerospace to Department of Physics, Graduate School, the Chinese Academies of Sciences, Beijing. Qian was promoted to Full Professor in 1989 and retired in 2005. In the 1990s, Qian served on the Academic Committee of the State Key Laboratory of Turbulence at Peking University, China, and attended regular advisory meetings there. Qian died in Beijing, China, on 30 November 2018 after a cerebral haemorrhage.

Qian was a typical traditional Chinese scholar. Unlike the meaning of his Chinese last name "Qian [Tsien]", which means money, he was not money hungry. Like the meaning of his Chinese first name "Jian", i.e. "frugal", one of the traditional Chinese gentleman's virtues, he was truly a gentleman with reserve and shyness. He was deservedly respected by all who knew him.

## Qian's academic pedigree

Nothing will come of nothing:
William Shakespeare's King Lear: Act 1 Scene 1

Johannes Martinus Burgers (1895–1981)(Figure 1, left) is regarded as one of the leading scientists in the field of fluid mechanics last century. Burgers was Professor of Aerodynamics and Hydrodynamics, Department of Mechanical Engineering and Shipbuilding, Technische Hoogeschool in Delft (now Technical University of Delft), Delft, Netherlands from 1918 to 1955 (https://history.aip.org/phn/11806008.html). Burgers became professor emeritus at the Institute for Physical Science and Technology at the University of Maryland at the end of a scientific career that extended over 65 years until his death in the summer of 1981 (Dorfman and Faller 1982).

Tchen was born on 8 December 1914 in Fenghua, Zhejiang Province. He did his B.B. in Mathematics and Physics at Aurora University in Shanghai in 1935 and M.S. in Civil Engineering in 1936. He went onto to study hydraulic engineering in the Netherlands, supported by the Boxer Rebellion Indemnity Scholarship in June 1939. Because of the War in Europe, he did not return to China. Under the supervision of Burgers, Tchen completed his Ph.D. Thesis entitled "Mean Value and Correlation Problems connected with Motion of Small Particles suspended in a turbulent fluid" in 1947. Tchen was awarded Guggenheim Fellow in 1959. He joined the Department of Mechanical Engineering, The City College of The City University of New York, U.S.A., in 1967 and retired in 1981. Ever since his PhD research (Tchen 1947), Tchen had continued to work on the statistical properties of homogeneous, isotropic turbulence (e.g. Tchen 1952, 1953, 1954, 1973, 1975). As a digression, it is unclear why Tchen, as a turbulence man, did not attend the Symposium on Turbulence at the Marseilles Symposium of 1961. However, we may cautiously infer that Tchen was not in favour of the dimensional analysis of turbulence as Tchen (1975, page 1, right column, second paragraph) wrote that "The dimensional method does not analyze the detailed dynamical processes of turbulence, and, therefore, its results are ambiguous as to the types of fluctuations (velocity or temperature), and as to the types of stratification (stable or unstable)".

What brought Qian into the field of small-scale turbulence? From a historical perspective, the kinetic theory of gases was founded by Fourier (1822), Clausius (1859), Maxwell (1860a), and Boltzmann (1872). Based on his B.Sc. thesis, Qian published his first paper in Chinese entitled "On the integral equations in the theory of mixed gas transport" in *Acta Physica Sinica* in 1964 (Qian 1964). From the references cited at the end of his paper, it can be seen that Qian had read Chapman and Cowling's (1952) book entitled "The Mathematical Theory of Non-Uniform Gases" (which was highly recommended by the British physicist Fowler (1939)), and Hirschfelder, Curtiss and Bird's (1954) book entitled "Molecular Theory of Gases and Liquids" (which in turn was highly recommended to aeronautical engineers by the British turbulence modeler Spalding (1954)). Probably Qian first learned about the Boltzmann equation and the variational approach from Hirschfelder, Curtiss and Bird (1954). Not surprisingly (like Hirschfelder in the U.S.A.), Qian's mathematical and physical background fitted with his work for the national military defense project before the Cultural Revolution in China.

It is clear Qian believed that his mathematical and physical knowledge about the dynamic theory of mixed gas transport could be directly applied to the studies of the physics of turbulence. It will be shown later that the mathematical approaches in Chapman and Cowling (1952) and Qian (1964) were used in his first paper (Qian 1983) and his Ph.D. research (Qian 1984b). It is also clear that Qian knew of Tchen's research interest in turbulence before he went to further study turbulence under Tchen's supervision at The City College of The City University of New York, U.S.A. in the late 1970s.

## Some physical backgrounds

> We are nothing without the work of others our predecessors, others our teachers, others our contemporaries. Even when, in the measure of our inadequacy and our fullness, new insight and new order are created, we are still nothing without others. Yet we are more.
>
> Julia Robert Oppenheimer (1904-1967), Reith Lecture, 20 December 1953 (Farmelo 2009)

### *Statistical mechanics*

Fourier (1822) published his famous book entitled "Théorie Analytique de la Chaleur (The Analytical Theory of Heat)". At Fourier's 250$^{th}$ birthday in 2018, *Nature* published Editorials' article by highlighting "Today, there is virtually no branch of science that is left untouched by his (Fourier's) ideas" (Editorials 2018). This is also true for turbulence.

Statistics Mechanics has grown out of the theory of the properties of matter in equilibrium. The following one is Liouville (1853, page 71) equation:

$$\frac{d^2 log\lambda}{dudv} \pm \frac{\lambda}{2a^2} = 0 \quad (1)$$

Qian (1984a, page 10, equation (1.9)) used the following form:

$$\frac{\partial P}{\partial t} + \bar{L}P = 0 \quad (2)$$

where $P$ denotes the probability distribution over an ensemble of numerous realizations of the turbulence; and $\bar{L}$ the Liouville operator.

The simplest statistical mechanics was used to illustrate the dynamical theory of gases (e.g. Clausius 1859; Maxwell 1860a). A differential equation for the distribution function describing Brownian motion was first derived by Langevin (1908), Fokker (1914) and Planck (1917). Based on Edwards (1964), Herring (1965) and Kraichnan (1971), Qian (1984a, page 11, equation (1.11)) used the following form of the Langevin (1908)–Fokker (1914)–Planck (1917) equation:

$$\sum_{j,m} A_{i,l,m} x_j \, x_m \simeq -\zeta_i x_i + f_i \tag{3}$$

where $-\zeta_i x_i$ is the dynamic damping force and is deterministic for the mode $i$, representing the average effect of nonlinear interaction on mode $i$; and $f_i$ a random force with white noise character.

*Homogeneous, isotropic turbulence*

Thomson (later Lord Kelvin) (1887a, page 272, line 18) coined the terminology "turbulence", which was briefly reviewed in Schmitt (2017).Thomson (later Lord Kelvin) (1887b, pages 345, 346 and 347) coined the terminology "homogeneous, isotropic turbulence" and proposed its general concept. However, the theory of Homogeneous, Isotropic Turbulence (HIT) was furthered mainly by Taylor (1935c, d, e, f; 1936), de Kármán and Howarth (1938), Kolmogorov (1941a, b, c), Oboukhov (1941), Heisenberg (1948a, b), Batchelor (1947), Corrsin (1949), Batchelor (1953), Saffman (1967), Kraichnan (1974), George (1992), and McComb (2014). According to Chandrasekhar (1949, page 335), Heisenberg's (1948) theory of turbulence is better than Kolmogorov's (1941a, c). According to Tchen (1954, page 4), Oboukhov's (1941) theory of turbulence is different from Heisenberg's (1948a, b). These observations are generally beyond the scope of this article.

Notably, at the Marseille Symposium in 1961, Batchelor (1961, page 87, paragraph 2, lines 4-5; paragraph 3, lines 1-2) raised the two questions: (a) "is it likely that the dynamical problem of homogeneous turbulence will ever be solved without the aid of hypotheses?"; (b) "if progress on the purely observational side is so rapid that a proper mathematical solution is unlikely to be needed". Batchelor did not think so. In the author's view, they are still open questions. Moffatt (2012) presented an introductory review of homogeneous turbulence at Turbulence Colloquium Marseille 2011.

The statistical dynamics of homogeneous turbulence was studied first by Thomson (1887b) and then by Kraichnan (1958), Edward (1964), and Edwards and McComb (1969). Turbulence remains today, a formidable, partially-solved problem that had defeated many brilliant physicists and mathematicians during the twentieth century (Warner 2017, page 250, lines 3-4).

*The kinematic viscosity*

Mentions and recognition of 'kinematic viscosity' since the occurrence of this idea could go back to at least Newton (1687), d'Alembert (1752), Navier (1823, page 414, 2nd set of equations), Stokes (1845, page 288, lines 6-7), Stokes (1851, page 17, 2nd paragraph, lines 1-2), Maxwell (1866, page 249, paragraph 2; page 254, bottom relation), Boussinesq (1877, page 45, equation (11)), Taylor (1915, page 14, line 3), Prandtl (1925, page 138, equation (9)). For example, Stokes (1845, page 288, lines 6-7) wrote "The amount of the internal friction of the water depends on the value of $\mu$". Stokes (1851, page 17, paragraph 2, lines 1-2) further defined:

$$\mu = \mu' \rho \tag{4a}$$

where the constant $\mu'$ may conveniently be called the index of friction of the fluid.

The viscosity of a body is the resistance which it offers to a continuous change of form,

depending on the rate at which that change is effected (Maxwell 1866, page 249, paragraph 2). Maxwell (1866, page 254, bottom) defined the following viscosity:

$$\mu = \frac{fa}{\mathrm{v}} \tag{4b}$$

where $\mu$ is the coefficient of viscosity; $f$ the tangential force; $a$ the distance between them; and $\mathrm{v}$ the velocity of the upper plane. The dimensions of the coefficient of viscosity are $L^{-1}MT^{-1}$ (Maxwell 1866, page 255, line 19).

By re-arranging equation (4a) above, Reynolds (1883, page 937, line 2 from the bottom) obtained $\frac{\mu}{\rho}$ or $\mu'$ and dimensionally defined it as "a quantity of the nature of the product of a distance and a velocity", which is the same as the coefficient of viscosity by Maxwell (1866). The symbol and term, '$\nu$ the (kinematic) viscosity', which replaced $\frac{\mu}{\rho}$ or $\mu'$, appeared in Lamb (1895, page 573, paragraph 2, line 17).

*The Reynolds number and "the Reynolds Number of Turbulence"*

The modern form of the Reynolds number actually can be found in Reynolds (1883, page 938, line 8 from the bottom); i.e.

$$[Re =]\frac{c\rho U}{\mu}, \text{ i.e. } Re = \frac{LU}{\nu} \tag{5a}$$

Ever since Thomson (1887b), and in particular Taylor (1921) who laid the foundations for the statistical theories of homogeneous, isotropic turbulence, the random motion of the turbulent fluids has been described by statistical properties, i.e. some characteristic factors and functions. However, "the Reynolds Number of Turbulence", namely $l\sqrt{\overline{u^2}}/\nu$, did not appear until in Taylor (1935c, page 422, bottom, lines 1-2), where $l$ is the Taylor's Lagrangian integral length scale [$L$ is used in Kolmogorov (1941a) and Batchelor (1961)]; $\overline{u^2}$ the mean square variation in one component of velocity; and $\nu$ the kinematic viscosity. Batchelor (1947, page 553, paragraph 3, line 9) defined the following Reynolds number of turbulence:

$$R_\lambda = \frac{u'\lambda}{\nu} \tag{5b}$$

Note that $\lambda$ is proportional to $\sqrt{\frac{l\upsilon}{u'}}$ (Taylor 1935c, page 443, paragraph 5, line 3). In memoriam of George Keith Batchelor, $R_\lambda$ the Taylor microscale Reynolds number, is used throughout.

Oboukhov (1962, page 78, equation (4)) proposed the following 'the internal Reynolds number':

$$\tilde{R}(r) = \tilde{\varepsilon}^{\frac{1}{3}} r^{\frac{4}{3}}/\nu \tag{5c}$$

where $r$ is the distance apart.

Kolmogorov (1962, page 84, bottom) proposed the following form of the Reynolds number:

$$Re = \frac{|u_\alpha(\mathbf{x}^{(0)}) - u_\alpha(\mathbf{x})||\mathbf{x}^{(0)} - \mathbf{x}|}{\nu} \tag{5d}$$

where $\alpha$=1,2,3.

By employing the results of a variable mean field theory in Effinger and Grossmann (1987), Lohse (1994) derived the Taylor-Reynolds and Reynolds number ($Re_\lambda$ and $Re$) dependence of the dimensionless energy dissipation rate ($c_\varepsilon$) for statistically stationary isotropic turbulence. The following relation is obtained (Lohse 1994, page 3223, equation (3):

$$Re_\lambda = \sqrt{15Re/c_\varepsilon(Re)} \tag{5e}$$

*Small-scale turbulence*

The term ‘isotropic turbulence’ was introduced first by Thomson (1887b) and then adopted by Taylor (1935). MacPhail (1940) provided an experimental verification of the isotropy of turbulence produced by a grid. Kolmogorov (1941a, page 10, paragraphs 1 and 2) defined ‘locally isotropic turbulence’, which is also called ‘local isotropy’ (Kolmogorov 1941c, page 15, line 1). There are two English translation versions of Kolmogorov (1941a): one by D. ter Harr published in *Soviet Physics Uspekhi* (Kolmogorov 1968) and the other by V. Levin republished in *Proceedings of the Royal Society of London* (Kolmogorov 1991). In his own words, the term “small region” refers to a region the linear dimensions of which are small compared with $L$ and the time dimensions small compared with $T = U/L$ (Kolmogorov 1968, page 734, right column, paragraph 3, lines 7-10), or the term “small domain”, to a domain, whose linear dimensions are small in comparison with $L$ and time dimensions – in comparison with $T = U/L$ (Kolmogorov 1991, page 10, paragraph 4, lines 7-9). Clearly, “small” refers to both the geometry and physics of turbulence. The word ‘small/smaller/smallest’ appeared more than 80 times while the term “the small domain $G$” more than 8 times in Batchelor (1947).

The term “small-scale turbulence” first appeared in Townsend (1947, page 552, paragraph 2, line 8) and then in (Oboukhov 1962, page 77, line 3) and Yaglom (1994, page 9, line 13), and even in the titles of papers of Lilly (1967); Stewart, Wilson and Burling (1970); Hunt and Vassilicos (1991a, b); Sreenivasan and Antonia (1997); Antonia and Burattini (2004); George (2013); and Antonia, Djenidi, Danaila and Tang (2017).

The term “the small-scale structure of fully developed turbulent flow” also appeared in the texts of Townsend (1948b, page 161, I. Introduction, line 1). The notion ‘the small-scale structure of turbulent motion’ appeared in Batchelor and Townsend (1949, page 238, Introduction, line 4); Pao (1965, page 1066, right column, lines 8- from the bottom); Qian (1994a, page 15, paragraph 2) and Sreenivasan and Antonia (1997, page 435, Abstract, lines 4 and 8) and in the titles of papers of Frenkiel and Klebanoff (1975); Kerr (1985); Ruetsch and Maxey (1992); and Kaneda and Morishita (2013). The term “small-scale motion” appeared in Batchelor and Townsend (1949, page 238, Introduction, lines 9 and 13).

The subtitle “The small-scale properties of the turbulence” appeared in Batchelor (1962, page 92, paragraph 4). The Fourier components for wave-numbers of magnitude $n \gg 1/L$ [where $L$ is an integral length representing the size of the energy-containing eddies] can be designated as comprising the “small-scale” components of the motion (Batchelor 1962, page 92, paragraph 4). Within this wavenumber range, the centre of dissipation is far from the center of energy, and the decay process involves transfer of energy, by inertial interaction, over a large range of wave-numbers (Batchelor 1962, page 92, paragraph 4). The term ‘the small-scale structure of turbulence’ appeared in Batchelor (1962, page 94, paragraph 1, line 17).

The term ‘small scales of turbulence’ also appeared in the title of a paper by Kim and Antonia (1993). The term ‘small-scale universality’ appeared the titles of papers of Schumacher, Scheel, Krasnov, Donzis, Yakhot and Srinivasan (2014) and Cerbus, Liu, Gioia and Chakraborty (2020).

To be consistent with the original term ‘small-scale turbulence’ in Townsend (1947, page 552, paragraph 2, line 8), the author adopts it throughout.

*Kolmogorov–Oboukhov normal scaling laws* (K41)

Andrei Nikolaevich Kolmogorov (1903 –1987) was a distinguished Russian mathematician (Figure 3, second). “For those interested in the turbulent motion of fluid, Kolmogorov – whom they think of as their Kolmogorov – will always be remembered for the theory of ‘local isotropy’, or universal equilibrium, of the small-scale components of fluid motion that he put forward in 1941”(Batchelor 1990, page 47). How did he become interested in turbulence? Kolmogorov said that “I became interested in turbulent liquid and gas flows at the end of the thirties. From the very beginning it was clear that the theory of random functions of many variables (random fields), whose development only started at that time, must be the underlying mathematical technique. Moreover, I soon understood that there was little hope of developing a pure, closed theory, and because of the absence of such a theory the investigation must be based on hypotheses obtaining in processing experimental data” [Kolmogorov (1985) quoted in Yaglom (1994) and Tikhomirov (1991, page 487)] [Note that the two versions are slightly different perhaps because of the translations.].

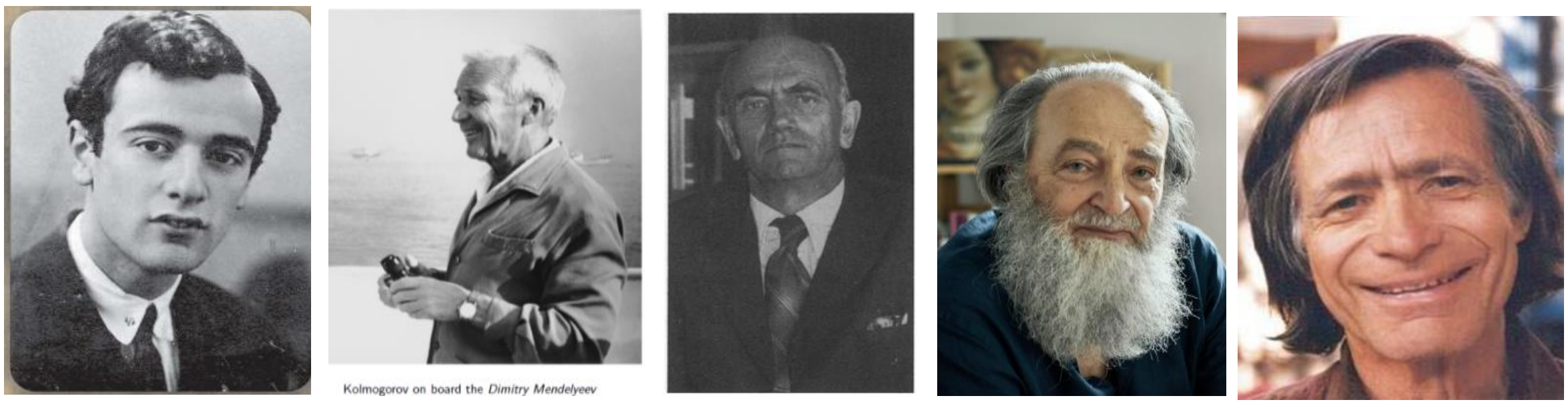

Figure 3 Lev Landau in the late 1920s (first) [Photo Credit: Ryndina 2004]; Andrey Nikolaevich Kolmogorov on board the *Dimitry Mendelyeev* (second) [Photo Credit: Shiryaev 1999, frontipiece] and Alexander Mikhailovich Oboukhov (third) [Photo Credit: Yaglom 1990]; Evgeny Alekseevich Novikov (fourth)[Photo Credit: Evgeny Alekseevich Novikov]; and Robert Harry Kraichnan (fifth)[Photo Credit: Chen, Eyink, Falkovich, Frisch, Orzag and Sreenivasan 2008].

In the author’s view, the vague term ‘for very large Reynolds numbers’, which appeared in the title of Kolmogorov (1941a), might have caused the possible subsequent ambiguous interpretations or even be misleading for both beginners and experts in the field of small-scale turbulence. Kolmogorov should have explicitly used ‘in the limit of Taylor microscale Reynolds number’ so that the conditions on which the laws derived can be clearly or precisely interpreted and understood, in particular, for our clear understanding of the discrepancy between the predictions by K41 and measured results.

Kolmogorov was not being very precise in his language – or at least the English translations thereof. When he says “very high Reynolds number" that in mathspeak is "in the limit of infinite Reynolds number"($Re \to \infty$). Or said another way "in the limit of zero viscosity". Remember (and most students have a lot of trouble with this), the limit of something does not have to be equal to that something evaluated at the limit. Many things are singular in the limit, and not equal to their limit. And this is an example of that. Very high Reynolds number is the same as saying "in the limit as the viscosity goes to zero". This is absolutely not the same as saying "the viscosity equals zero."(William K. George with Personal Communication, 28 June 2020).

Since a fundamental theory is still lacking, it is not possible to tell if this scaling is approximate or exact, nor to know to which observable quantities it can be applied. There is no unambiguous evidence for departures from this scaling in the energy spectrum (Nelkin 1994, page 153, the 2nd paragraph from the bottom).

*Kelvin-Richardson-Onsager cascade*

The early physical mechanism of the distribution of turbulent motion was described in Thomson (later Lord Kelvin) (1887b, page 346, Sections 10 and 11), and further described in Richardson (1922, page 66, paragraph 3, lines 7-9), on which was based by Kolmogorov (1941a, page 12, footnote †† to the second hypothesis; 1941c, page 16, line 8 from the bottom). The word 'cascade', as a physical mechanism to account for the stepwise distribution of energy in turbulence, appeared in Onsager (1945, Abstract, line 21). In the author's view, the appropriate term 'Kelvin-Richardson-Onsager cascade' should be used in the field of small-scale turbulence.

According to Kolmogorov (1941a, page 12, footnote †† to the second hypothesis) and especially Kolmogorov (1941c, page 16, line 8 from the bottom), the graphical sketch of the physical background can be re-constructed and three regions inferred in Figure 4.

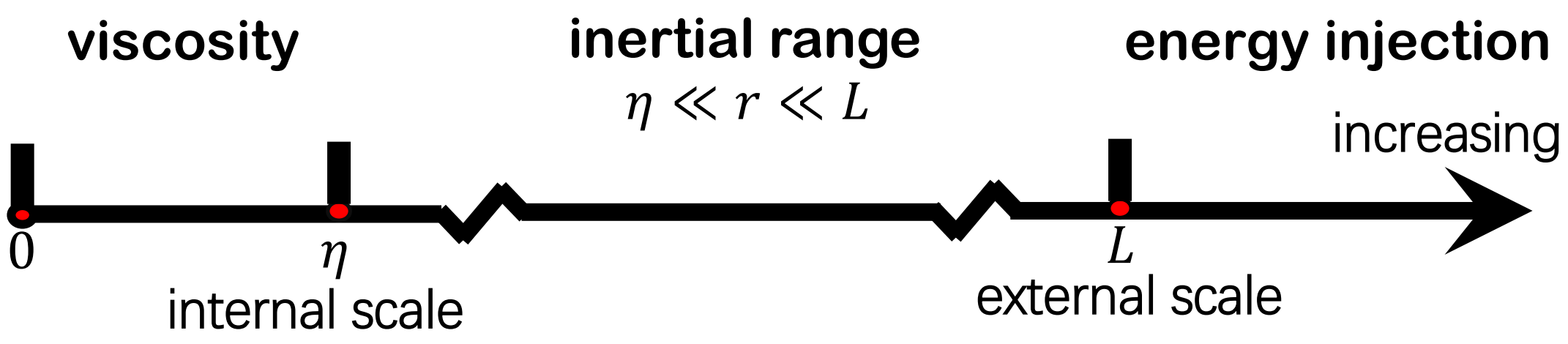

Figure 4 Superimposed on the mean motion are turbulent pulsations of various length scales. It also shows Richardson's idea of the existence in the turbulent flow of vortices on all possible scales between the external scale ($L$)(the Taylor's Lagrangian integral scale) and the internal scale ($\eta$)(the Kolmogorov local scale of turbulence). Three regions: a region dominated by viscosity at the very smallest scales; an intermediate inertial range characterized by zero dissipation; and a region of energy injection (or energy-containing) at the largest scales.

Based on the physical idea shown in Figure 4, Kolmogorov developed his own statistical theory of small-scale turbulence and published 5 papers on turbulence (Kolmogorov 1941a, b, c, 1942, 1962). Alexander Mikhailovich Oboukhov (1918–1989)(Figure 3, third) was a student of Kolmogorov's who published several papers on turbulence (Oboukhov 1941, 1942, 1949a, 1949c, 1962). According to Yaglom (1990, p. vii), the spectral representation of the velocity field itself in Oboukhov (1941) was also considered and used to derive the universal '-5/3 power law'. The major findings of Kolmogorov (1941a, c) and Oboukhov (1941) include:

(i) *The Kolmogorov local scale of turbulence*

Based on the first hypothesis and the dimensional arguments, the following Kolmogorov local scale of turbulence is obtained in Kolmogorov (1941a, page 12, line 14) and Kolmogorov (1941c, page 16, line 6):

$$\eta = \lambda = \nu^{\frac{3}{4}} / \varepsilon^{\frac{1}{4}} \quad (6)$$

where $\eta$ or $\lambda$ is the scale of finest pulsations (Kolmogorov 1941a, page 12, bottom, lines 1-2) or the local scale of turbulence (Kolmogorov 1941c, page 16, line 6). In this paper, to be consistent with the original usage of "the local scale of turbulence" (Kolmogorov 1941c, page 16, line 6), the author adopts the term "the Kolmogorov local scale of turbulence" throughout.

(ii) *Kolmogorov's* $\frac{2}{3}$ *law*

Kolmogorov (1941a, p. 13, equation (23); 1941c, equation (9)) obtained the following equation:

$$B_{dd}(r) \approx C\bar{\varepsilon}^{\frac{2}{3}}r^{\frac{2}{3}} \tag{7}$$

In Kolmogorov's (1941c, foot note) own words, Oboukhov (1941) found the above relation by computing the balance of the energy distribution of pulsations over the spectrum.

(iii) *The Kolmogorov equation*

de Kármán and Howarth (1938, page 206, equation (51)) was rederived in the similar form (Kolmogorov 1941c, page 15, equation (3)). Based on it, in virtue of the condition $(d/dr)B_{dd}(0) = B_{ddd}(0) = 0$, Kolmogorov (1941c, page 15, equation (5)) obtained the following relationship:

$$6\nu \mathrm{d}\frac{B_{dd}}{\mathrm{d}r} - B_{ddd} = \frac{4}{5}\bar{E}r \tag{8a}$$

The above equation can be rewritten as

$$B_{ddd} = -\frac{4}{5}\bar{E}r + 6\nu\mathrm{d}\frac{B_{dd}}{\mathrm{d}r} \tag{8b}$$

This equation is called the Kolmogorov equation (Monin and Yaglom 1975; Frisch 1995; Qian 1998a; Moisy, Tabeling and Willaime 1999), which is an exact relationship between the third-order structure function ($B_{ddd}$) and the second-order structure function ($B_{dd}$) in the limit of Taylor microscale Reynolds number ($Re \to \infty$). Note that the Kolmogorov equation is also called the Kármán-Howarth-Kolmogorov equation (Gotoh, Fukayama and Nakano 2002, page 1171, VI; Kanedo, Yoshino and Ishihara 2008, page 2).

Evgeny Alekseevich Novikov is a Russian mathematician (Figure 3, fourth). As a historical note, in George Keith Batchelor's view, Novikov is "a first rate scientist and mathematician who has made important contributions to turbulence and other areas of fluid mechanics" [George Keith Batchelor's Telegram Steven A. Orzag at MIT dated 14 September 1983, The Batchelor Archive, The Wren Library, Trinity College Cambridge, U.K.]. Novikov (1963) and Novikov and Stewart (1964) proposed the random-force method in the Lagrangian description of turbulence. As an extension of Novikov (1963) to the Eulerian velocity field, Novikov (1965, page 1292, equation (3.11)) obtained the following equation relating the third order and second order velocity structure functions with external correction function:

$$D_3(r) - 6\nu\frac{dD_2(r)}{dr} = -\frac{4}{5}\varepsilon r\left[1 - \frac{5}{14}\left(\frac{r}{L}\right)^2 + O\left(\frac{r}{L}\right)^4\right] \tag{8c}$$

where $D_3(r)$ refers to $B_{ddd}$; $D_2(r)$ to $B_{dd}$. When $r \ll L$, the second and third terms on the right hand side disappear, equation (8c) goes back to the Kolmogorov equation as shown in equations (8a) and (8b).

(iv) *Kolmogorov's* $\frac{4}{5}$ *law*

The following relationship can be easily derived from equation (6b):

$$B_{ddd} \sim -\frac{4}{5}\bar{\varepsilon}r \tag{8d}$$

The above $\frac{4}{5}$ law (Kolmogorov 1941c, page 19, Relation (7), in Russian) is one of the only exact results of the theory of three-dimensional homogeneous, isotropic turbulence in the limit of large Reynolds numbers ($Re \to \infty$).

(v) *The Kolmogorov–Oboukhov inertial range spectrum*

A Fourier transform yields the form of the equilibrium spectrum (Batchelor and Townsend 1947, page 240, equation (2.1)):

$$E(k) = A\epsilon^{2/3}k^{-5/3} \quad (9)$$

where $A$ is an absolute constant, $k$ is a wave-number in the range in which the effects of viscous forces on the spectrum first become comparable with the effects of inertia forces. The above -5/3 law is actually the spectral equivalent of the 2/3 law in Kolmogorov (1941a, page 303, Equation (23))(Monin and Yaglom 1975, page 355), i.e. the energy spectrum's equivalent of second-order structure function. Equation (5) is also called the Kolmogorov law or Kolmogorov inertial range law (Qian 1983, pages 2101-2102) or the Kolmogorov–Oboukhov inertial range spectrum (Barenblatt and Chorin 1998; Arena and Chorin 2006), the Kolmogorov–Oboukhov scaling laws (Barenblatt and Chorin 1998, page 445, Abstract, line 5).

(vi) *The skewness factor* ($S$)

The skewness factor ($S$) of the probability distribution of the velocity difference $(u'_d - u_d)$ is written as (Kolmogorov 1941c, page 15, equation (8)):

$$S = B_{ddd} : B_{dd}^{\frac{3}{2}} \quad (10a)$$

Kolmogorov did not relate it to the physics of small-scale turbulence. However, the skewness is a measure of vortex stretching (Frisch 1995, page 156, 3rd paragraph, line 1).

Batchelor (1947, page 545, equation (5.6)) rewrote the skewness factor $S(r)$:

$$S(r) = \frac{\overline{(u'_d - u_d)^3}}{[\overline{(u'_d - u_d)^2}]^{\frac{3}{2}}} = \frac{B_{ddd}(r)}{[B_{dd}(r)]^{\frac{3}{2}}} \quad (10b)$$

For large $r$, Kolmogorov (1941c, page 16, relation (10) defined

$$C = (-4/5S)^{\frac{2}{3}} \quad (11a)$$

Batchelor (1947, page 545, paragraph 2, line 5) rewrote it:

$$S(r) = -\frac{4}{5}C^{-\frac{3}{2}} \quad (11b)$$

Interestingly, $S(r)$ may be constant throughout the range $0 \ll r \ll L$, but there is no theoretical support for this speculation (Batchelor 1947, page 554, paragraph 3, lines 6-7).

*The 'flattening factor' or flatness factor*

The terminology 'flattening factor' first appeared in Townsend (1948a, page 561, line 6). Its mathematical expression is as follows (Townsend 1948a, page 561, line 7):

$$\overline{\left(\frac{\partial u}{\partial x}\right)^4} \Big/ \left[\overline{\left(\left(\frac{\partial u}{\partial x}\right)\right)^2}\right]^2 \quad (12a)$$

Townsend (1948a, page 561, lines 4-7) wrote that "Thus $S(0)$, the skewness factor of $\partial u/\partial x$, is an absolute constant independent of the flow, and this invariancy holds for the mean value of all non-dimensional functions of $\frac{\partial u}{\partial x}$, in particular for the 'flattening factor'".

Batchelor and Townsend (1949, page 248, paragraph 3, line 3) redefined the flattening factor ($\alpha_n$):

$$\alpha_n = \overline{\left(\frac{\partial^n u}{\partial x^n}\right)^4} \Big/ \left[\overline{\left(\left(\frac{\partial^n u}{\partial x^n}\right)\right)^2}\right]^2 \quad (12b)$$

*The inertial subrange (or range)*

In his original papers Kolmogorov (1941a, b, c) did not mention ‘inertial range’. The so called ‘inertial range’, independent from viscosity, is partially hidden in Kolmogorov (1941a, page 12, footnote †† to the second hypothesis) and Kolmogorov (1941c, page 16, line 8 from the bottom). In the author’s view, the term ‘the inertial range’ was first interpreted from Kolmogorov (1941a) by Batchelor and Townsend (1949, page 240, Section 2) who wrote that “In his original paper, Kolmogoroff [Kolmogorov] (1941a) postulated that when the Reynolds number of the turbulence is high enough there is a subrange of the equilibrium range in which conditions are not affected by viscosity. The physical idea behind this postulate is that the motion associated with a small range of wave-numbers has its own Reynolds number, and at smaller values of $k$ for which this characteristic Reynolds number is very large the motion is dominated by inertia forces”.

The term ‘the inertial subrange’ clearly first appeared in Batchelor and Townsend (1949, page 241, paragraph 1, lines 2–3) “…, and the form of the triple-velocity correlation found from the same postulates about the inertial subrange,...”, and also as the subtitle of Section 6.5 in Batchelor’s monograph (Batchelor 1953, page 121). The term ‘inertial range’ appeared in Tchen (1954, page 4, right column, paragraph 3, lines 9-10) and the title of Yakhot, She and Orszag (1989). Apparently, the term ‘the inertial subrange’ or ‘the inertial range’ has been used ever since. It can refer to (a) the range where the viscous effect is negligible (Kolmogorov 1941a, Batchelor and Townsend 1949), and corresponds to $T(k) = 0$ in a stationary turbulence (Monin and Yaglom 1973), and (b) the inertial wave-number range is defined as the range where $E(k) \sim k^{-5/3}$. Notably, (a) is the original idea of Kolmogorov (1941a) and refers to the real ‘inertial range’; (b) is the so called ‘inertial range’, actually ‘scaling range’. However, as Qian pointed out later, the ‘scaling range’ is not the same as Kolmogorov’s ‘inertial range’.

K41 was supported by laboratory experimental evidence (Townsend 1948a, b). Corrsin (1949) provided an experimental verification of local isotropy and he was credited for the first serious experimental verification of the concept of local isotropy. Stewart and Townsend (1951) measured the double and triple velocity correlation functions and of the energy spectrum function in the uniform mean flow behind turbulence-producing grids of several shapes at mesh Reynolds numbers between 2000 and 100000. They used those results to assess the validity of the various theories which postulate greater or less degrees of similarity or self-preservation between decaying fields of isotropic turbulence. They found that (a) the conditions for the existence of the local similarity considered by Kolmogoroff [Kolmogorov] and others are only fulfilled for extremely small eddies at ordinary Reynolds numbers, (b) the inertial subrange in which the spectrum function varies as $k^{-5/3}$ is non-existent under laboratory conditions; (c) spectrum function is best represented by an empirical function such as $k^{-a\ logk}$ within the range of local similarity, (d) all suggested forms for the inertial transfer term in the spectrum equation are in error (Stewart and Townsend 1951, page 359). K41 was then supported by tidal-channel turbulence field measurements (Grant, Moilliet and Stewart 1959, page 809, Fig. 1; Grant, Stewart and Moilliet 1962) and a round air jet (Gibson 1962).

In his review of Kolmogorov’s inertial theories, Kraichnan (1974, page 306, paragraph 3, lines 1-6) argued that “Kolmogorov’s 1941 theory has achieved an embarrassment of success. The ‘$-5/3$’–spectrum has been found not only where it reasonably could be expected but also where it reasonably could be expected but also at Reynolds numbers too small for a distinct inertial range to exist and in boundary layers and shear flows where there are substantial departures from isotropy, and such strong effects from the mean shearing motion that the stepwise cascade appealed to by Kolmogorov is dubious".

George and Gamard (1999) used Near-Asymptotics to explain the wind tunnel data of Mydlarski and Wahrhaft (1996) which showed a clear dependence of the inertial range spectral power on $R_\lambda$. In the same vein, by matched asymptotic expansions, Lundgren (2002, 2003, 2005) proved that Kolmogorov (1941a, c) is a statistical solution of the Navier-Stokes equation. Jiménez (2004) presented a short historical account of turbulence theory before Kolmogorov, his contributions, and his inspirational influence on the understanding of turbulence in fluids. However, K41 was questionable (e.g. George 1992, Long 2003, George 2013).

Edwards and McComb (1971) developed a local energy transport equation for isotropic turbulence, and they showed the resulting equation yields the Kolmogoroff distribution for the case of the infinite Reynolds number limit. Edwards and McComb (1972) further applied it to a simple turbulent shear flow. McComb (1974) obtained a new equation for the turbulent effective viscosity. Furthermore, McComb (1976) applied it to derive the inertial-range spectrum. McComb (1978) derived second-order equations for the turbulent velocity-field correlation and propagator functions. Based on the local-energy-transfer (LET) theory (McComb 1978), McComb and Shanmugasundaram (1984) and McComb, Shanmugasundaram, and Hutchinson (1989) studied decaying isotropic turbulence and velocity-derivative skewness and two-time velocity correlations of isotropic turbulence. McComb and Yoffe (2017) further studied the local energy transfer (LET) theory of homogeneous turbulence.

Using a time-ordered fluctuation-dissipation relation, McComb and Kiyani (2005) studied Eulerian spectral closures for isotropic turbulence. McComb (2009) studied scale-invariance and the inertial-range spectrum in three-dimensional stationary, isotropic turbulence, and he obtained the energy spectrum in the limit of infinite Reynolds number with the specific prefactor.

*Intermittency in the turbulent fluctuations and small-scale intermittency of turbulence*

What is the small-scale intermittency of turbulence? What is its physical mechanism? As shown in Figure 5, Arthur Fage (1890–1977) spent the whole of his academic life at the National Physical Laboratory in Teddington where turbulence then was the arcane topic to which Fage was to devote some 20 years of research (Collar 1978). Fage used a hot-wire anemometer and an ultramicroscope for turbulence studies (Fage and Johansen 1927; Fage and Townend 1932). Those findings by Fage and Townend (1932) stimulated G.I. Taylor (Collar 1978, page 42, paragraph 1, lines 15-18). From a historical perspective, Fage pioneered in (1) the experimental investigations of the changes of turbulent characteristics with Reynolds number and (2) the decay of turbulence behind a narrow prism set across a square pipe. Fage's apparatus, a hot-wire anemometer, had directly inspired Hugh Dryden, indirectly Stanley Corrsin, and Albert Alan Townsend. Stanley Corrsin (1920–1986) was an American aeronautical engineer. Albert Alan Townsend (1917-2010) and George Keith Batchelor (1920–2000) were the British applied mathematicians and fluid dynamicists. Corrsin (1942, 1943) started his turbulence work at Caltech before moving to the Johns Hopkins University when the new Aeronautical Department was started after the war by Francis Clauser (also at Cal Tech). Corrsin wrote to Batchelor when he had just been in England for about five months (Corrsin's Letter to Batchelor September 5, 1945; Batchelor's letter to Corrsin 17th October 1945, the Batchelor Archive, the Wren Library, Trinity College Cambridge, U.K.). According to Townsend (1948b, page 173, lines 2-7), Corrsin (1943) already observed "intermittency" in his investigation of flow in an axially symmetrical heated jet of air. He explained that "his oscillographic observations [Corrsin 1943, page 49, Figure 27 and Figure 28] of intermittency in the turbulent fluctuations as due to a phenomenon similar to transition in a laminar boundary layer, and classified the flow into a fully turbulent core, an intermittent 'transition' region and a laminar core." The notion 'not continuously turbulent' in Corrsin (1943, page 27, paragraph 2, line 7) might be interpreted as 'intermittent' by Townsend (1948b, page 173, lines 2-7).

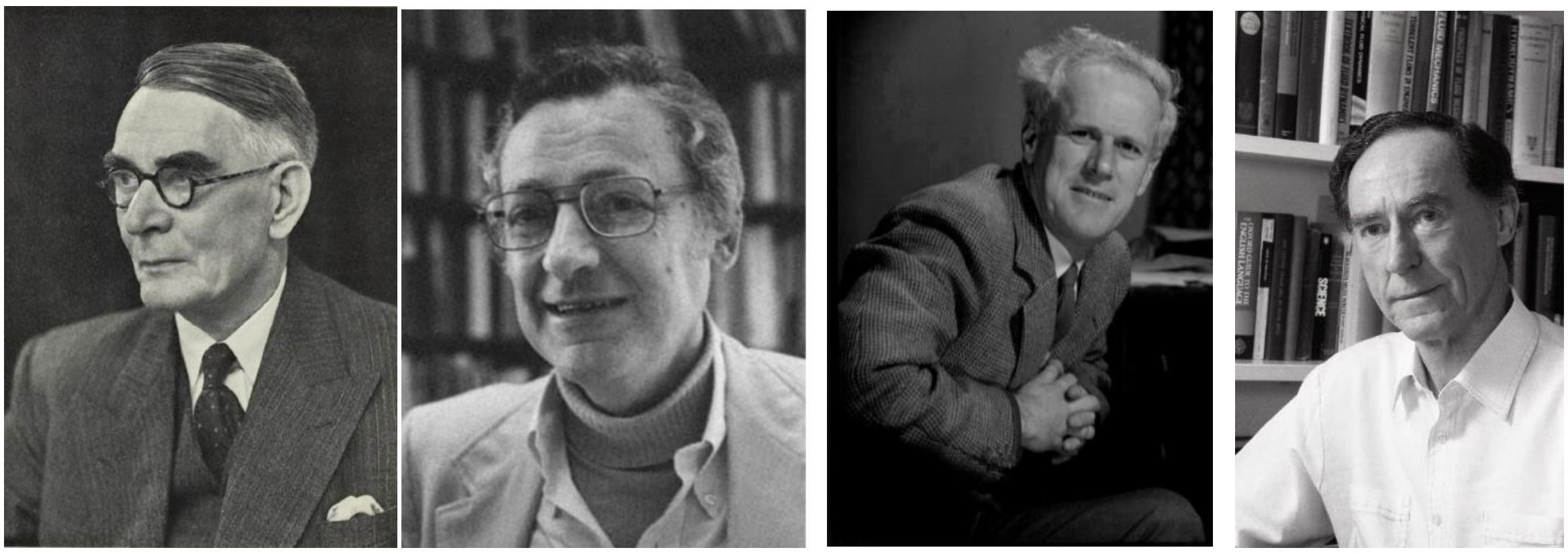

Figure 5 Arthur Fage (1890-1977)[Photo by Lafayette of London: Collar 1978]; Stanley Corrsin (1920–1986) (left) [Photo Credit: https://www.nae.edu/29545.aspx]; Albert Alan Townsend (1917–2010)(middle)[Photo by Antony Barrington Brown, the National Portrait Gallery of London, U.K.] and George Keith Batchelor (1920–2000)(right)[Photo Credit: the Wren Library, Trinity College Cambridge, U.K.]

As shown in Figure 6a, it is interesting to see that the similar oscillograms were used/obtained by Fage and Johansen (1927, Plate 7), Dryden and Kuethe (1929, page 380, FIGURE 18), Corrsin (1943, page 49, Figure 27). Furthermore, Corrsin (1943), Townsend (1948b) and Batchelor and Townsend (1949)(Figure 6b), which were among those early papers to test K41, interpreted their 'intermittency' based on the similar oscillograms.

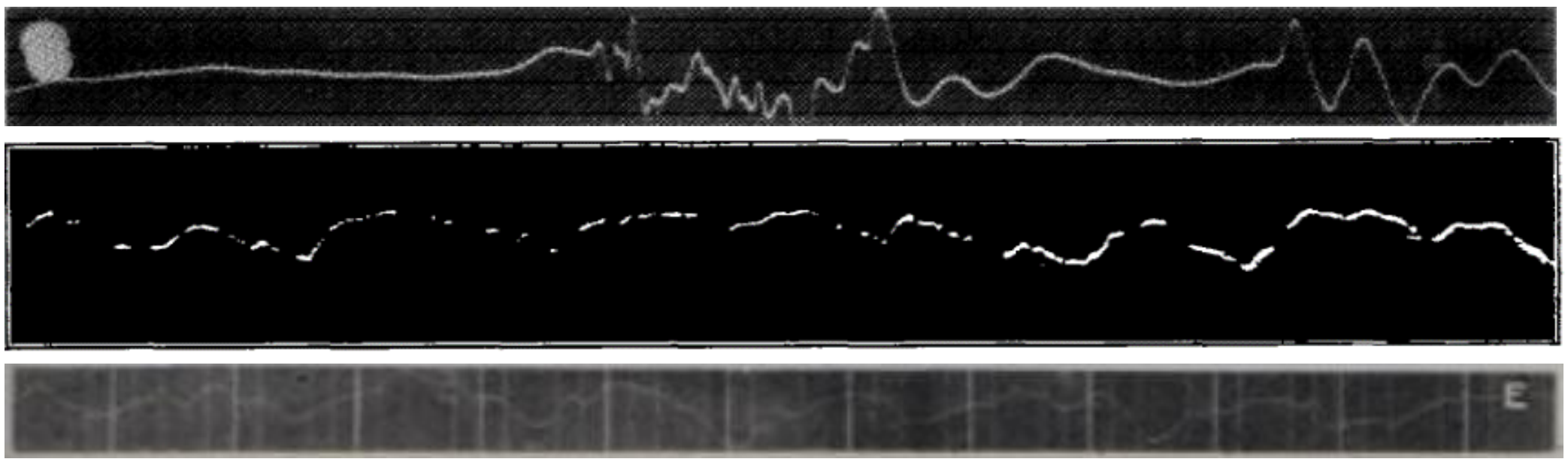

Figure 6a The snapshots of oscillograms used in Fage and Johansen (1927, Plate 7)(bottom); Dryden and Kuethe (1929, page 380, FIGURE 18) (middle); and Corrsin (1943, page 49, Figure 27)(top).

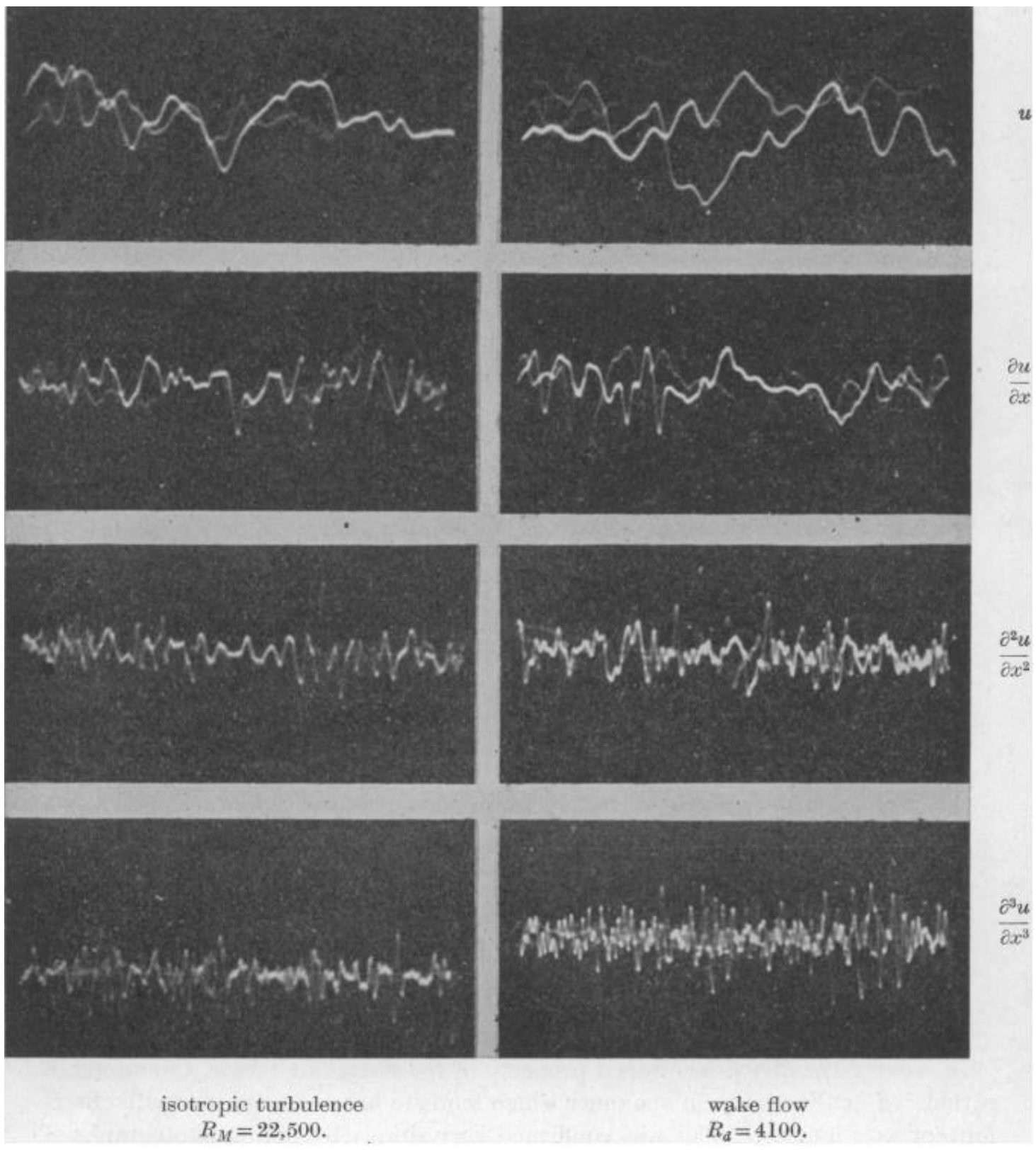

Figure 6b Oscillograms of fluctuations of $u$, $\partial u/\partial x$, $\partial^2 u/\partial x^2$ and $\partial^3 u/\partial x^3$ for two cases: left: isotropic turbulence at $R_M$=22,500 and right: wake flow at $R_d$=4100 (Batchelor and Townsend 1949, page 250, Figure 6).

Following up Corrsin (1943), Townsend (1948b) first studied the small-scale intermittency of turbulence when he proved local isotropy of K41 in the turbulent wake of a cylinder. The word 'intermittent' appeared 6 times in Townsend (1948b, page 161, Abstract, line 10; page 172, bottom, lines 2 and 9; page 173, paragraph 1 lines 12 and 14; page 173, bottom, line 2). The word 'intermittency' or the term 'intermittency factor' appeared more than 4 times in Townsend (1948b, page 161, Abstract, line 12; page 170, Figure 9's caption; page 171, Figure 10's caption; page 172, Figure 11's caption). The term 'intermittency factor' was expressed by using $\gamma$ to denote the mean fractional duration of turbulent flow at any given point (Townsend 1948b, page 172, lines 3-4). Townsend (1948b, page 172) defined the following relations:

$$S_0(u) = \gamma^{-\frac{1}{2}} S_0'(u);\ T_0(u) = \gamma^{-1} T_0'(u);\ T_0(v) = T_0(w) = \gamma^{-\frac{1}{2}} T_0'(v)$$

where $S_0(u)$ is the skewness factor; $T_0(u)$, $T_0(v)$, and $T_0(w)$ the flattening factors; $S_0'(u)$, $T_0'(u)$, and $T_0'(v)$ are the values obtained in steady homogeneous turbulent flow.

The word "intermittent" appeared 2 times in Batchelor and Townsend (1949, page 249, bottom, line 2, last word; page 252, line 7, 1st word). Note that this finding is different from Footnote 1 in McComb (2014, page 144). The following five major findings can be found in Batchelor and Townsend (1949, page 238, Abstract): (i) the probability distributions of $\partial u/\partial x$, $\partial^2 u/\partial x^2$ and $\partial^3 u/\partial x^3$ show that the energy associated with large wave-numbers is very unevenly distributed in space (Figure 6); (ii) there appear to be isolated regions in which the large wave-numbers are 'activated', separated by regions of comparative quiescence; (iii) this spatial inhomogeneity becomes more marked with increase in the order of the velocity derivative, i.e. with increase in the wave-number; (iv) it is suggested that the spatial inhomogeneity is produced early in the history of the turbulence by an intrinsic instability, in the way that a vortex sheet quickly rolls up into a number of strong discrete vortices; (v) thereafter the inhomogeneity is maintained by the action of the energy transfer.

The inference of the relationship between the flattening factor and the intermittency of turbulence can also be seen in Batchelor and Townsend (1949, page 249, bottom) who wrote that "In view of the above-mentioned property of the flattening factor, the inference is that $\partial^n u/\partial x^n$ fluctuates in a manner which tends to become more markedly intermittent as $n$ increases". Batchelor and Townsend (1949, page 252, paragraph 1) further discussed that the fluctuations in flattening factors appear to be intrinsic to the equilibrium range of wave-numbers, i.e. the fluctuations are small in the region of the smallest wave-numbers of equilibrium range and becomes increasingly large at larger wave-numbers. Batchelor and Townsend (1949, page 252, paragraph 2) gave an immediate suggestion that space is divided into a number of regions each of which has its own value of the energy dissipation $\epsilon$ and in each of which the energy spectrum has the universal form (1.8). Clearly, it is Batchelor and Townsend (1949) who first linked the flattening factor with the intermittency of turbulence and then the energy dissipation $\epsilon$.

In summary, based on Batchelor and Townsend (1949), the intermittency of turbulence, mathematically, refers to (a) the spatial uneven distributions of the probabilities of the velocity derivatives, the first order ($\partial u/\partial x$), the second order ($\partial^2 u/\partial x^2$), the third order ($\partial^3 u/\partial x^3$) ; (b) the spatial fluctuations of the flattening factors; physically, it refers to (i) the spatial unevenly distributions of the turbulent energy associated with large wave-numbers; (ii) the spatial fluctuations of the energy dissipation $\epsilon$ within each of a number of regions with the universal form of the energy spectrum. Although Batchelor and Townsend (1949) did not specify it, the occurrence of the intermittency of turbulence is the $R_\lambda$-dependent, i.e. its effect depends on $R_\lambda$.

In the author's view, there could be two interpretation options for those so-called 'intermittency' data in Batchelor and Townsend (1949, page 250, Figure 6) (Figure 6b): (i) the intermittency correction should be made to K41; (ii) it may be due to the finite $R_\lambda$ effect of turbulence without any corrections to K41. It is hard for the author to understand why Corrsin, Townsend, Batchelor, Landau, Kolmogorov and Oboukhov did not recognize the following two points: (a) the mesh Reynolds number ($R_M$=22,500) and the cylinder (diameter) Reynolds number ($R_d$=4100) in Batchelor and Townsend (1949, page 250, Figure 6) were finite ones and not really in the limit of infinite Taylor microscale Reynolds numbers ($Re \rightarrow \infty$); (b) the intermittency of turbulence did exist but might be due to the finite $R_\lambda$ effect of turbulence. Apparently, it will be shown that these have been fueling controversy debate ever since.

Frenkiel and Klebanoff (1975, page 137, Introduction, paragraph 2, lines 7-10) drew our attention to two important points: (i) 'intermittency' introduced by Batchelor and Townsend (1949) actually refers to 'small-scale intermittency', which should be distinguished from the intermittencies associated with the outer boundaries of jets, wakes, and boundary layers; (ii) Batchelor and Townsend (1949) studied grid turbulence at low turbulence Reynolds numbers. 'Small-scale intermittency' was also explicitly highlighted in the paper title of Benzi and Vulpiani (1980) and Sreenivasan (1995a).

One may argue that so-called 'intermittency' by Corrsin (1943) and particularly Townsend (1948b) was different from that by Batchelor and Townsend (1949). Corrsin (1943) first discovered the outer intermittency (i.e., the existence of a stochastic interface at the edge of a free shear flow). Townsend (1948b) found a way of quantifying intermittency by means of an intermittency factor. The next most important step was the Corrsin and Kistler (1952) NACA report. The next most important thing was the characterization of the interface as a stochastic fractal (Sreenivasan and Meneveau 1986). The first discovery of small-scale intermittency in grid turbulence was due to Batchelor and Townsend (1949). This is very different from the intermittency discussed above. After some attempts by Corrsin's group, along with a few other speculative results by Tennekes and Saffman, the next most important step was its characterization by multifractals (Meneveau and Sreenivasan 1998), following Mandelbrot (1982) and the Chicago work on dynamical systems (Halsey *et al.* 1986).

*Kolmogorov–Oboukhov anomalous scaling laws* (K62)*: intermittency corrections to* K41

What motivated Kolmogorov to refine K41, i.e. making intermittency corrections to K41? Lev Landau (Figure 3, first), a Russian Nobel Laureate in Physics, was also interested in the problem of turbulence (Landau 1944). Landau made a remark on K41 (in a footnote in the 1944 Russian version of Landau and Lifshitz (1959)), i.e. the universality $C_K[A]$ in equation (5). The essence of Landau's point is that $C_K[A]$ is not invariant to the composition of sub-ensembles because the left-hand side of equation (5) is an average while the right-hand side is the $2/3$ power of an average. It can also be inferred that the rate of energy dissipation is intermittent, i.e. spatially inhomogeneous, and cannot be treated as a constant (Arenas and Chorin 2006, paragraph 2).

In his own words (Kolmogorov 1962 [K62], page 82, paragraph 2), Kolmogorov was motivated by the fact that "Landau noticed that they [Kolmogorov and Oboukhov] did not take into account a circumstance which arises directly from the assumption of the essentially accidental and random character of the mechanism of transfer of energy from the coarser vortices to the finer: with increase of the ratio $L: l$, the variation of the dissipation of energy $\epsilon$ should increase without limit." To the best knowledge of the author, Landau's remark was inferred as that K41 fails to describe intermittency effects (e.g. She, Jackson and Orzag 1991).

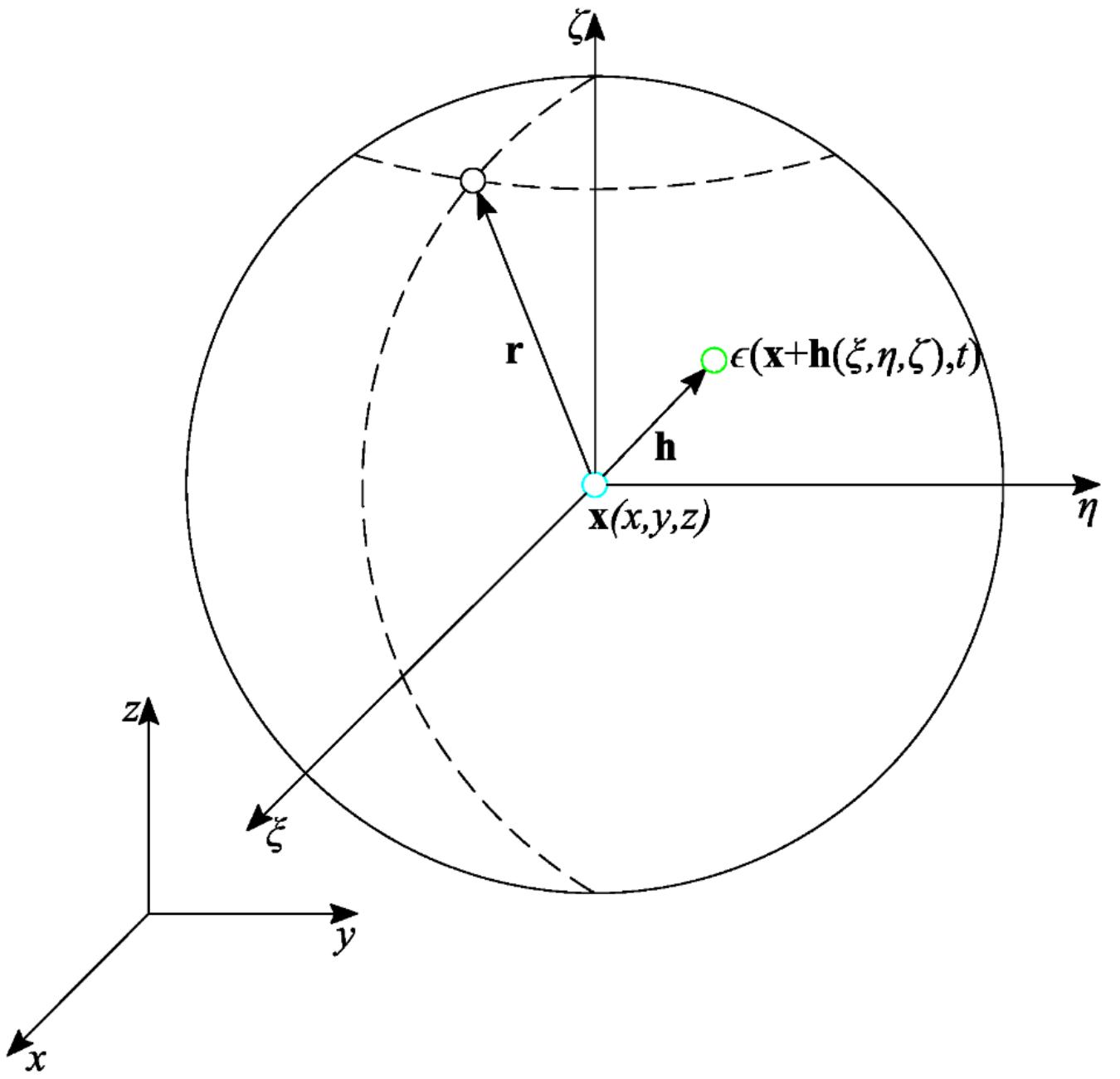

Figure 7 A local dissipation $[\epsilon_r(\mathbf{x}, t)]$ averaged over a sphere of radius $(r)$.

Kolmogorov (1962) proposed that the similarity hypothesis should be modified to use a local dissipation $[\epsilon_r(\mathbf{x}, t)]$ averaged over a sphere of radius $(r)$(Figure 7), instead of the mean dissipation $(\epsilon)$(Kolmogorov 1962, page 82, bottom):

$$\epsilon_r(\mathbf{x}, t) = \frac{3}{4\pi r^3} \int_{|h|\le r} \epsilon(\mathbf{x}+\mathbf{h}, t)\, d\mathbf{h} \tag{13}$$

He further assumed the probability distribution of the dissipation ($\epsilon$) to be log-normal, and calculated the resulting scaling exponent ($p$)(Kolmogorov (1962, page 456, equation (10)):

$$\overline{|\Delta_{dd}(r)|^p} = C_p(\mathbf{x}, \mathrm{t})\ (\mathrm{r}\bar{\varepsilon})^{\frac{p}{3}} \left(\frac{L}{r}\right)^{\frac{1}{2}kp(p-3)} \tag{14}$$

Clearly, the anomalous scaling exponent, i.e. $\frac{1}{2}kp(p-3)$, appears in equation (14). Thus, K62 also refers to the anomalous scaling law. The scaling exponent ($p$) was often inferred as the intermittency exponent ($\eta$) (McComb 1990) or the intermittency parameter (e.g. Wang, Chen, Brasseur and Wyngaard 1996). The 'anomalous scaling exponent' may be inferred as 'multiplicative cascades' or 'multiplicative cascade' may be implicit in the 'anomalous scaling exponent'(e.g. Jiménez 2000).

Based on the grid-turbulence experimental data, Kuo and Corrsin (1971) found that the square of the signal associated with large wave-numbers may be approximated by a log-normal probability distribution for amplitudes when probabilities fall between 0.3 and 0.95, in limited agreement with the theory of Kolmogorov (1962). The extensive data in Anselmet, Gagne, Hopfinger and Antonia (1984, page 86, Figure 14) were ever interpreted as the strongest experimental evidence supporting K62. Anselmet, Gagne, Hopfinger and Antonia (1984) has been one of the most-cited papers in the field of small-scale turbulence after Kolmogorov (1962). Numerous various types of turbulent intermittency models have been proposed (e.g. Novikov and Stewart 1964; Novikov 1971; Frisch, Sulem and Nelkin 1978; Siggia 1981; Kerr 1985; She, Jackson and Orszag 1988; She, Jackson and Orszag 1990); She 1991a, b; She and Orszag 1991; She, Jackson and Orszag 1991; Lohse and Grossmann 1993; Dubrulle and Graner 1996; Wang, Chen, Brasseur and Wyngaard 1996; Birnir 2013; Dubrulle 2019). For the early reviews, see Borgas (1992) and Sreenivasan and Antonia (1997). According to Eyink and Sreenivasan (2006, page 87, Abstract, lines 14-15), Onsager's private notes of the 1940s contains the modern physical picture of spatial intermittency of velocity increments, explaining anomalous scaling of the spectrum.

Robert Harry Kraichnan (Figure 3, fifth) was an important figure in the field of small scale turbulence. EUROMECH Colloquium 512 Small Scale Turbulence and Related Gradient Statistics was dedicated to Kraichnan [http://www.euromech512.polito.it/dedication/robert_kraichnan/]. Kraichnan (1991, page 70, paragraph 3) summarized the following three possible sources of intermittency corrections to K41: (a) non-Gaussian statistics of the energy-containing scales that influence the statistics of the inertial-range cascade; (b) fractal-like [multiplicative] buildup of intermittency during the cascade; (c) intermittency effects intrinsic to the dissipation range. He further argued that at the finite Reynolds numbers of experiments and simulations, separation of these various effects is a non-trivial problem, insufficiently recognized in the literature. Arenas and Chorin (2006) found that (a) it is OK for low order structure functions not to make the intermittency corrections; (b) the scaling for high order structure functions must depend on the Reynolds number; and (c) it is necessary to explain the behavior of higher structure functions.

*The controversy about K62*

The controversy about K62 began with Novikov and Stewart (1964) and Novikov (1971), and then Kraichnan (1974), Saffman (1977), Qian (1986a), and Sreenivasan (1999). For example, Novikov (1971) noted that this part of the argument is incorrect. The central limit theorem (CLT) only applies to the central part of the distribution of the logarithms, not to the tails, and everything interesting for the log-normal has to do with the tails (e.g. moments). In his brief review of Novikov (1971), Jiménez (2004) did mention that Novikov (1971) challenged Kolmogorov's (1962) log-normal argument on the grounds that it relied on the application of the central-limit theorem to extreme events. Jiménez (2004, page 603, lines 3-6) wrote that "However, even if *flawed* in detail, Kolmogorov's 1962 paper had an immense influence, and intermittency continues to be an active research field which has generated some of the most beautiful theoretical results in turbulence theory".

Notably, the strong word '*flawed*' was used above. Did Novikov (1971) and Jiménez (2004) really start doubting about Kolmogorov (1962)? Jiménez (2000) gave a more detailed discussion about intermittency and cascades. This particular issue is well known, and is discussed in many places (for example, the book by Frisch 1995). The argument in Kolmogorov (1962) was that the energy cascade is multiplicative (probably approx. OK), so that the logarithms of the inertial-range energy flux are additive (true), and that, applying the central limit theorem, the distribution of the logarithms should be normal (and the dissipation therefore log-normal)(Javier Jiménez with Personal Communication, 15 March 2020). K62 is not "fully" flawed (few things are). It introduces intermittency, the detailed Kolmogorov similarity and multiplicative processes. All of them remain useful. He just extracted the wrong conclusions (Javier Jiménez with Personal Communication, 16 March 2020).

To testify the refined similarity hypothesis of Kolmogorov (1962), a number of studies have been undertaken to measure the correlation coefficients between the velocity difference across a distance $r$ ($|\Delta u_r|$ (or $\Delta u_r$) ) and the local average dissipation over the scale $r$ ($\epsilon_r$ [or $(r\epsilon_r)^{1/3}$] ) of the high-Reynolds-number turbulence. Three intermittency models, i.e. (a) log-normal (Kolmogorov 1962); (b) $\beta$ model (Frisch, Sulem and Nelkin 1979); (c) She and Leveque's (1994), are used to examine how the correlation coefficients change with $r$ in the inertial range of Kolmogorov (1962). Qian (1996a) found that the experimental data of correlation coefficients contradict consequences of Kolmogorov (1962) for three intermittency models. Qian (1996a) also re-examined whether the experimental data of the correlation coefficients (Stolovitzky, Kailashnath and Sreenivasan 1992; Praskovsky 1992; Thoroddsen and Van Atta 1992; Zhu, Antonia and Hosokawa 1995) really supported the refined similarity hypothesis of Kolmogorov (1962). He concluded it did not. Two possible findings can be inferred: (a) $V = \Delta u_r/(r\epsilon_r)^{1/3}$ depends on $(r\epsilon_r)^{1/3}$ in the inertial range captured in the experiments while $\epsilon_r$ is represented by its 1D surrogate; (b) this disagreement may reflect the problem with the intermittency models and even the anomalous scaling law.

*The renormalization group*

Forster, Nelson and Stephen (1976, 1977) pioneered renormalization work on fluid motion (NOT turbulence). McComb (1982) and McComb and Shanmugasundaram (1983) did the first nontrivial application to turbulence and this was later developed into the two-field theory with conditional averaging. There are number of problems in science which have, as a common characteristic, that complex microscopic behavior underlies macroscopic effects (Wilson 1983). Fully developed turbulent fluid flow is listed in the problems of the category where fluctuations persist out to macroscopic wavelengths, and fluctuations on all intermediate length scales are important too (Wilson 1983). Yakhot and Orszag (1986, 1987) and Yakhot, Orszag and She (1989) simply took Forster, Nelson and Stephen (1976, 1977) and made ad hoc modifications to it in order to claim a theory of turbulence. Their work aroused a great deal of controversy in the 1990s but found applications in engineering modelling. In his review article, McComb (1995) discussed renormalization group for fluid turbulence. The work done by Qian (1983, 1990) was more sympathetic to it than in McComb (1990). A review of this subject was made in Zhou (2010).

*A few historical notes related to K41 and K62*

Firstly, we need to re-look at how Batchelor was thinking of K41. Batchelor (1947) first introduced the work of Kolmogorov (1941a, c) to the West. Why did Batchelor himself stop working on it after the Marseilles symposium of 1961? It was mainly because he concluded that the study of turbulence had reached an impasse. At the Marseilles Symposium, Batchelor (1962, pages 92-95) used 3 pages to comment on The small-scale properties of the turbulence. Batchelor (1962, page 92, paragraph 4, lines 13-14) did express his more concern with the validity of K41 in the light of current ideas and the available data. In his own words, Batchelor (1962, pages 93-94) wrote:

> So far as formal deduction is concerned, the theory is as much of a hypothesis as it was at the beginning. Despite the extreme simplicity and generality of the hypothesis, no deductive analytical argument has yet shown it to be true or false, and I know of no line of argument which gives any promise of doing so. However, the rational and physical content of the theory have been given a great deal of thought, and I think it is fair to say that it grows more and more appealing. It was an indication of the physical plausibility of the theory that von Weizsäcker (1948), and Onsager (1945), and I understand Oboukhoff [Oboukhov] (1941) also, should have independently conceived the same basic idea, or one very close to it, and since then the ideas of the theory have gained wide acceptance. The universal similarity theory of the small-scale components has an intrinsic naturalness and "rightness", and stands as one of the landmarks in the development of fluid mechanics. This is not to say that I cannot conceive of the theory being wrong; I am asserting that, relative to the state of knowledge in 1941 (and also in 1961), the universal similarity theory makes possible a great jump forward in our understanding of turbulence, and that, if it should prove to be wrong in whole or in part, the reasons for this would be almost as interesting as the theory itself.

Clearly, in the author's view, Batchelor was very cautious about the universal similarity theory, i.e. he was not so sure whether it was right or wrong.

Secondly, Julian C.R. Hunt shared his following memories with the author:

> I discussed Kolmogorov theory with George K. Batchelor in 1991, which is reported in Hunt and Vassilicos (1991). The relevant point on page [pages] 186-188, was that in Batchelor (1953), he did not include the very important result on third moments structure function -which had no arbitrary constant. When I asked him about this omission in his 1953 book [Batchelor 1953], though it had been included in this 1947 paper [Batchelor 1947], and was included in Landau and Lifshitz (1960)[1959], GKB said to me that "between 1947 and 1953 the results of experimental studies indicated that high Re for turbulence was unlikely to be measurable. He now admits he was too pessimistic". The other omission in K41 was the omission of reference to LF Richardson's name as the author of the physics of cascade. But this was corrected in K62 (Personal Communication with Julian C.R. Hunt, 01 December 2019)

It can be inferred from what Batchelor said to Hunt that Batchelor was fully aware of the finite Reynolds number turbulence measured in the laboratory and field before he wrote his famous monograph (Batchelor 1953). However, it is not so easy to understand what he meant by saying 'he was too pessimistic'. Was Batchelor pessimistic about K41 because of the finite Reynolds number turbulence data measured? In other words, was Batchelor pessimistic about the fact that the finite Reynolds number turbulence data measured could not prove K41 is correct since K41 is valid in the limit of infinite Reynolds number?

As a digression, in conversation with the author Hunt ever shared the following anecdote:

> After Batchelor found Kolmogorov's papers (French versions) on turbulence from the library, he happily went to the DAMTP and showed them to Taylor. Surprisingly, Taylor said to Batchelor that he was not interested in them. Nevertheless, Batchelor was not discouraged by Taylor's attitude towards Kolmogorov's papers and even did not want to mention this in his book about Taylor after Hunt's comment (Personal Communication with Julian Hunt, Trinity College Cambridge, U.K., 2 December 2019).

As another digression, in conversation with the author Keith Moffatt recalled that "I [Keith], as a newly-appointed Assistant Editor for the *Journal of Fluid Mechanics,* edited the English of Kolmogorov's (1962) JFM paper, which was imperfectly-translated from its original Russian version." (Personal Communication with Keith Moffatt, Trinity College Cambridge, U.K., 17 December 2019)

Thirdly, in an interview (Leong 2006, page 20, left column, paragraph 6), Moffatt expressed his views on turbulence that "Progress in turbulence at the fundamental level is extremely slow. You sometimes take one step forward and two backwards! This applies even the most fundamental theoretical development in turbulence, the theory of Kolmogorov (1941) [1941a, c] which essentially boils down to inspired dimensional analysis. Even Kolmogorov recognized a fundamental flaw in his theory, and he published a revision (his updated thoughts) in 1962, some 20 years later. At that stage, he himself undermined his own theory! One of the 'firmest' foundations of turbulence from that point on became very shaky. This is typical of the history of the subject."

Finally, the year of 1991 was the 50th anniversary of the publications of Kolmogorov (1941a, c). In 1991, three members of the editorial committee of *Proceedings of the Royal Society of London A*, Julian C.R. Hunt, Owen M. Phillips, David Williams, were asked to organize a special issue of *Proceedings of the Royal Society of London* to commemorate the famous papers by Andrei Nikolayevich Kolmogorov 1931 and 1941 on probability theory and turbulence respectively (Hunt, Phillips and Williams 1991). Kolmogorov was a Foreign Member of the Royal Society. The intention in this commemorative volume is to have papers that explain some of the implication of and developments from these and others of Kolmogorov's papers. George Keith Batchelor was among the 13 possible contributors, including V.I. Arnold, Dinkin, D.G. Kendall, R. Temam, G.K. Batchelor, O.M. Phillips (Ocean Turbulence), Hunt & Moffatt (Structure of turbulent eddies – Kolmogorov's theory and its modern interpretation), Yaglom (Atmospheric Turbulence), Gibson/Van Atta (Laboratory Turbulence), D. Williams, and Mandelbrot (Fractals) [a file dated ca. 1990, the Batchelor archive, the Wren Library, Trinity College Cambridge, U.K.].

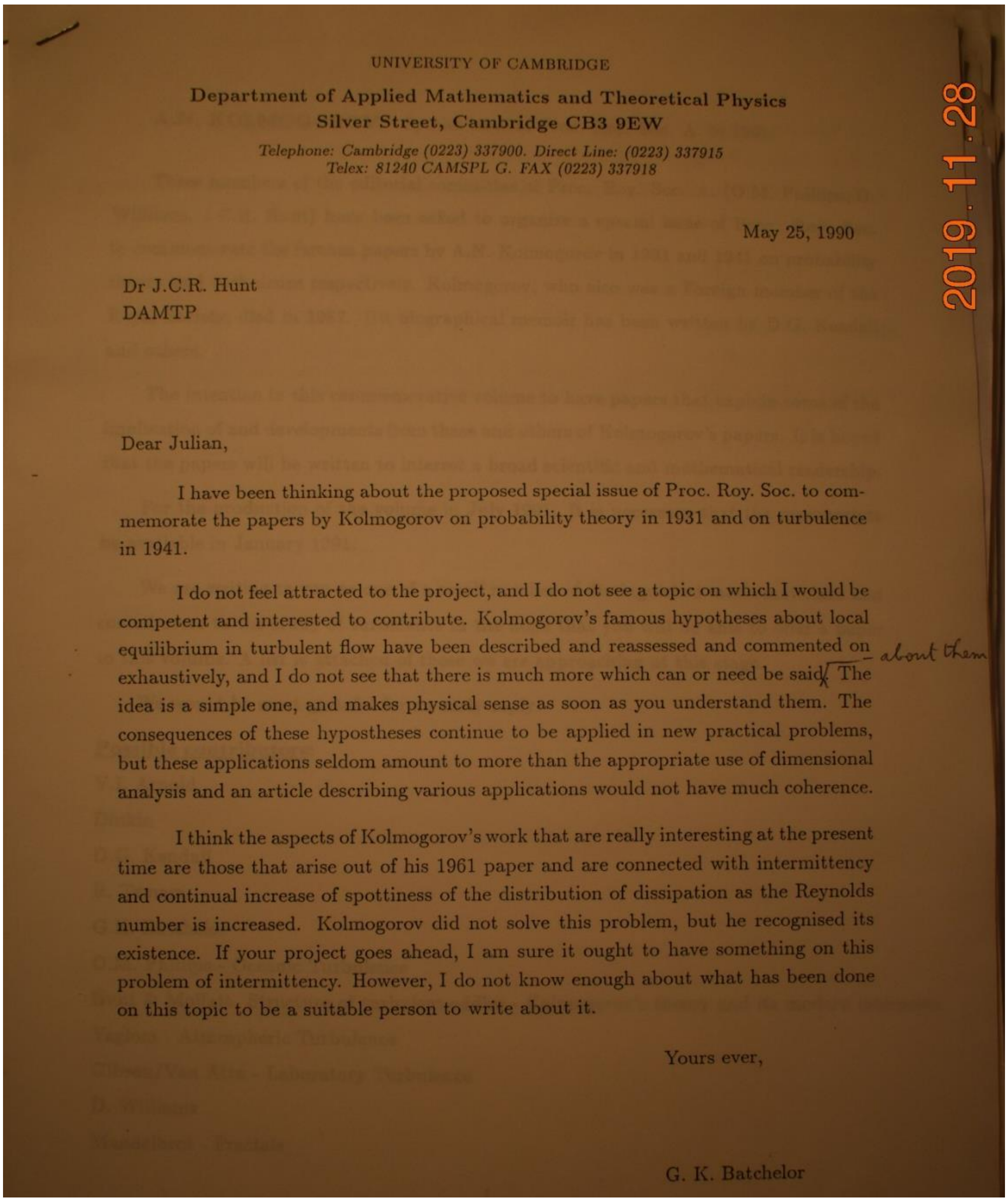

UNIVERSITY OF CAMBRIDGE

Department of Applied Mathematics and Theoretical Physics
Silver Street, Cambridge CB3 9EW

*Telephone: Cambridge (0223) 337900. Direct Line: (0223) 337915*
*Telex: 81240 CAMSPL G. FAX (0223) 337918*

May 25, 1990

Dr J.C.R. Hunt
DAMTP

Dear Julian,

I have been thinking about the proposed special issue of Proc. Roy. Soc. to commemorate the papers by Kolmogorov on probability theory in 1931 and on turbulence in 1941.

I do not feel attracted to the project, and I do not see a topic on which I would be competent and interested to contribute. Kolmogorov's famous hypotheses about local equilibrium in turbulent flow have been described and reassessed and commented on exhaustively, and I do not see that there is much more which can or need be said about them. The idea is a simple one, and makes physical sense as soon as you understand them. The consequences of these hypostheses continue to be applied in new practical problems, but these applications seldom amount to more than the appropriate use of dimensional analysis and an article describing various applications would not have much coherence.

I think the aspects of Kolmogorov's work that are really interesting at the present time are those that arise out of his 1961 paper and are connected with intermittency and continual increase of spottiness of the distribution of dissipation as the Reynolds number is increased. Kolmogorov did not solve this problem, but he recognised its existence. If your project goes ahead, I am sure it ought to have something on this problem of intermittency. However, I do not know enough about what has been done on this topic to be a suitable person to write about it.

Yours ever,

G. K. Batchelor

Figure 8 Batchelor's letter to Hunt dated May 25, 1990. Copyright: the Wren Library, Trinity College Cambridge, U.K.

Hunt sent his invitation letter to Batchelor [Batchelor's letter to Hunt dated May 25, 1990, the Batchelor archive, the Wren Library, Trinity College Cambridge, U.K.]. However, as shown in Figure 8, Batchelor declined it in his replying letter to Hunt dated May 25, 1990 by saying that "I do not feel attracted to the project, and I do not see a topic on which I would be competent and interested to contribute. Kolmogorov's famous hypothesis about local equilibrium in turbulent flow have been described and reassessed and commented on exhaustively, and I do not see that there is much more which can or need be said about them. The idea is a simple one, and makes physical sense as soon as you understand them. The consequences of these hypotheses continue to be applied in new practical problems, but these applications seldom amount to more than the appropriate use of dimensional analysis and an article describing various applications would not have much coherence."

It surprised the author that Batchelor declined Hunt's invitation. However, his reason was acceptable. Nevertheless, as shown in Figure 8, he also kindly gave Hunt some constructive suggestion about K62 that "I think the aspects of Kolmogorov's work that are really interesting at the present time are those that arise out of his 1961 paper and are connected with intermittency and continual increase of spottiness of the distribution of dissipation as the Reynolds number is increased. Kolmogorov did not solve this problem, but he recognized its existence. If your project goes ahead, I am sure it ought to have something on this problem of intermittency".

Please note that 'his 1961 paper' actually refers to Kolmogorov (1962). As shown before, 'intermittency' was inferred and recognized based on the experimental data by Batchelor and Townsend (1949). Why did Batchelor think that Kolmogorov recognized the existence of intermittency? Does it have to do with the inertial range (Kolmogorov 1962) versus dissipative scales (Batchelor and Townsend 1949)? It is true that Kolmogorov (1962) did recognize "continual increase of spottiness of the distribution of dissipation as the Reynolds number is increased." However, Batchelor and Townsend (1949) did suggest that 'space is divided into a number of regions each of which has its own value of the energy dissipation and in each of which the energy spectrum has the universal form [of the energy spectrum]'. What Kolmogorov (1962) recognized is very much similar to what Townsend (1948b) and Batchelor and Townsend (1949) already did. If so, why did Batchelor still suggest Hunt to have something on this problem of intermittency which Townsend (1948b) and Batchelor and Townsend (1949) already had? Was the intermittency recognized by Townsend (1948b), Batchelor and Townsend (1949) and Kolmogorov (1962)? Is there any difference about the concept of intermittency by them? Or did they recognize the same concept of intermittency of turbulence?

There are a few points relating to the term 'intermittency': (a) the term "intermittency" first appeared in Townsend (1948b) and Batchelor and Townsend (1949), as presented before; (b) 'the variation of the dissipation of energy' was discussed by Kolmogorov (1962); (c) Kolmogorov (1962) did not use the word 'intermittency' at all; (d) it was clearly shown in Batchelor's replying letter to Hunt that Batchelor seemed to have attributed the 'intermittency (of turbulence)' to Kolmogorov (1962); (e) this also implied that the 'intermittency' by Townsend (1948b) and Batchelor and Townsend (1949) might be different from the 'the variation of the dissipation of energy' by Kolmogorov (1962), i.e. mathematically, the 'intermittency' refers to the probability distributions of the velocity derivatives by Batchelor and Townsend (1949), while 'the variation of the dissipation of energy' to the probability distributions of the energy dissipation by Kolmogorov (1962). In a sense, this also led to be ambiguous as to what K62 corrected for and as to whether K62 was right or flawed.

As a postscript it may be mentioned that Moffatt (2002, page 21, paragraph 3, lines 5-8) highlighted the following major contribution by Batchelor, i.e. "Among these, for example, is the problem of intermittency, which was first identified by Batchelor and Townsend (1949) and which perhaps contributed to that sense of frustration that afflicted Batchelor (and many others) from 1960 onward". However, based on the literature review above, it was Townsend (1948b) who first drew our attention to the problem of intermittency of turbulence.

## Qian's major work on small-scale turbulence between 1983 and 1996

......a mind forever Voyaging through strange seas of Thought, alone.

William Wordsworth (1770—1850) had imagined Newton, The Prelude/Book III, 'Residence at Cambridge', 1805

From the previous section, it can be seen that the statistical properties of small-scale turbulence of both K41 and K62 were already intensively studied by numerous people around the world before the 1970s as summarized in the two volumes entitled Statistical Fluid Mechanics. Mechanics of Turbulence by Monin and Yaglom (1971, 1975). However, both K41 and K62 have been controversial since the 1970s. Why did Qian become interested in small-scale turbulence in 1979 when he went on to The City College of The University of

New York, U.S.A.? In Qian's own view, if you are interested in turbulence, you must read the two 'Red Treasure Books', i.e. Monin and Yaglom (1971, 1975). Clearly, Qian first learned about Kolmogorov's work on turbulence from Monin and Yaglom (1971, 1975) and then became interested in Statistical Mechanics of Turbulence. From the literature cited in his papers, one can see that Qian also read the books on turbulence by others, e.g. Leslie (1973), Hinze (1975) and McComb (1990).

Despite "the wrenching experience of the Cultural Revolution" (Day 2010), it was not an easy task for a Chinese physicist (turbulence man) as a single author had contributed 10 papers to *Physics of Fluids* (Qian 1983, 1984b, c, d, e, 1985, 1986a, b, c, 1988) alone within 12 years after the Cultural Revolution. From the contents and the dates of their submissions, it is clear that the papers Qian (1983, 1984b, c, d, e) are an outgrowth of his PhD thesis (Qian 1984) and Qian (1985, 1986a) are partially the outgrowth of his PhD Thesis (Qian 1984a).

In this section, the author attempts to summarize Qian's major work on the physical understandings of small-scale turbulence between 1983 and 1997. A few sub-topics have been classified and then roughly presented in the chronological order. However, Qian's work on the finite $R_\lambda$ effect of turbulence and finite $R_\lambda$ turbulence will be presented and discussed in the next separate section. Notably, the sub-topic entitled *Qian's work on small-scale intermittency of turbulence using closure approaches* is presented at the end of this section so as to also serve as a transitional sub-topic leading to the separate section regarding his work on *the finite $R_\lambda$ effect of turbulence and finite $R_\lambda$ turbulence*.

*A perturbation variation approach to the numerical study of isotropic turbulence*

What did motivate Qian to propose a perturbation variation approach to the closure problem of turbulence theory in the 1980s? There are possibly several reasons: (i) Qian had his great interest in the perturbation variation approach and obtained a sound background about it when he was an undergraduate student at Peking University in the 1950s/1960s; (ii) a direct-interaction (DI) approximation approach by Kraichnan (1959, 1964, 1965, 1971, 1972, 1977) and Kraichnan and Herring (1977) was done in a Lagrangian framework, however, its mathematics is very complicated, an alternative approach might be beneficial; and (iii) the principle of maximum entropy approach by Edwards (1964), Herring (1965), Edwards and McComb (1969), and Herring and Edwards (1979), which gave a rather poor value of the Kolmogorov constant ($K_0$), is questionable in application to stationary turbulence.

Certainly, during his PhD research, Qian (1983) proposed a perturbation variation approach to solve the closure problem of isotropic turbulence theory so as to derive the Kolmogorov's $k^{-5/3}$ law in an Eulerian manner. Complex Fourier components of the velocity field have been conventionally used as modal parameters for describing the dynamic states of turbulence. Instead, Qian (1983) used a complete set of independent real parameters and dynamic equations. He also applied classical statistical mechanics to study the statistical behavior of the turbulence. By using a perturbation based on the famous Langevin-Fokker-Planck model, Qian obtained an appropriate stationary solution of the Liouville equation. In his own words, Qian obtained the following closed set of equations for determining the two unknowns $q(k)$ and $\eta(k)$(Qian 1983, page 2101, Equation (5)):

$$q(k)\eta(k) = 2\pi \int^{\Delta} \int dpdr \frac{kpr}{[\eta(k)+\eta(p)+\eta(r)]^2} \times \{b(k,p,r)[q(r) - q(p)] \times [2\eta(p) + \eta(r) + b^*(k,p,r)q(k) \times [q(r) - q(p)\eta(p)]]\} \quad (15)$$

Qian (1983, page 2101, Equation (45)) obtained the following equation:

$$\varepsilon = \int_0^\infty dk' \int_0^k dp \int_{p*}^{p+k'} drS(k'|p,r) \quad (16)$$

Qian treated the dynamical damping coefficient $\eta$ as an optimum control parameter to minimize the error of the perturbation solution. A convergent integral equation has replaced the divergent response equation of Kraichnan's (1958) direct-integration approximation. Thus, the closure problem can be solved without appealing to a Lagrangian formulation. Qian obtained a value for the Kolmogorov constant of 1.2, which is supported by the experimental result of Gibson and Schwartz (1963).

The variational method proposed by Qian (1983) was criticized by McComb (1990, page 294, paragraph 4). However, Qian (1996c) made an effort to defend his method and argued that "It is proved that the turbulence entropy, defined in the (maximal) entropy method of turbulence closure problem, does not attain a maximum value or stationary turbulence, and a solution for the basic equations of the entropy method does not exist" was published as COMMENT in Journal of Physics A: Mathematical and General. Both Edwards and McComb (1969) and Qian (1983) developed Liouville (1853) equation-based turbulence theories. Edwards and McComb (1969) proposed maximal entropy principle while Qian (1983) proposed a perturbation approach to obtain an equation for the damping coefficient. The debate between the variational method of Qian and the entropy method of Edwards (not McComb) is due to the fact that from a physics point of view, in thermodynamics it is usually a competition between entropy and energy. In the case of an ideal gas, entropy is in control. In the case of turbulence, energy is in control. This has been clarified by W. David McComb (Personal Communication with W. David McComb, 21 March 2020).

Bazdenkov and Kukharkin (1993) also discussed the perturbation-variation method of Qian (1983). They argued that Qian's (1983) method is based on a clear physical idea but it is not self-consistent. The procedure to obtain the equation for the dynamic damping coefficient does contain arbitrariness, which leads to the dependence of this equation on the choice of variables. This ambiguity is illustrated by numerical evaluations of the Kolmogorov constant in two-dimensional and three-dimensional cases. They obtained and analyzed the equation for the dynamic damping coefficient which is invariant with respect to a change of variables. They discussed the principle inevitability of arbitrariness in closure methods. In response to Bazdenkov and Kukharkin (1993), Qian (1995a) discussed the issue of dropping the random force and the arbitrariness of choosing the basic variable in the variational approach to the turbulence closure problem. Qian (1995a) was published in *Acta Mechanics Sinica*. It is unclear why Qian did not publish his Reply in *Physics of Fluids*. Qian argued that: (i) according to the mean-square estimation method, the random force ($f_i$) should be dropped in the error expression of the Langevin-Fokker-Planck model; (ii) $f_i$ is not neglected, its effect has been taken into account by the variational approach; (iii) optimization of the perturbation solution of the Liouville equation, the Langevin-Fokker-Planck model requires that the basic variable is as near to Gaussian as possible in (iv) the velocity, instead of the vorticity, should be chosen as the basic variable in the three-dimensional turbulence (v) although the Langevin-Fokker-Planck model and the zero-order Gaussian term of probability density function imply whiteness assumption (zero correlation time of $f_i$), the higher-order non-Gaussian terms of the probability density function correspond to the nonwhiteness of turbulence dynamics, and the variational approach does calculate the nonwhiteness effect properly.

Nevertheless, despite this, the perturbation approach to the closure problem of turbulence theory, proposed in Qian (1983), was successfully used by Qian (a) to evaluate the flatness factor, indicating the degree of intermittency of turbulence; (b) to study a passive scalar field convected by isotropic turbulence; and (c) to study the spectrum dynamics of a turbulent passive scalar in the viscous-convective range.

*One-dimensional model of turbulence*

As presented in the earlier section, Burgers supervised Tchen's Ph.D. research while Tchen supervised Qian's Ph.D. thesis. It was a happy coincidence that Qian (1984b) solved the initial-value problem of a forced Burgers' (1948) equation by the use of the Fourier expansion method. Its solutions reach a steady state of laminar flow without randomness. Qian argued that Burgers turbulence is not really the turbulence at all. Qian further proposed simulating the Navier-Stokes turbulence and did numerical experiments on this one-dimensional turbulence. Qian obtained Kolmogorov's $k^{-5/3}$ inertial range turbulence energy spectrum. Kolmogorov constant ($k_0$) ranges from 0.5 to 0.65.

Burgers' (1948) equation can be written as

$$\frac{\partial}{\partial t}u(t,x) + u(t,x)\frac{\partial}{\partial x}u(t,x) = \nu\frac{\partial^2}{\partial x^2}u(t,x) \quad (17)$$

The one-dimensional model of turbulence is written as (Qian 1984b, page 1961, equation (31)):

$$\frac{d}{dt}U(k) = -iW_m(k) + P(k) - \nu(k)U(k) \quad (18)$$

where $-iW_m(k)$ is a modified advection term, which transfers the energy from one mode to the other but conserves the total energy of all modes; $P(k)$ the pressure term.

For studying the inertial range dynamics, equation (13) becomes (Qian 1984b, page 1962, equations (43a, b, c)

$$\frac{d}{dt}U(1) = i\omega U(1) \quad (19a)$$

$$\frac{d}{dt}U(k) = -iW_m(k) + P(k),\ 2 \ll k \ll k_d \quad (19b)$$

$$\frac{d}{dt}U(k) = -iW_m(k) + P(k) - \nu_d(k-k_d)^n U(k),\ k_d < k \ll k_c \quad (19c)$$

where $k_d$ is the Kolmogorov wavenumber.

Qian (1984b) concluded that equations (14a, b, c) can simulate the essential feature of the cascade transfer of energy in a turbulence and a Kolmogorov's $k^{-5/3}$ inertial range turbulence energy spectrum. By solving the two integral equations (Qian 1984e, page 2968, equations (16) and (17)), Qian (1984e, page 2968, equation (18)) derived the Kolmogorov $k^{-5/3}$ law. The Burgers equation, which has a $k^{-2}$ inertial-range spectrum instead of $k^{-5/3}$, is the commonly used 1-D model of turbulence. Can we get the inertial-range spectrum by applying the variational approach to the Burgers equation? Qian (1984e) argued that the method of statistical mechanics and the variational approach cannot be applied to the Burgers equation because Qian (1984b) had already found that solutions of the Burgers equation are not ergodic stochastic processes.

*Two-dimensional turbulence*

Qian (1984b) extended Qian's (1983) closure approach to the vorticity dynamics of two-dimensional turbulence. He found that the $k^{-3}$ enstrophy-cascade range is more characteristic of two-dimensional turbulence (Qian 1984b, page 2416, equation (61):

$$E(k) = 1.82(lng - 1.23)^{-2/3}\chi^{2/3}k^{-3},\ (g \geq 10) \quad (20a)$$

where $g$ is a localization factor.

Inertial ranges in two-dimensional turbulence has been studied by Kraichnan (1967). The nonequilibrium statistical closure method in Qian (1983, 1984d) was used by Qian (1986c) to study the inverse energy cascade in two-dimensional turbulence. Qian (1986c, page 3610, equation (20)) obtained the following inverse energy cascade inertial range energy spectrum:

$$E(k) = 2.58g^{0.244}\epsilon^{2/3}k^{-5/3} \quad (20b)$$

where $\epsilon$ is the rate of energy cascade.

*Cascade model of turbulence*

Qian (1988) proposed a simple low-order dynamic system to model the energy cascade process of a fully developed hydrodynamic turbulence. Results show that the main structural properties of this dynamic system are the same as those of the normalized Navier-Stokes equation. He obtained the Kolmogorov's $k^{-5/3}$ spectrum as a statistical property of chaotic trajectories along the strange attractor.

*The analytical theory of a turbulent passive scalar*

Qian (1985) applied statistical mechanics to the study of a passive scalar field convected by isotropic turbulence. He derived the inertial-convective-range spectrum (Qian 1985, page 1299, Abstract, line 10):

$$F(k) = 0.61\chi\epsilon^{-1/3}k^{-5/3} \quad (21)$$

where $\chi$ is the dissipation of the scalar variance and $\epsilon$ the dissipation rate of the energy.

Qian (1985) developed the statistical-mechanics theory of the passive scalar field convected by turbulence. Qian (1986b) extended this theory to the case of a small molecular Prandtl number. The so-called equation-error method is used to solve the set of governing integral equations. He obtained the following scalar-variance spectrum for the inertial range (Qian 1986b, page 3586, Abstract, lines 4-5):

$$F(k) \sim k^{-5/3}/[1 + 1.21x^{1.67}(1 + 0.353x^{3.32})] \quad (22)$$

where $x$ is the wavenumber scaled by Corrsin's (1951) dissipation wavenumber.

Qian (1989) extended Qian's (1985) non-equilibrium statistical-mechanics theory of a passive scalar field convected turbulence to the two-dimensional case. To this end, Qian derived a set of closed equations of special dynamics and then applied it so as to obtain the scalar-variance spectrum in the inertial-convective range

$$G(k) = CA^{-1/3}\eta\varepsilon^{-1/3}k^{-5/3} \quad (23a)$$

corresponding to Kolmogorov's energy spectrum

$$E(k) = A\varepsilon^{2/3}k^{-5/3} \quad (23b)$$

In summary, Qian (1983) established Liouville's equation-based isotropic turbulence theory using non-equilibrium statistical mechanics and a perturbation variation approach. And it is compatible with the Kolmogorov spectrum. Qian (1990a) applied his theory to further confirm Batchelor's $k^{-1}$ spectrum (Batchelor 1959, page 126, equations 4.8 and 4.9). Qian (1990a) applied his equations to the study of the spectral dynamics of a turbulence passive scalar in the viscous-convective range and further confirmed Batchelor's $k^{-1}$ spectrum for the scalar for Pr>>1 and Pr <<1.

The analytical theory of a turbulent scalar by Qian (1985, 1990) was extended to the case of large Prandtl number (Qian 1995b). The following scalar variance spectrum in the viscous range is obtained (Qian 1995b, page 109, equations (22.1)-(22.2)):

$$F(k) = 4.472(\nu/\varepsilon)^{1/2}\chi k^{-1}H(x), x \equiv (k/k_b)^2 \quad (24a)$$

$$H(x) = 0.7687\exp(-3.79x) + 0.2313\exp(-11.13x) \quad (24b)$$

$$k_b \equiv (\varepsilon/\nu\mu^2)^{1/4} \quad (24c)$$

where $k_b$ is the Batchelor wavenumber, $\varepsilon$ and $\chi$ are the energy and variance dissipation rates, respectively, $\mu$ the scalar diffusivity. Borgas, Sawford, Xu, Donzis and Yeung (2004) also numerically tested equations (24a, b) and found agreement with Kraichnan (1968).

*The kinetic energy spectrum versus the bump phenomenon*

Hill (1978) described the 'bump' phenomenon in the scalar spectrum. The perturbation approach to the closure problem of turbulence theory in Qian (1983, 1984a, b) is used by Qian to study the universal equilibrium range of turbulence. He derived two integral equations for two unknown functions, taking the viscous dissipation into account. A simple model of the bump phenomenon (Figure 9), which is between the inertial and dissipation ranges, can be expressed as (Qian 1984c, page 2230, equation (26)):

$$F(x) = F(0)(1 + Bx^{\beta})\exp(-Cx^{\gamma}) \quad (25a)$$

When $(\beta, \gamma)$=(1,1), (2/3, 2) and (2/3, 4/3), equation (15) can be rewritten (Qian 1984c, page 2231, equations (27a, b, c)

$$F(x) = F(0)(1 + Bx)\exp(-Cx) \quad (25b)$$

$$F(x) = F(0)(1 + Bx^{2/3})\exp(-Cx^{2}) \quad (25c)$$

$$F(x) = F(0)(1 + Bx^{2/3})\exp(-Cx^{4/3}) \quad (25d)$$

where $F(x)$ is not a monotonically decreasing function.

Equation (15c) is equivalent to Saffman's (1963) formula for the energy spectrum of the dissipation subrange. The transition from the inertial subrange to the dissipation subrange is quite slow, occurring over a wide range of $k$ due to the fact that the viscous dissipation approaches zero slowly as $k/k_d$ approaches zero. Qian (1994b) assumed the following turbulent energy transfer function

$$\Pi = \Pi(\mathrm{k}) = \Pi_1 k^{\alpha} \quad (26)$$

By using the nonequilibrium statistical closure method, Qian (1994b) proved the following turbulent energy spectrum

$$E(k) = C(\alpha)\Pi^{2/3}k^{-5/3} \quad (27)$$

where $C(\alpha)$ is a dimensionless coefficient which depends on the exponent $\alpha$. It was found that Kolmogorov's law corresponds to the special case of $\alpha$ =0 and $\Pi = \epsilon$.

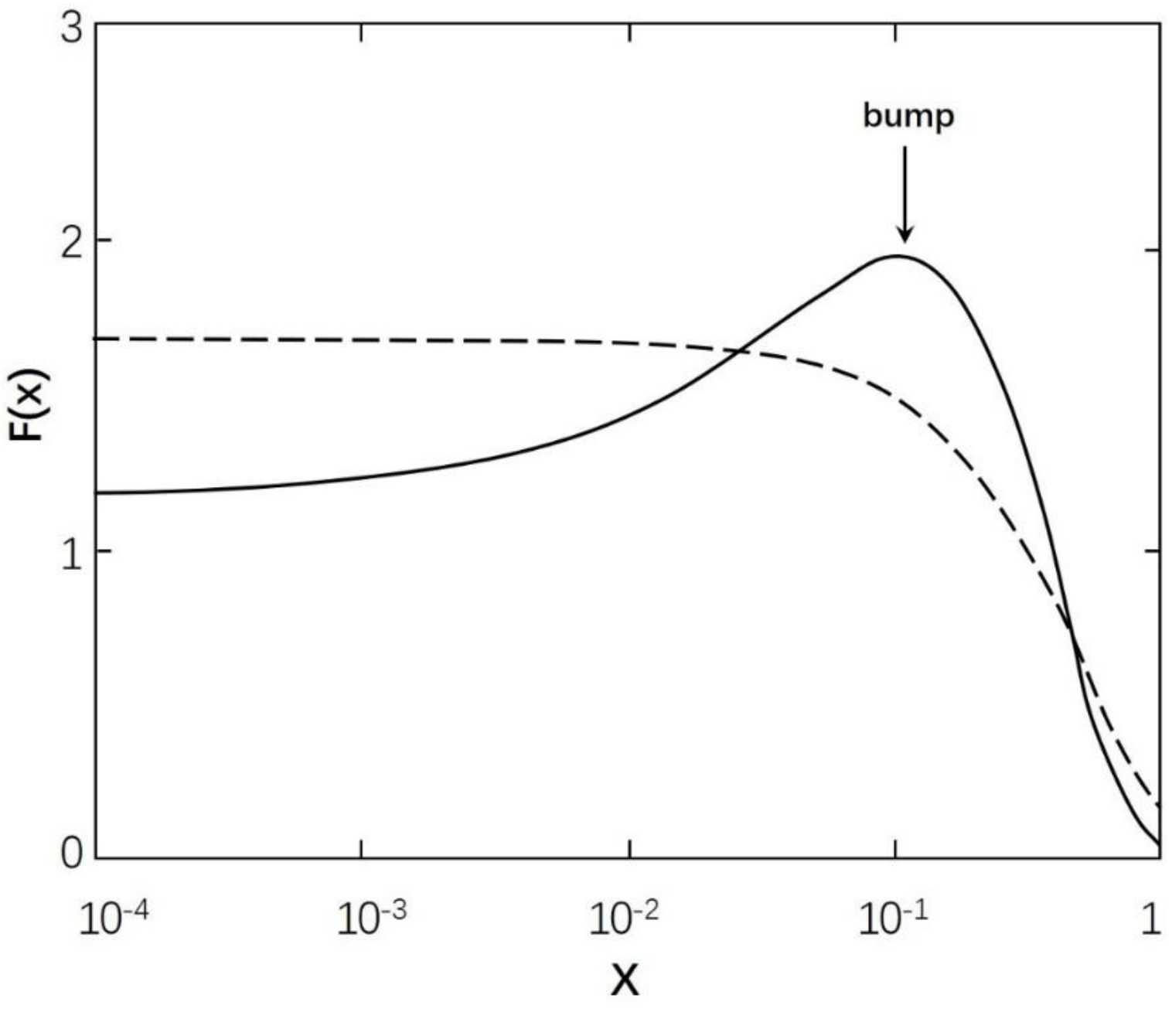

Figure 9 Group of the bump phenomena (Qian 1984c, page 2232, FIG. 1).

*The renormalization group method vs the Kolmogorov constants of turbulence*

The Yakhot–Orszag relation is made between the turbulent Prandtl number ($P_t$), the Kolmogorov constant ($K_o$), and the Batchelor constant ($Ba$). However, Qian (1990b) revised it as

$$Ba = \mathrm{C}K_o P_t \quad (28)$$

The Kolmogorov constant ($K_o$) can be obtained theoretically and also by direct numerical simulation and large-eddy numerical simulation. The former real one is appreciably smaller than the latter pseudo one. By using the non-equilibrium statistical mechanics (Qian 1984c), Qian (1993a) analysed the two possible reasons for this discrepancy: (a) the viscous effects dies away slowly as the wavenumber decreases and persists outside the narrow range near the Kolmogorov wavenumber; (b) the resolution of the numerical simulations is limited. He suggested that the resolution of simulations must be higher than $128^3$.

The original early version of Qian (1996b) was rejected by *Physical Review* E and a revised version was rejected by *Journal of Fluid Mechanics*. Akiva Yaglom, who was invited by George Keith Batchelor, took on the review of this manuscript [Yaglom's letter to Batchelor dated April 11, 1995. Copyright: the Wren Library, Trinity College Cambridge, U.K.].

In his letter to Batchelor dated April 11, 1995, Yaglom wrote that

> My preliminary opinion about the [Qian's] paper was based of [on] the fact that several months ago I received the same paper for review from Sreeni (K.R. Sreenivasan) who is now one of editors of Phys. Rev. [Physical Review] E. My review was negative but included the remarks that mayby [maybe] the paper can be transformed in a short publicable [publishable] note containing an empirical equation for the velocity spectrum describing its form in the high-wave-number range adjacent to the inertial range and including the bump between inertial and dissipation ranges discovered in some recent experimental works. (Later I discovered that such equation has been already found by She and Jackson in their paper of 1993 [She and Jackson 1993]; …...) However, the paper sent by you was longer than its first version which I saw earlier; therefore I did not expect that it is worth publication. Now I read the manuscript and compared it with the previous version; it contains practically all the old text withou [without] any changes but with some new additions. Only one remark from several [comments] made in my previous review was taken into account – I recommended to use also the data by Saddoughi and Veeravally [Saddoughi and Veeravally 1994] which were not mentioned in the first version of the paper. However, I wrote that it is necessary to contact the authors and to ask them to send numerical data of their measurements; this was not done and I cannot understand how it was possible to treat the data given in a form of figures with a cloud of scattered experimental points. My feeling is that Qian's paper is not only poor but also his treatment of data is quite questionable.

Qian's responses to Yaglom's comments [Qian's reply to the comments by Referee A. Note that Batchelor's handwritten notes to Yaglom dated 22 June 1995. Copyright: the Wren Library, Trinity College Cambridge, U.K.]:

> (1) The three methods (method[s] No.3, 4, & 5) of determining $K_o$ proposed by the author are different from Pao's [1965] method. Similar to the traditional method which is considered most "reliable" by Referee A, Pao's [1965] method neglects the bump phenomenon and overestimates the value of $K_o$. The three methods (No.3, 4, & 5) take into account the bump phenomenon, and are more reasonable. The conclusion derived by author's bump models does not depend upon whether we use the high-wavenumber spectral data, which are rejected in the traditional method, or not. Similar to the traditional method, with the fourth method, we reject the spectral data in the range $k/k_d > 0.1$, the resultant $K_o$ is still much lower than the $K_o$ predicated by the traditional method as well as Pao's [1965] method.
>
> (2) It is true that the high-wavenumber spectra are less precise and we have no final model of them. However, it does not mean we know nothing about them. In fact, we have some well-grounded results of theories, numerical simulations and experiments, for example, $F(k/k_d)$ decaying exponentially in the dissipation range, the bump phenomenon, and so on. There is no reason to deny the effort of applying these well-grounded results to improve the estimation of $K_o$.
>
> (3) The conclusion that the traditional method assuming (8) overestimates $K_o$, is not dependent of the particular mathematical form of describing bump phenomenon, but is an important consequence of the following well-grounded results about universal equilibrium range: $F(x)$ approaches $K_o = F(0)$ in the inertial range ($x = k/k_d \to 0$) , decays exponentially in the dissipation range, and has a $k-1$ bump between. The 3-D bump model (11) is just a simple mathematical expression of these well-grounded results. Referee B said, "the author claims that the experimentalists have consistently overestimated the value of $K_o$ by assuming a form of the energy spectrum which may not be valid over the range of wavenumbers captured in the experiments. This is a valid point."

(4) Referee A said, "the inertial wave-number range is defined as the range where $E1(k) \sim k^{-5/3}$". The author does not follow this definition, but adopts the definition given by Monin & Yaglom 1: the inertial range is the range where the viscous effect is negligible, and corresponds to $T(k) = 0$ in a stationary turbulence. As explained in this paper, *the approximate $k^{-5/3}$ range $b < k/k_d < d$ captured in experiments is not the same as the inertial range. This point is the crux of the matter.* If Referee A can agree with the author on this point, then he will accept the opinion of Referee B that the main conclusion of this paper "is a valid point" and "most of this paper is clear and easy to understand."

In the author's view, as highlighted in Italics in Qian's reply (4), the key point is that "the inertial range is the range where the viscous effect is negligible, and corresponds to $T(k) = 0$ in a stationary turbulence".

the approximate $k^{-5/3}$ range $b < k/k_d < d$ $\neq$ the inertial range

This key point led to the discovery of the finite Taylor-Reynolds number effect of turbulence by Qian (1997). In a sense, the key point in Qian (1996b) is a turning point in the study of turbulence.

Yaglom's letter to Batchelor and Qian's replies to Referee A (Yaglom) are fortunately kept/deposited in the Batchelor archive, the Wren Library, Trinity College Cambridge, U.K. Eventually, Yaglom and Batchelor declined Qian's submission. However, it is clear that Batchelor did take Qian's submission and his Reply to Referee A very seriously, otherwise, they would not have been kept in Batchelor's personal file. It is evidently that both Yaglom and Batchelor did not fully recognize the merits of Qian's (1984c) simple model of the bump phenomena and Qian's key point in this submission, i.e. the approximate $k^{-5/3}$ range $\mathrm{b} < k/k_\mathrm{d} < \mathrm{d}$ captured in experiments is not the same as the inertial range. Although Qian attributed this to the primary bump phenomena between the inertial and dissipation rages, and a secondary 'bump' near the energy containing range.

Qian (1993b) compared three methods of defining eddy transport coefficient, i.e. the optimization method developed by Qian himself, the distant-interaction by Kraichnan (1987) and energy-dissipation approximations. Again, based on the non-equilibrium statistical mechanics closure method (Qian 1983, 1984c, 1985), by treating eddy viscosity and eddy diffusivity as optimal control parameters, Qian (1993b) developed an optimization method. He found that (a) the distant-interaction approximation is simple but has larger error; (b) the optimization method, implying a 'vorticity–transfer' approximation, is physically as plausible as an energy-dissipation approximation; (c) the optimization method minimizes the error of Boussinesq's model and gives reasonable turbulent Prandtl number; and (d) the optimization method is intrinsic to the essence of the Boussinesq's model.

*Qian's work on small-scale intermittency of turbulence using closure approaches*

What motivated Qian to study small-scale intermittency of turbulence? Clearly, Qian was strongly motivated by consistency and uniqueness questions raised by both the K41 and K62 inertial range theories by Kraichnan (1974). Most existing closure methods discard the high-order terms and fail to address the intermittency phenomenon. However, since the flatness factor ($F$) is related to the fourth moment of a turbulent velocity field and indicates the degree of intermittency of turbulence, is it possible to develop a closure theory that would enable evaluation of the flatness factor directly from the Navier–Stokes equation by means of the general method of statistical mechanics? The answer is yes; i.e. Qian (1986a) extended Qian's (1983, 1984c) theory to evaluate the flatness factor ($F$) by taking into account the corresponding higher-order terms in the perturbation solution of the Liouville's (1853) equation. It was found that the flatness factor ranged from 9 to 15 for the infinite isotropic turbulence.

More importantly, Qian (1986a) also found: (a) the intermittency phenomenon does not necessarily negate the Kolmogorov's $k^{-5/3}$ inertial range spectrum; (b) the Kolmogorov's $k^{-5/3}$ law and the high degree of intermittency can coexist as two consistent consequences of the closure theory of turbulence; and (c) the K41 theory, as Kraichnan (1974) argued, cannot be disqualified merely because the energy dissipation rate fluctuates.

As shown in Qian (1986a, page 2169, FIG. 1), the flatness factor ($F$) approaches a constant (15) as $R_\lambda \to \infty$. Using nonequilibrium statistical mechanics approach, Qian (1994a) studied the skewness factor of the turbulent velocity derivative and obtained a relationship between the skewness factor ($S$) and the Reynolds number ($R_\lambda$). It was found that the skewness factor ($S$) becomes as constant (-0.515) as $R_\lambda \to \infty$. The work done by Qian (1986a, 1994a) suggested that the boundedness of both the flatness factor ($F$) and the skewness factor ($S$) of the velocity derivative does exist as $R_\lambda \to \infty$.

Qian (1986a, page 2170, equation (52)) reflects the conflict between the K41 and K62:

$$F \sim R_\lambda^h \qquad (29)$$

where $F$ is the flatness factor; $h$ a constant ranging from 0.2 to 1.5. Equation (29) shows that the flatness factor ($F$) significantly changes with $R_\lambda$.

What did 'conflict' really mean? In Qian's own words, "The 1941 [K41] theory does not deny the weak dependence of $F$ on $R_\lambda$, it asserts that $F$ asymptotically approaches a constant as $R_\lambda$ approaches infinity. On the contrary the 1962 [K62] theory asserts that $F$ increases infinitely with $R_\lambda$ according to (52)[29] (Qian 1986a, page 2170, left column, paragraph 2, lines 30-40)".

Implication of those interesting findings (a), (b), (c) and Equation (29) above by Qian (1986a) is two-fold: on the one hand, the theory of isotropic turbulence which Qian (1983, 1984a, 1984c) developed was useful and correct, on the other hand, they led to the further study of the finite Reynolds number effect by Qian (1997, 1998b, 1999, 2000, 2001, 2002, 2003, 2006a, b), which will be presented and discussed in the next separate section.

As a digression, Wouter Bos found Qian (1986a) original. In Bos's own words, the question, e.g. to what extent closure can reproduce higher-order statistics, was one of Kraichnan's (1990) principal concerns in small-scale turbulence theory. Qian (1986a) attempted to describe intermittency and higher-order moments using closure approaches. Bos wanted to take into account intermittency in closure (as for instance measured by the flatness of the velocity gradients as function of the Reynolds number)(Bos, Rubinstein and Fang 2012; Bos and Rubinstein 2013). Qian (1986a) actually did, what Bos had planned to do, and obtained the results Bos had anticipated (constant flatness at very high Reynolds numbers (Personal Communication with Wouter Bos, 15 July 2020). Okamura (2018) gave a balanced overview of the recent statistical theories of turbulence and also included the work of Qian (1983, 1984, 1994).

## Qian's work on the finite $R_\lambda$ effect of turbulence between 1997 and 2006

……Nature never did betray
The heart that loved her:
Wordsworth (1894, page 196, lines 36-37)

The controversy surrounding K41 basically amounts to the following: intermittency corrections versus “finite Reynolds number effect” (e.g. McComb, Yoffe, Linkmann and Berera 2014, page 1, left column, bottom). This section focuses on the roles of the finite $R_\lambda$ effect and the finite $R_\lambda$ turbulence in the general context of small-scale turbulence, more specifically, its possible solution to the controversy surroundings K41 due to the occurrence of the intermittency of turbulence and our further understanding of the dynamics of finite $R_\lambda$ turbulence. To highlight Qian’s contribution in an appropriate position in the field of the finite $R_\lambda$-related study, it shall be helpful to have a brief historical overview and a state-of-the art review.

*Overview*

Since both the Reynolds number ($Re$) and the Taylor microscale Reynolds number ($R_\lambda$) are the kinematic viscosity ($\nu$)–dependent, which goes back to d'Alembert (1752), Navier (1823), Stokes (1845), Saint–Venant (1851), Boussinesq (1877, page 7; page 45, equation (11)) and Taylor (1915, page 13, line 3), in a broad sense, there has been a long history in light of the identification of the importance of the effect of finite but small values of $\nu$ or finite $Re$ or $R_\lambda$ in general. The quantification of the turbulent viscosity was attempted by Boussinesq (1877, page 46, equation (12)). Prandtl (1905, 1927) studied the motion of fluids of very small viscosity. The dimensional analysis of the Boussinesq’s (1877) turbulent viscosity partially led to the development of the mixing length by Prandtl (1925, page 137, Section II, paragraph 3, line 5). The von Kármán–Howarth (1938) equation is an exact consequence of the Navier-Stokes equation, and is valid for any Reynolds number. As for the identification and quantification of the effect of finite but small values of $\nu$ in the study of small-scale turbulence, one may recall many closure theories including Obukhov (1946), Heisenberg (1948a,b), Kovasznay, Kibens and Blackwelder (1970), Orszag (1970), and Kraichnan (1971). For example, Heisenberg (1948b, page 404, line 5 from the bottom) had studied the intensity $F(k)$ of the small eddies ‘In the case of finite but small values of $\nu$...’. According to Chandrasekar (1949, page 337, line 5), “the solution of Heisenberg’s [Heisenberg 1948b] equation also shows that, for any finite Reynolds number, no matter how large, we must get departures from the $k^{-5/3}$ –law when $k$ approaches $k_s$...”. The terms ‘finite viscosity’ and ‘finite-viscosity effects’ appeared on the first page of Chapter 6 Kolmogorov’s (1941) revisited (McComb 2014, page 143), i.e. ‘...any deviations from K41 are due to finite viscosity (McComb 2014, page 143, paragraph 3, lines 4-5)’ and ‘...finite-viscosity effects in explaining deviations from K41 (McComb 2014, page 143, paragraph 4, lines 4-5).’ Those two terms are somehow inconsistent with Sections 6.4 and 6.5 Finite-Reynolds-number effects on K41 (McComb 2014, pages 161-181). Notably, since both the Reynolds number ($Re$) and the Taylor microscale Reynolds number ($R_\lambda$) are inversely proportional to the kinematic viscosity ($\nu$), ‘finite viscosity’ and ‘finite-viscosity effects’ may not be physically the same as ‘Finite-Reynolds-number effects’.

As for the $Re$– or $R_\lambda$–dependent study and the finite $R_\lambda$–related study, are they the same or different? Why? It can be inferred from the above equation (5e), i.e. Lohse (1994, page 3223, equation (3)), that the $Re$- or $R_\lambda$-dependent studies are mutually related to each other. The finite $R_\lambda$–related study is different from them. Generally, based on the literature available, the $Re$- or $R_\lambda$-dependent study may include the following topics: (i) the number of orders of eddies possessing local isotropy and contributing to the motion within $G$ versus $R_\lambda$; (ii) the skewness factor ($S$) of turbulent velocity derivative versus $R_\lambda$; (iii) the flatness factor ($F$) of turbulent velocity derivative versus $R_\lambda$; (iv) the double velocity correlation function [$f(r,t)$] versus grid/mesh $Re$; (v) the triple correlation function [($k(r,t)$] versus mesh $Re$; (vi) the turbulence level versus $Re$; (vii) the numerical coefficients in the correlation function of energy dissipation versus $R_\lambda$; (viii) the parameters in the energy spectrum versus $R_\lambda$; (ix) the intermittency of turbulence versus $R_\lambda$; (x) the turbulent length scales (the integral length scale and the dissipation length scale) versus $R_\lambda$; (xi) the Kolmogorov constant $C_K$ versus $R_\lambda$; (xii) the dimensionless energy dissipation rate ($c_\varepsilon$) versus $R_\lambda$ and $Re$; (xiii) the range over which the self-similar range versus $R_\lambda$; and (xiv) the Kolmogorov function, $K(r) = -S_3/r\epsilon$, versus $R_\lambda$. The finite $R_\lambda$-related study may include the following topics: (xv) the second-order structure function ($D_{LL}$) versus the finite $R_\lambda$; (xvi) the third-order structure function ($D_{LLL}$) versus the finite $R_\lambda$; (xvii) the width of the inertial range of finite $R_\lambda$ turbulence versus $R_\lambda$; and (xviii) (the anomalous values of ) the scaling exponents $\xi_p$ $[\xi_n]$ of $p$th $[n$th$]$ order structure function versus the finite $R_\lambda$. Note that both $\xi_p$ and $p$ were used in Qian (2006), however, to avoid their confusions with the scaling exponent ($p$) in Oboukhov (1962).

*Others' work on the $Re$– or $R_\lambda$–dependent turbulence*

In the two English translation versions of Kolmogorov (1941a), the terms "at very high Reynolds numbers" (by D. ter Haar) and "for very large Reynolds numbers" (by V. Levin) appeared in the paper titles, respectively. It is clear that Kolmogorov himself was fully aware of the importance of Reynolds number in his local structure of turbulence. In a broad sense, one can also infer that it was Kolmogorov (1941a, c) who first studied the $R_\lambda$–dependent small-scale turbulence. In his discussion of the requirement of high Reynolds number for the existence of locally isotropic turbulence (Kolmogorov 1941a, c), Batchelor (1947, page 536, paragraph 3, lines 13-14) argued that 'the number of orders of eddies possessing local isotropy and contributing to the motion within $G$ increases with Reynolds number'. The term 'low Reynolds states of turbulence' also appeared in Batchelor and Townsend (1948b, page t527, Introduction, paragraph 1, line 9).

After Thomson (1887b) and Taylor (1921) pioneered the statistical theory of turbulence, the $R_\lambda$–dependent or $R_\lambda$–related study has become how the statistical properties of turbulence change with $R_\lambda$, which can be finite or infinite. For example, just before and after Batchelor (1947) examined in detail Kolmogorov (1941a, c)[K41], to put K41 to the test, extensive studies had been undertaken by George Keith Batchelor and his associates at Cavendish Laboratory, Cambridge, U.K., e.g. Batchelor and Townsend (1947), Townsend (1947), Batchelor and Townsend (1948a, b), Townsend (1948a, b), Batchelor and Townsend (1949a, b), Stewart (1951), and Stewart and Townsend (1951). Since those studies mainly involved the measurements of double and triple velocity correlation functions and of the energy spectrum at different mesh Reynolds numbers, in a sense, they can be cautiously regarded as the $R_\lambda$-dependent study. It is interesting to note that all the mesh Reynolds numbers are *finite* ones, and thus they can also be regarded as the finite $R_\lambda$-related study.

Based on both theoretical and experimental studies, Batchelor and Townsend (1947) found that (a) a negative contribution to $d\overline{\omega^2}/dt$ arises from the effect of viscosity; (b) the skewness factor is approximately constant during decay, with the same value at all Reynolds number; (c) the viscous contribution is always the greater but the contributions tend to equality as the grid Reynolds number increases; (d) the double velocity correlation function tends to a cusp at the origin as the Reynolds number increases indefinitely; (e) there is a dynamical similarity (Kolmogorov 1941a). Townsend (1947) measured the double and triple correlation derivatives in isotropic turbulence. Stewart (1951) studied the effect of mesh Reynolds number on the triple velocity correlation at mesh Reynolds numbers 5300, 21200 and 42400. To assess the validity of the various theories which postulate greater or less degrees of similarity or self-preservation between decaying fields of isotropic turbulence, Stewart and Townsend (1951) studied the effects of mesh Reynolds number, which vary from 2000 to $10^5$, on the double and triple velocity correlation functions and the energy spectrum. They found that the conditions for the existence of the local similarity considered by Kolmogorov and others are only fulfilled for extremely small eddies at ordinary Reynolds numbers, and that the inertial subrange in which the spectrum function varies as $k^{-5/3}$ ($k$ is the wave-number) is non-existent under laboratory conditions. Furthermore, the spectrum function within the range of local similarity is best represented by an empirical function such as $k^{-alogk}$, and all suggested forms for the inertial transfer term in the spectrum equation are in error.

Pao (1965, page 1067, right column, lines 17-23) observed that "Obukhov[47][Oboukhov 1962] and Kolmogorov[48][Kolmogorov 1962] recently refined the original Kolmogorov's hypothesis; they took into consideration the effects due to the intermittency of the high-frequency turbulent fluctuation and the finite Reynolds numbers. One cannot tell whether these effects are present from the accuracy of the measurements reported so far". If Pao is right, the effect due to 'the finite Reynolds numbers' was already studied in Kolmogorov (1962)? Is this really true?

There are important differences between Oboukhov (1962) and Kolmogorov (1962)(Djenidi and Antonia 2020). Oboukhov was trying to correct for the finite Reynolds number effect (Djenidi and Antonia 2020). It was unclear what Kolmogorov was trying to correct for. It is clear however that Kolmogorov assumes the 4/5 law, which is valid strictly for an infinitely large Reynolds number (one could then surmise that Kolmogorov's corrections to K41 may be due to the effect of intermittency when the Reynolds number is infinitely large but I am not at all sure about this). No one questions the presence of intermittency but one should recognise that its effect will depend on the Reynolds number. It is unclear why Pao chose to separate the two effects (Robert A. Antonia with Personal Communication, 1 July 2020). The author generally agrees with Djenidi and Antonia (2020) since Oboukhov (1962) and Kolmogorov (1962) used the different Reynolds numbers as shown in equations (5c) and (5d).

The term 'the very high Reynolds number turbulence' appeared in Batchelor (1961, page 87, line 8 from bottom), while the opposite term, 'low Reynolds number turbulence', appeared in Hanjalic and Launder (1976). In his review of the spectral data for the wake, jet and atmosphere at high wavenumbers, Champagne (1978) attributed this to 'the Reynolds number dependence' of the normalized spectra.

Using a wind tunnel, Kistler and Vrebalovich (1966, page 37, Abstract, lines 2-3; page 42, Figure 2) studied the effect of Reynolds number on the turbulence level ($\tilde{u}/u_0$ and $\tilde{v}/u_0$). They concluded that the spectral curve should follow along the $-5/3$ slope for at least a factor of two in energy, then a value of $(Re_L)_{min}$=300 is obtained, where $(Re_L)_{min}$ is the minimum Reynolds number, at which an inertial subrange might be expected. According to Crowdy and Tanveer (2014, page 390, paragraph 2, lines 5-7), Philip Saffman thought seriously about many of these assumptions and identified some fundamental questions that he believed needed to be resolved before serious progress could be made to turbulence, including (i) the independence of Reynolds number of the dissipation rate; (ii) the dependence of the inertial range, small eddies and intermittency on Reynolds number. Although (i) was studied by Sreenivasan (1998) and Kaneda, Ishihara and Yokokawa (2003), it still remains an important open theoretical problem. Notably, McComb, Berera, Salewski and Yoffe (2010) reinterpreted Taylor (1935) dissipation surrogate, and they suggested that $U^3/L$ is a more appropriate measure of peak inertial flux than of dissipation. McComb, Berera, Yoffe and Linkmann (2015) derived a model for the Reynolds-number dependence of the dimensionless dissipation rate from the dimensionless dissipation rate. McComb and Fairhurst (2018) extended the techniques used by McComb, Berera, Yoffe and Linkmann (2015) to calculate the dimensionless dissipation rate for stationary isotropic turbulence, to the case of free decay. Yoffe and McComb (2018) further studied onset criteria for freely decaying isotropic turbulence.

Double velocity correlation function was first proposed by Taylor (1921) and then used by von Kármán and Howarth (1938) and Batchelor (1947). Correlations functions are self-preserved (Batchelor 1948). High $R_\lambda$ measurments of the longitudinal (second-order) structure function $D_{\mathrm{II}}(r)$ are reported in the literature, e.g. $R_\lambda$=4,300 (van Atta and Chen 1970); $R_\lambda$=966 (Antonia, Satyaprakash and Chambers 1982); $R_\lambda$=616 (Mestayer 1982); $R_\lambda$=852 (Anselmet, Gagne, Hopfinger and Antonia 1984).

Notably, Sato, Yamamoto and Mizushina (1983) proposed empirical equations for the double velocity correlation function $f(r,t)$ and the triple velocity correlation function $k(r,t)$, which describe the statistical structure and dynamic behavior of isotropic turbulence, for 10≤ $R_\lambda$ ≤$10^4$. As shown in Figure 10, Sato, Yamamoto and Mizushina (1983, page 277, Fig. 5) calculated the second order structure functions normalized with the Kolmogorov length scale for for 1≤ $R_\lambda$ ≤$10^4$. Sato, Yamamoto and Mizushina (1983, page 277, Fig. 6) obtained one dimensional energy spectrum functions normalized with the Kolmogorov length scale for 1≤ $R_\lambda$ ≤$10^4$. Based on their own empirical equations in Sato, Yamamoto and Mizushina (1983) and the turbulent energy transfer functions in Obukohov (1946), Kovasznay (1948), Heisenberg (1948b) and Pao (1965), Sato, Yamamoto and Mizushina (1984, page 210, Fig. 1) made a comparison of calculated turbulent energy transfer function $W(k)$ for $R_\lambda$=3,000, 300 and 30. As for the quantification of the effect of Reynolds number on the third-order structure function ($D_{LLL}$) or its equivalence, the energy transfer function, $S(k)$ or $W(k)$, in $k$-space, where $dS(k)/dk = -T(k)$ or $dW(k)/dk = -T(k)$, one may find it in Sato, Yamamoto and Mizushina (1984, page 209, equation (2)). Here any of $k(r,t)$, $T(k)$ and $W(k)$ is simply called "triple moment". As shown in Sato, Yamamoto and Mizushina (1984, page 210, Table 1), one can quantify the dependence of triple moment(s) on $Re$ and wave number ($k$ or $r$) by using a closure hypothesis and a model of the energy spectrum $E$.

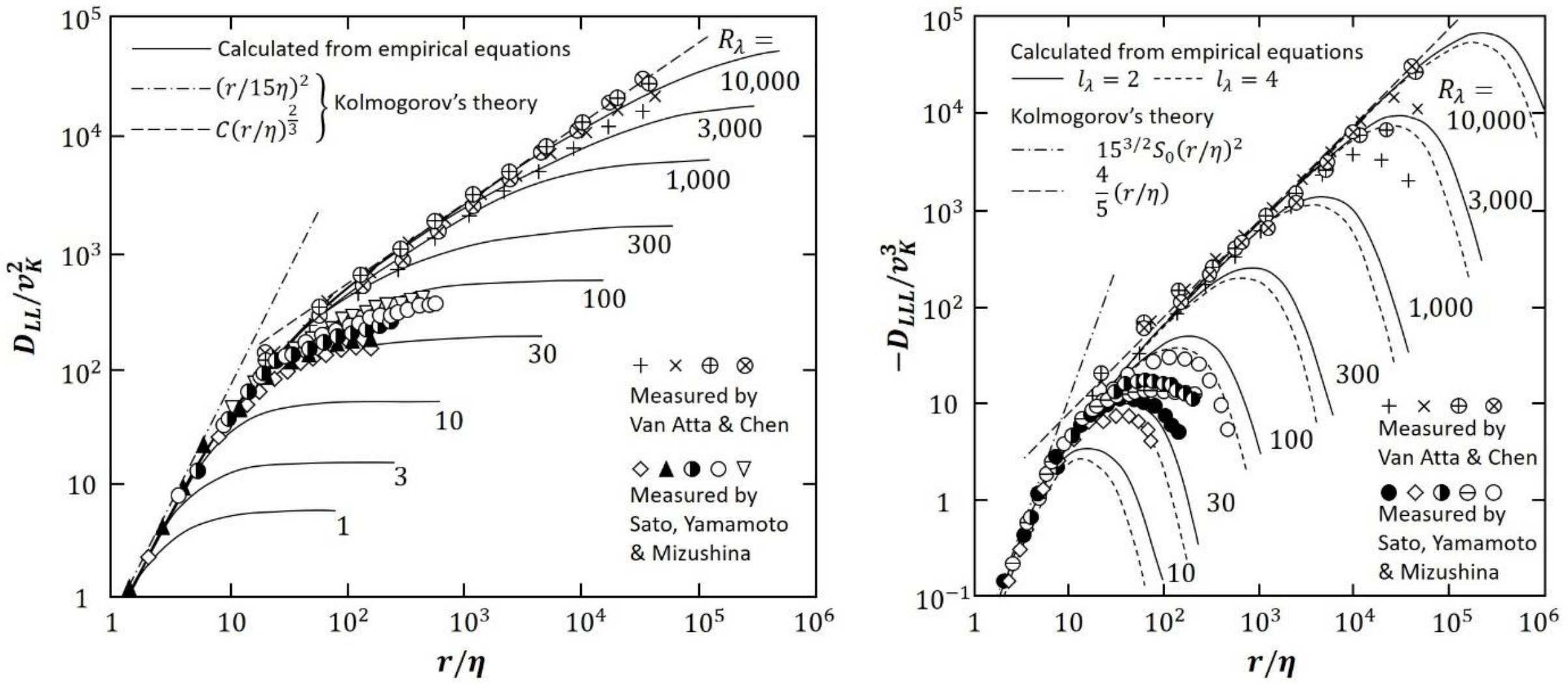

Figure 10 Second-order structure functions (left) and third-order structure functions (right) normalized with Kolmogorov scale (Sato, Yamamoto and Mizushina (1983, page 277, Fig. 5; page 278, Fig. 8).

The term 'the Reynolds number dependence' appeared in Champagne (1978), Antonia,

Chambers, and Satyaprakash (1981), and Antonia, Satyaprakash and Chambers (1982). Using the hypotheses of Kolmogorov and Oboukhov for the probability density and variance of the dissipation fluctuations, van Atta and Antonia (1980) examined the influence of fluctuations in the rate of local turbulent energy dissipation on higher-order (the $n$-th order) structure functions for small separation distances ($r$) and on moments of turbulent velocity derivative and studied $R_\lambda$-dependent skewness and flatness factors of turbulent velocity derivatives. Rey and Rosant (1987) studied the effect of Reynolds number on the determination of isotropic velocity and temperature turbulent length scales. In the case of small Reynolds numbers, with significant spectral transfer, no variation using -5/3 power law in the inertial range can be found.

Based on field measurements, Praskovsky and Oncley (1994) examined the relationship among the Kolmogorov constant, intermittency exponent and the Taylor microscale Reynolds number. The Kolmogorov constant is found to be weakly dependent on $R_\lambda$. The term 'Reynolds number effects' appeared in Gad-el-Hak and Bandyopadhyay (1994), George (1992; 1995, Section 2) and Gamard and George (1999). Gad-el-Hak and Bandyopadhyay (1994) reviewed Reynolds number effects in wall-turbulent flows and tried to address the question as to what are the Reynolds number effects on the mean and statistical turbulence quantities and on the organized motions? By using high-resolution, direct numerical simulations of three-dimensional incompressible Navier-Stokes equations, Martinez, Chen, Doolen, Kraichnan, Wang and Zhou (1997) studied the energy spectrum in the dissipation range. They examined the possible values of the two parameters in the energy spectrum and their dependence on $R_\lambda$. Based on the Kolmogorov equation derived by Novikov (1993), Moisy, Tabeling and Willaime (1999) studied the Kolmogorov function, $K(r) = -S_3/r\epsilon$, versus $r/\eta$, at $R_\lambda$= 120, 300 and 1170. In the author's view, they actually studied low $R_\lambda$ turbulence. Mi, Xu and Zhou (2013) studied in detail Reynolds number influence (or the effect of Reynolds number) on statistical behaviors of turbulent in a circular free jet. Other terms, 'the Reynolds-number effect' and 'Reynolds-number dependent', also appeared in Mi, Xu and Zhou (2013).

*Others' early notions of the Reynolds number effects and finite Reynolds number*

As shown before, in a sense, Kolmogorov (1941a, c) were already aware of the finite $R_\lambda$. To the best knowledge of the author, the term 'Reynolds number effects' first appeared in Corrsin (1942, page 5, line 10). The term 'a Reynolds number effect' appeared in Corrsin (1942, page 14, paragraph 2, lines 7-8). However, Corrsin did not relate it to small-scale turbulence at that time, instead to experimental values of the exponent in the theoretical relation derived by de Kármán and Howarth (1938, page 213, line 4) for the rate of decay of isotropic turbulence in a uniform air stream.

In his famous interpretation of Kolmogorov's (1941a, c) similarity hypotheses, Batchelor (1947, page 541, paragraph 1, lines 5-8) wrote that "the small eddies have finite characteristic Reynolds number and therefore dissipate a certain proportion of their energy, this proportion being unity when the characteristic Reynolds number is low enough for the motion to be entirely laminar." Batchelor was also aware of 'finite characteristic Reynolds number'. As presented before, the term 'the finite Reynolds numbers' also appeared in Pao (1965, page 1067, right column, paragraph 1, line 21). The terms 'finite Reynolds number' and 'Finite Reynolds number turbulence' appeared in Kraichnan (1991, page 66, paragraph 4), viz. "The Kolmogorov theories have profoundly shaped and illuminated thinking about turbulence. But, in one respect, this influence perhaps has been unfortunate: relatively little attention has been devoted to the prediction of turbulence statistics at finite Reynolds number. Finite Reynolds number turbulence has a rich and interesting structure. Moreover, it is likely that the question of intermittency corrections to K41 can be resolved only when a detailed understanding of the dynamics at finite Reynolds numbers has been achieved.”

The term “the finite Reynolds number” appears in George (1992, page 1496, right column, paragraph 2, lines 5-6; page 1497, paragraph 1, lines 4-5 and lines 10-11). The term ‘finite Reynolds number effects’ appeared in Sreenivasan’s conclusion that “It appears that there is indeed intermittency in both dissipative and inertial ranges. However, this conclusion should not mask potential uncertainties from an experimental perspective: finite Reynolds number effects, finite shear effects,…..”(Sreenivasan 1995, page 552, Conclusions, lines 1-5). The term ‘Finite Reynolds number turbulence’ appeared in Kraichnan (1991, page 66, paragraph 4); and ‘finite Reynolds number effects’ in Sreenivasan (1995a, page 552, Conclusions, lines 1-5). The term ‘finite Reynolds numbers’, appeared in Sreenivasan (1995b, page 2783, left column, line 4). The average value of the Kolmogorov constant is 0.53 (Sreenivasan 1995b, page 2782, FIG. 3). Sreenivasan (1995) emphasized that “This is at best an asymptotically valid result, a full assessment of its validity in shear flows at finite Reynolds numbers has not been made.”

*Grossmann and Lohse’s work on finite Reynolds numbers and finite size effects*

A number of studies have been seeking to improve on the K41 analysis by means of matching solutions valid in the inertial range to those valid in the viscous dissipation range (e.g. Effinger and Grossmann 1987). According to McComb (2014, Chapter 6, Section 6.4, page 161), Effinger and Grossmann’s (1987) work is also within the category of Finite-Reynolds-number effects. However, the term ‘Finite-Reynolds-number’ never appeared in Effinger and Grossman (1987), but the term ‘an effective viscosity’, which is finite, was used.

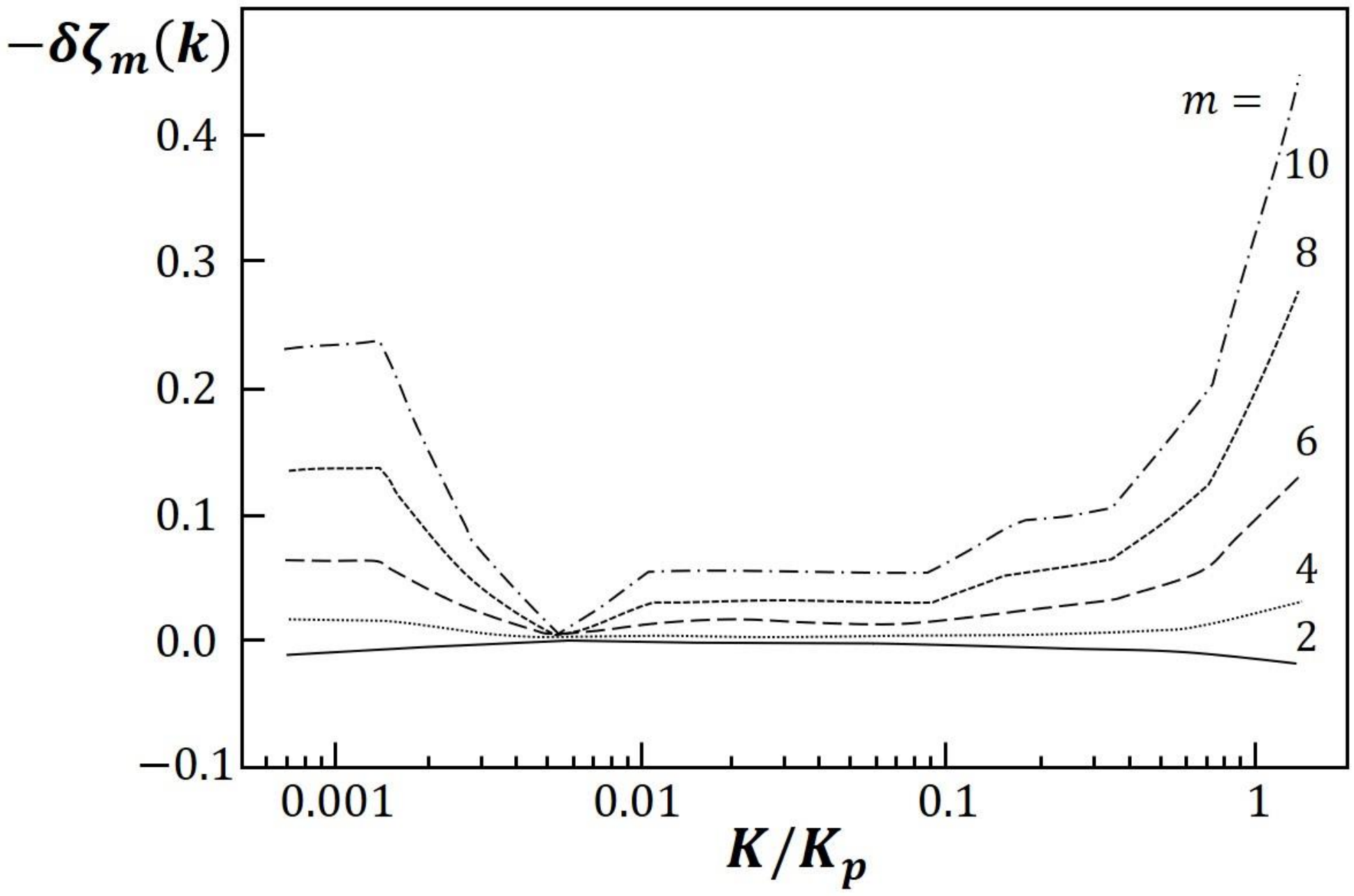

Figure 11 Scale resolved intermittency corrections $-\delta\zeta_m(k)$(Grossmann and Lohse 1994b, page 2788, FIG. 3).

The term ‘finite Reynolds numbers’ appeared in Grossman and Lohse (1993, Abstract). Grossman and Lohse (1993) showed that the $r$-scaling exponents $\mu(q)$ of the moments of the locally averaged energy dissipation are nonzero even for a Gaussian probability distribution of the velocity gradients, which casts – at least for finite Reynolds numbers – additional doubts on the use of the refined similarity hypothesis in K62 and O62 leading to $\zeta_m = m/3 - \mu(m/3)$ for the scaling exponents of the velocity structure functions. They used the presumed universality (i.e. independence of $\mathrm{Re}_\lambda$) of the $\mu(q)$ to predict the degree of the non-Gaussian character of the $\partial_1 u_1$ probability distribution. They explicitly evaluate the $\mathrm{Re}_\lambda$-dependence of its stretching $\beta$ together with the $\mathrm{Re}_\lambda$-scaling exponent of the hyperflatnesses $F^{(q)}$, which are in agreement with recent experiments. Grossmann and Lohse (1994a) calculated the deviations $\delta\zeta_m$ ("intermittency corrections") from classical (K41) scaling $\zeta_m = m/3$ of the $m$th moments $< |u(p)|^m >$ in high Reynolds number turbulence. They introduced the notion of scale resolved intermittency corrections $\delta\zeta_m(p)$ which are extremely small in the inertial subrange. It is found that there is no inertial subrange intermittency in the large $Re$ limit. By extending the numerical simulations of She, Chen, Doolen, Kraichnan, and Orszag (1993) of highly turbulent flow with $\mathrm{Re}_\lambda$=45,000, Grossmann and Lohse (1994b) studied universality in fully developed turbulence (Figure 11).

The terms 'Finite Size Corrections' and 'finite size effects' appeared in Grossmann, Lohse, L'vov and Procaccia (1994, page 432). Numerical simulations have been used to decide whether classical [normal] or anomalous scaling could be expected in high Re flows. However, the range of scales for which power law behavior is observed at such values of $\mathrm{Re}_\lambda$ may be insufficient to distinguish between classical [normal] and anomalous scaling. Furthermore, one can have corrections to scaling due to finite size effects at low to moderate Re, and it is hard to take these into account if their expected Re dependence is unknown. To this end, Grossmann, Lohse, L'vov and Procaccia (1994) examined these corrections to scaling analytically and numerically. The following power law is obtained by Grossmann, Lohse, L'vov and Procaccia (1994, page 433, the power law (4)):

$$\delta\zeta_m(\mathrm{Re}) = c_m \mathrm{Re}^{-\beta_m} \quad (30)$$

where $\delta\zeta_m$ refers to the deviations, $\zeta_m$ to the scaling exponents, $c_m$ and $-\beta_m$ two constants. Their results are shown in Figure 12.

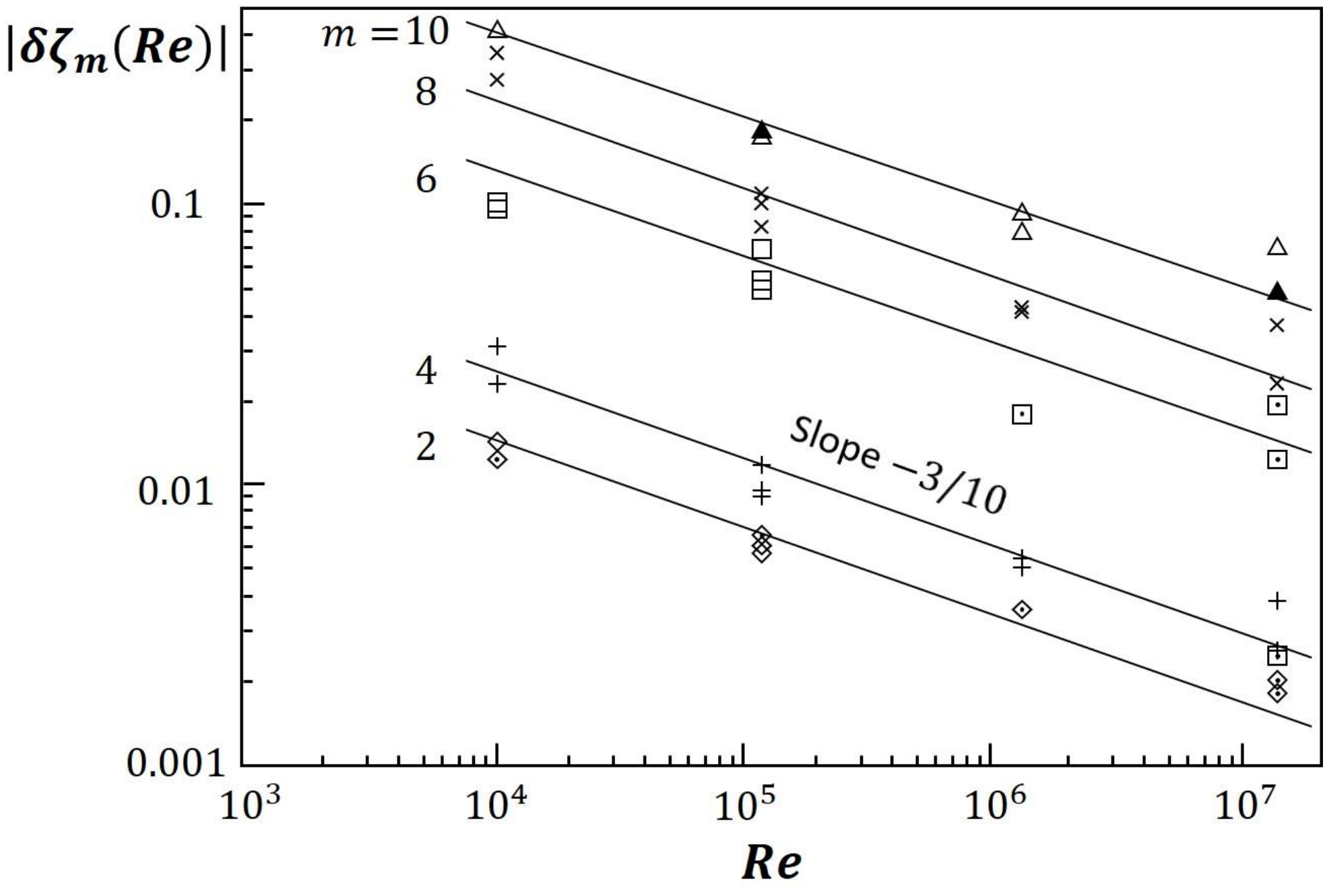

Figure 12 Results from the reduced wave vector set approximately for $|\delta\zeta_m(\mathrm{Re})|$(Grossmann, Lohse, L'vov and Procaccia 1994, page 433, FIG. 1).

What is the physics of Finite Size Corrections or finite size effects? The author contacted the second author of Grossmann, Lohse, L'vov and Procaccia (1994), Detlef Lohse, about it. Lohse confirmed that 'finite size corrections' meant Finite Reynolds number (Detlef Lohse with Personal Communication, 31 July 2020). According to Vainshtein and Sreenivasan (1994, page 3085, right column, paragraph 2, lines 4-6), Grossmann, Lohse, L'vov and Procaccia (1994) suggested that the observed deviations from K41 could arise partially from artifact of finite Reynolds numbers.

d'Alembert (1752) proposed the paradox that the drag force is zero on a body moving with constant velocity relative to the fluid. A problem closely related to d'Alembert's paradox in the context of small-scale turbulence is that of the normalized dissipation rate. It is a classic problem, and the importance of the finite Reynolds number (FRN) effect has been well-known (e.g. Lohse 1994). As for the FRN effect on the second order correlation structure function ($D_{LL}$), or equivalently the spectrum $E(k)$, one may refer to e.g., Eq. (9) in Lohse and Muller-Groeling (1995). The FRN effect on the third order correlation $D_{LLL}$ can be readily estimated by $D_{LL}$ or $E(k)$, and the (generalized) Kármán-Howarth equation.

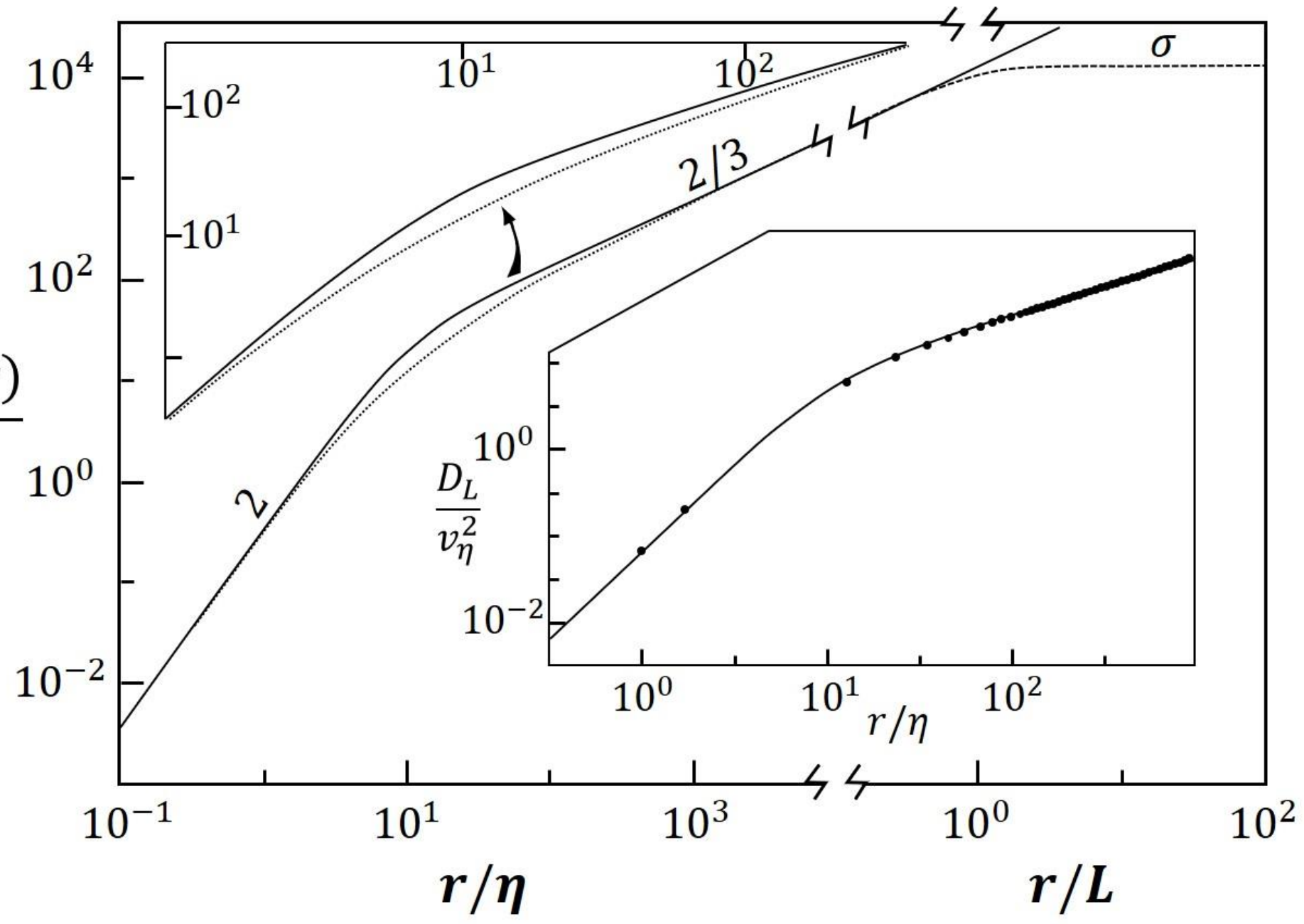

Figure 13 Velocity structure functions $D(r)$, calculated from Benzi *et al.*'s data (dashed), and its Batchelor fit (1)(solid) for $\zeta = \frac{2}{3}$ (Lohse and Muller-Groeling 1995, page 1748, FIG. 1).

As shown in Figure 13, the second crossover at $r = L$ is described by a Batchelor type transition (Lohse and Müller-Groeling (1995, page 1749, equation (9)):

$$D(r) = 2\langle u^2\rangle r^2 \cdot (r_d^2 + r^2)^{-1+\zeta/2} \cdot (L^2 + r^2)^{-\zeta/2} \qquad (31)$$

*Qian's work on the finite $R_\lambda$ effect of the third-order structure function* ($D_{LLL}$)

It is not so easy to determine the year when Qian became interested in the finite $R_\lambda$ effect of turbulence. The study of the finite $R_\lambda$ effect of turbulence by Qian may roughly begin in 1986 (e.g. Qian 1986a), when he studied the intermittency of turbulence and 1990 (e.g. Qian 1994a). Using non-equilibrium statistical mechanics closure model, Qian (1994a) found that the skewness factor ($S$) of the velocity derivative of isotropic turbulence approaches a constant of -0.515 when the Taylor microscale Reynolds number is very high. Qian (1994a, page 14, Fig. 1) showed the skewness factor ($S$) of the velocity derivative of isotropic turbulence versus $R_\lambda$ only for $10 \leq R_\lambda \leq 10^3$. Qian (1994a, page 15, paragraph 2, lines 1-3) concluded that "The issue, whether the skewness $S$ and the flatness $F$ approach a finite constant as $R_\lambda \rightarrow \infty$, is essentially related to the problem of universality of small-scale structure for high-$R_\lambda$ turbulent flows". The major finding of Qian (1994a) was that the skewness $S$ and the flatness $F$ become universal constants independent $R_\lambda$ of when $R_\lambda$ is high enough, probably implying the universality of small-scale structure of high-$R_\lambda$ turbulent flows. Clearly, those important findings stimulated Qian to study the finite Reynolds number effect (Qian 1997).

What is the finite Reynolds number effect of turbulence? Qian (1997, page 339, last paragraph) definition is that "The finite Reynolds number effect mainly refers to the situation that within the scaling range found in experiments and simulations some small-scale statistics deviate from the prediction of idealized inertial-range models of infinite Reynolds number. A concrete measure of the finite Reynolds number effect depends upon which small-scale property is studied". Firstly, Qian reviewed the concept of inertial range, which plays a central role in the statistical physics of turbulence, proposed by Kolmogorov. Secondly, Qian found the following (Qian 1997, page 337, paragraph 1):

> (a) the idealized model of inertial range corresponds to the asymptotic case of infinite Reynolds number; (b) field observations, laboratory experiments, and numerical simulations are made for turbulent flows at the finite Reynolds number; (c) when the Reynolds number is high enough, small-scale statistics within some small-scale range of finite Reynolds number turbulence can be described by the idealized model of the inertial range; (d) the finite Reynolds number effect approaches zero as the Reynolds number becomes higher and higher; (e) the finite Reynolds number effect may persist while the Reynolds number approaches infinity; (f) it is important to know quantitatively how fast the finite Reynolds number effect approaches zero as the Reynolds number becomes higher and higher.

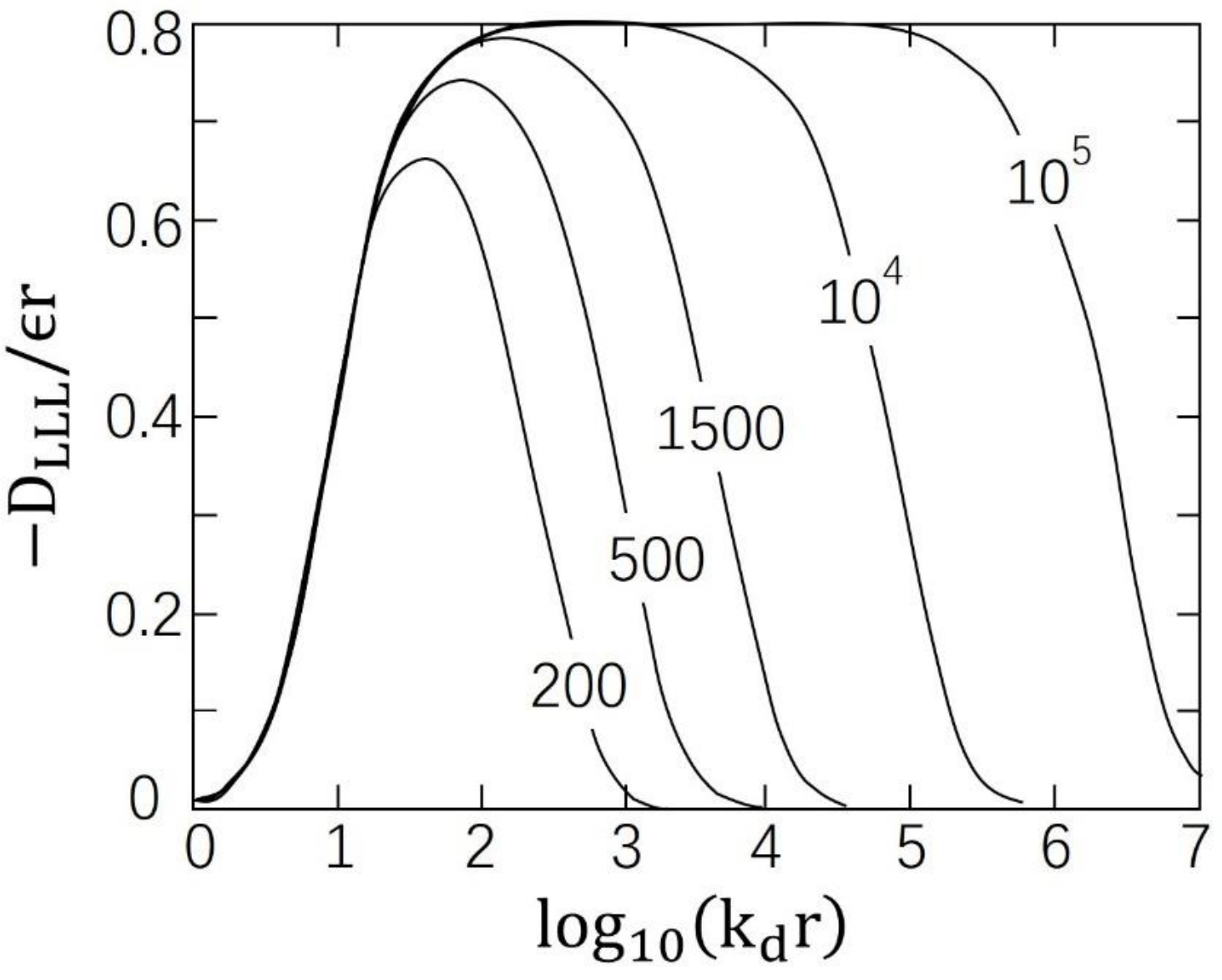

Figure 14 $-D_{LLL}(r)/\epsilon r$ against $\mathrm{Log}_{10}(k_d r)$ for the finite $R_\lambda$ turbulence at $R_\lambda$=200, 500, 1500, $10^4$ and $10^5$. $k_d$ is the Kolmogorov wave number (Qian 1997, page 337, FIG. 1).

Most importantly, as shown in equations (6b) and (6d), by eliminating or neglecting both the large-scale effect and the viscous effect in de Kármán and Howarth (1938, page 206, equation (51))(Kolmogorov 1941c, page 15, equations (3) and (4)), the 4/5 law was obtained (Kolmogorov 1941c, page 15, equation (7). It was Qian (1997, page 337, left column, paragraph 2, lines 11-14) who first recognized that "In fact, the viscous effect and the large-scale effect cannot be neglected within the scaling range observed in experiments and simulations, and the scaling range is not the same as Kolmogorov's inertial range."

*Qian's work on the width of the inertial range and the high value of* $R_\lambda$

As a historical note first, the largest value of $R_\lambda$ attained in a well-defined turbulent flow appears to be that in Laufer's (1954) pipe, $R_\lambda \approx 250$, Laufer does find some evidence for an isotropic inertial subrange (Corrsin 1958, page 13, paragraph 4). It seems worthwhile to set up a turbulence flow with still higher values of $R_\lambda$ then those attained by Laufer (1954), perhaps greater than or equal to 500 (Corrsin 1958, page 13, last paragraph, lines 1-3). Clearly, Corrsin (1958) made estimates and found that $R_\lambda$ of the order 1000 would be needed for universal spectral form to exist, but did not assign a precise value. Corrsin (1958) did not know about the bottleneck effect and thus his estimates are not really correct, however, he seemed to be aware of what 'high $R_\lambda$' means.

It is clear that those facts above motivated Qian's study of the $R_\lambda$ effect, especially (a) the quantitative understanding of the finite $R_\lambda$ effect; and (b) the difference between the inertial range and the scaling range. To study this, since there is an exact inertial range relationship for the third-order structure function ($D_{LLL}$), Qian used it to investigate the finite $R_\lambda$ effect, in particular to address how the width of the inertial range of finite $R_\lambda$ turbulence changes with $R_\lambda$. Qian (1997) found that (a) there is no inertial range when $R_\lambda \leq 2000$; (b) $R_\lambda$ should be higher than $10^4$ so as to have an inertial range wider than one decade; (c) the so-called 'inertial range' obtained in experiments and simulations is just a scaling range and not the same as Kolmogorov's inertial range; (d) the finite $R_\lambda$ effect should be taken into account within such a scaling range when a comparison is made between experimental and simulated results and theoretical ones of the inertial range statistics. As shown in Figure 14, the 4/5 law can only be attained when $R_\lambda$ is sufficiently large ($10^5$), which was supported by Antonia and Burattini (2006). Extrapolation of the results in Mydlarski and Warhaft (1996, page 320) suggests that the power 5/3 will occur in this flow for $R_\lambda \sim 10^4$. This value is consistent with Qian (1997).

The third bump model is written as (Qian 1997, page 339, equation (8b, c))

$$F(x) = (1 + Bx^\alpha g)\exp(-Cx^\beta) \quad (32a)$$

$$g = [1 + C_1 Z + C_2 Z^2 + \cdots + C_m Z^m]^2 \ , \ Z = x^\gamma \quad (32b)$$

The parameters $B$, $\alpha$, $C$, $\beta$, $C_1$, $C_2$, …, $C_m$ and $\gamma$ are adjusted to minimize the error of the spectrum form of the von Kármán-Howarth equation for stationary turbulence (Qian 1997, page 338, equation (6a)).

By adopting Qian's (1986) model, the following energy spectrum within the energy-containing range is obtained Qian (1997, page 339, equation (11)):

$$E(k) = K_o \epsilon^{2/3} k^{-5/3} F(k/k_d)/[1 + (k_0/k_d)^{n+5/3}] \quad (33)$$

Qian (1997, page 342, bottom/last 5 lines) wrote that "At present we are far from understanding the finite Reynolds number effect. This paper represents a preliminary effort to understand the finite Reynolds number effect in the case of the third-order structure function $D_{LLL}(r)$".

By use of the rms fluctuating velocity $u'$ and $R_\lambda$, Sato, Yamamoto and Mizushina (1983) proposed an empirical equation for the double velocity correlation in the de Kármán-Howard equation. The triple correlation function $[k(r,t)]$ in Sato, Yamamoto and Mizushina (1983, 1984) is equivalent to the third order structure function $D_{LLL}\ (r)$ in Qian (1997) apart from a normalization factor. However, the method of calculation $k(r,t)$ in Sato, Yamamoto and Mizushina (1983, page 277, Section 3.2) was empirical, while the method of calculation $D_{LLL}\ (r)$ in Qian (1997, pages 338 and 339) more mathematically and physically sensible and justifiable. In Qian (1997, page 337, FIG. 1.), the third-order structure functions are normalized with the Kolmogorov wave number, and thus identical to Sato, Yamamoto and Mizushina (1983, page 278, Fig. 8) in which the third-order structure functions were normalized with the Kolmogorov length scale. Note that Qian's result was for $200 \leq R_\lambda \leq 10^5$. Although the second order and the third order structure functions for $1 \leq R_\lambda \leq 10^4$ were obtained in Sato, Yamamoto and Mizushina (1983, page 277, Fig. 5; page 278, Fig. 8) (Figure 10), the results were not clearly interpreted with regard to the controversy surrounding the K41. $k(r,t)$ and $T(k)$ (or $W(k)$) are equivalent to each other in the sense that one is readily known from the other by a simple transformation (cf. Yamamoto and Mizushina 1984, page 209, equation (4)); Qian 1997, page 338, equation (2)). Like Sato, Yamamoto and Mizushina (1983, 1984), Qian (1997) also used a closure hypothesis and a model(s) of the spectrum $E$ for the estimate/quantification of triple moment(s), but the closure hypothesis and the model of $E$ Qian used, which are more convincing, are different from those used in Sato, Yamamoto and Mizushina (1984).

Benzi, Tripiccione, Baudet, Massaioli and Succi (1993) proposed Extended Self Similarity to measure the anomalous scaling exponents in turbulence, i.e. accounting for the intermittency in turbulence. Apparently, after Qian (1997) proposed the finite Reynolds number effect, he was naturally motivated to re-examine the Extended Self-Similarity (ESS) proposed by Benzi, Tripiccione, Baudet, Massaioli and Succi (1993) and Benzi, Biferale, Ciliberto, Struglia, and Tripiccione (1996). After a thorough theoretical re-analysis of the relative scaling of $D_{LL}(r)$ versus $-D_{LLL}(r)$, where $D_{LLL}(r) = \langle \Delta u_r^3 \rangle$ is the third-order structure function, Qian (1998a) found that (a) in the log-log plot of $D_{LL}(r)$ versus $-D_{LLL}(r)$, an approximate relative scaling law is observed over the range $4 < r/\eta < 10^3$, which is the ESS range adopted by Benzi, Tripiccione, Baudet, Massaioli and Succi (1993) while they obtain $S_2 = 0.7$; (b) the local slope of the log-log plot is not constant and has a remarkable bump in the ESS range, so that the relative scaling law is not valid in a strict sense; and (c) the ESS scaling exponent $S_2$ is greater than the real inertial range scaling exponent $\zeta_2$ for both normal scaling ($\zeta_2 = 2/3$) and anomalous scaling ($\zeta_2$=0.7), i.e. $S_2 \gg 0.7$ for $\zeta_2 = 2/3$ and $S_2 > 0.7$ for $\zeta_2 = 0.7$. Inertial range scaling is robust and does not tend to K41, as far as the analysis of data [is concerned](Personal Communication with Luca Biferale, 02 April 2020).

Qian (1998a) further argued that the experimental and numerical results ($S_2 = 0.7$) have previously been interpreted as clear evidence of anomalous scaling ($\zeta_2 = 0.7$) of $D_{LL}(r)$ based upon the assumption that $\zeta_2 = S_2$. However, Qian's results show that the data actually favour the Kolmogorov 2/3 law ($\zeta_2 = 2/3$) rather than anomalous scaling ($\zeta_2 = 0.7$). Although Qian did not explicitly explain why, in the author's view this can be best explained by taking the finite Reynolds number effect into account.

As shown in Figure 15, the log-log plot of $D_{LL}(r)$ against $-D_{LLL}(r)$ over the extended self similar range $4 < r/\eta < 10^3$ for the third bump model equation (Qian 1997, page 339, equation (8b, c)), $K_o = 1.2$, and equation (10a) of Qian (1997, page 339), an approximate relative scaling law is clearly observed. As shown in Figure 16, there is a remarkable bump in the EES range, suggesting that the relative scaling law is not really valid.

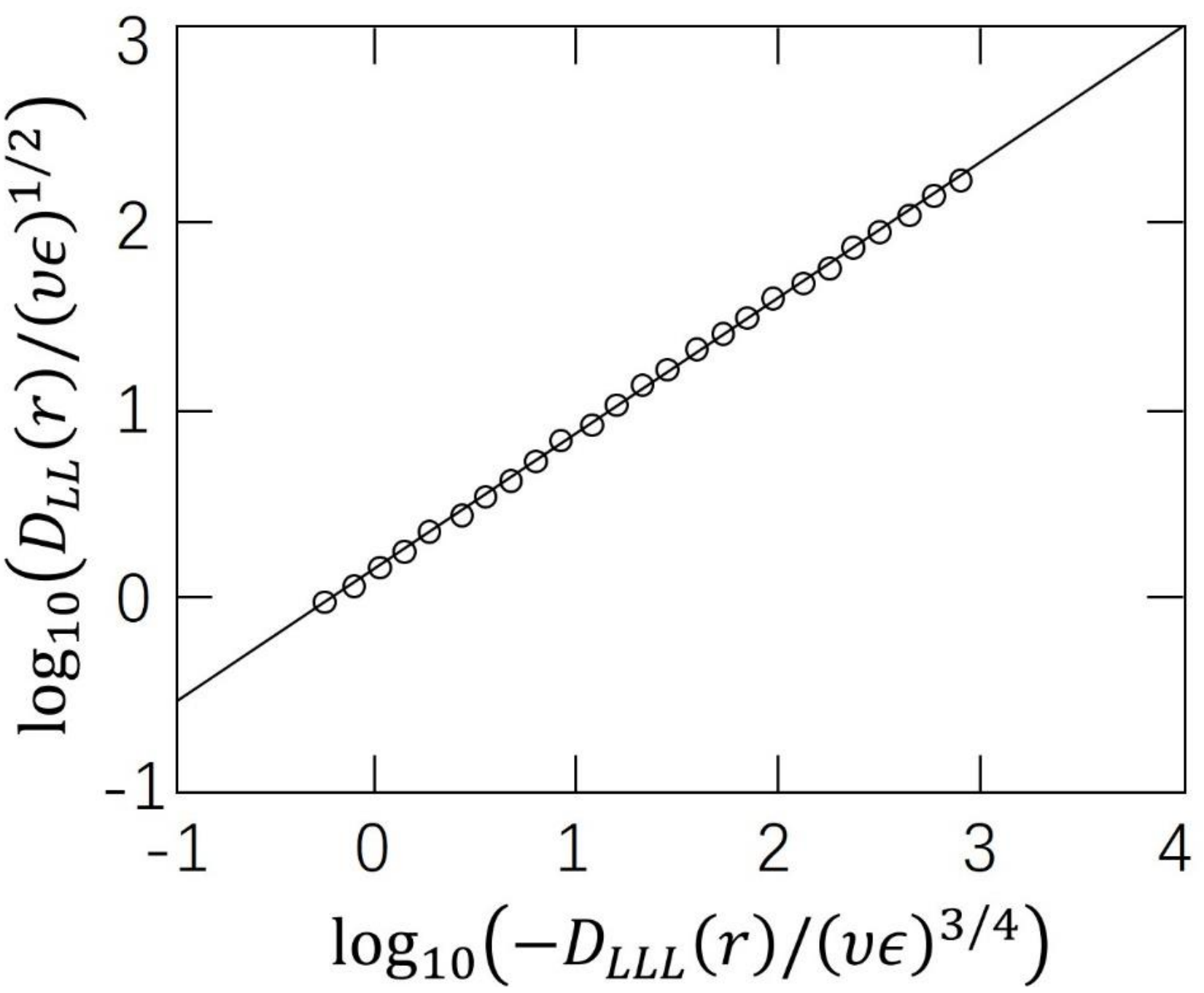

Figure 15 $D_{LL}(r)$ against $-D_{LLL}(r)$ over the extended self similar range $4 < r/\eta < 10^3$ for the third bump model equation (Qian 1997, page 339, equation (8b, c)), $K_o = 1.2$, and equation (10a) of Qian (1997, page 339)(Qian 1998a, page 3201, Figure 9).

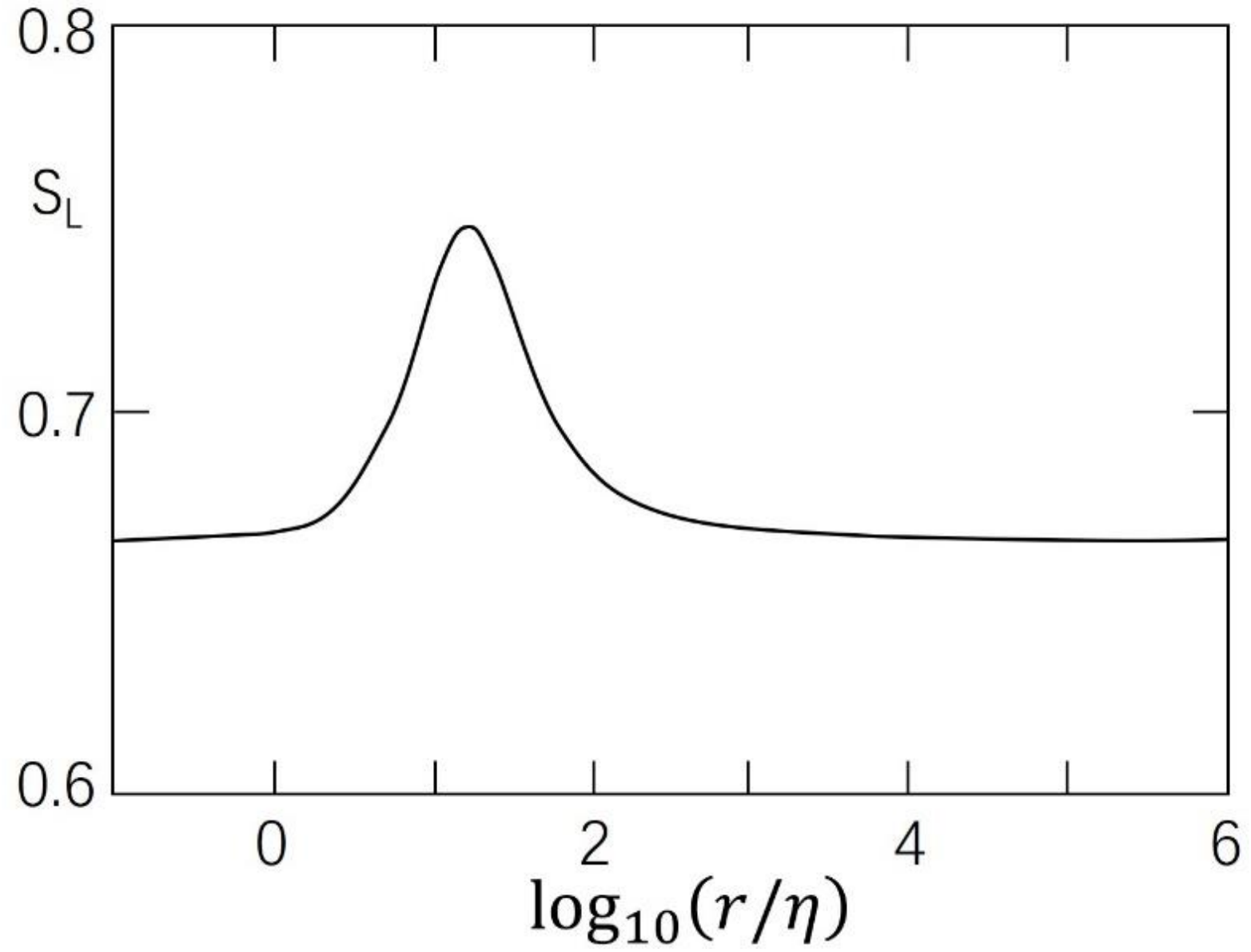

Figure 16 The local slope $S_L$ against of the third bump model equation (Qian 1997, page 340, equation (13)), $K_o = 1.2$, and equation (10a) of Qian (1997, page 339) (Qian 1998a, page 3201, Figure 10).

Qian (1998b) proposed a non-Gaussian model of the probability density function of $|\Delta u_r|$ to study how the scaling exponents $S_p$ of the structure function $\langle|\Delta u_r|^p\rangle$ of finite $R_\lambda$ turbulence depends upon $R_\lambda$. The model can predict both anomalous scaling ($S_p < p/3$ for $p > 3$) observed in experiments at a finite $R_\lambda$ and normal scaling ($S_p = p/3$) while $R_\lambda$ is very high. In contrast to the prevailing multiscaling models available in the literature, the non-Gaussian probability density function model suggests that the anomalous scaling observed in experiments is a finite $R_\lambda$ effect, and the normal scaling is valid in the real Kolmogorov inertial range corresponding to an infinite $R_\lambda$.

Apparently, after the idea of the finite $R_\lambda$ effect of turbulence was conceptually proposed (Qian 1997, 1998a, b), Qian attempted to quantify this effect by using some specific terms. To this end, Qian (1999) used the third-order structure function to study the finite Reynolds number (FRN) effect of turbulence, i.e. the deviation of turbulence statistics observed at finite $R_\lambda$ from predictions of the Kolmogorov theories. He found that the finite $R_\lambda$ effect decreases as $CR_\lambda^{-\mu}$, when $R_\lambda$ is high, and $\mu \leq 6/5$. In memoriam of Qian Jian's contribution, the finite $R_\lambda$ effect of turbulence is expressed by the symbol $Q_e$, in which $Q$ refers to the first letter of 'Qian', the subscript $e$ to the first letter of 'effect'.

In a sense, Lohse and Müller-Groeling (1995, page 1749, equation (9)), Qian (1997, page 339, equation (11)) and Qian (1998b, page 7327, equation (12)) are an extension of Batchelor's (1951, page 371, eq. (7.10)) approximation (McComb 2014, Chapter 6, pages 162-163).

*Qian's work on the decay law of the finite $R_\lambda$ effect of turbulence*

By using the exact spectral equations, Qian derived the decay law of the finite $R_\lambda$ effect as follows (Qian 1999, page 3409, equations (6); page 3412, equation (29)):

$$\delta = C_\delta R_\lambda^{-\mu} \text{ and } \phi = C_\phi R_\lambda^{-\mu} \text{ if } R_\lambda \gg 1 \tag{34}$$

$$\mu = 6m/(3m+4) \tag{35}$$

$$C_\delta = \left[1 + \frac{4}{3m}\right] D,\ C_\phi = 7.069\left(\frac{4}{3} + m\right) D \tag{36}$$

$$D = C_2^{\frac{3m}{(3m+4)}} \frac{\left(\frac{3mC_1}{4}\right)^{\frac{4m}{(3m+4)}}}{B^{\frac{4m}{(3m+4)}}} \tag{37}$$

where the shortest distance ($\delta$) and the curvature ($\phi$) are the proper measure of the finite $R_\lambda$ effect. The constant and decay exponent are able to be determined for typically fully developed turbulent flows.

It should be mentioned that the two relations (34) above are very closely related to [$\delta\zeta_m(\mathrm{Re}) = c_m \mathrm{Re}^{-3/10}$] as shown in Grossmann, Lohse, L'vov and Procaccia (1994, Abstract, line 4), [$\delta\zeta_m^{(app)} = c_m \mathrm{Re}^{-3/10}$] in Grossmann, Lohse, L'vov and Procaccia (1994, page 435, equation (23)), and are similar to [$\delta\zeta_m(\mathrm{Re}) = c_m \mathrm{Re}^{-\beta_m}$] (Grossmann, Lohse, L'vov and Procaccia 1994, page 433, the power law (4)).

*The Kolmogorov–Novikov–Qian equation*

It is feasible that the finite $R_\lambda$ effect can be quantitatively predicted. Qian (1999, page 3411, equation (24)) obtained the following equation:

$$-\frac{D_{LLL}(r)}{\varepsilon r} = 0.8 - C_1\left(\frac{r}{r_c}\right)^m - C_2\left(\frac{r}{\eta}\right)^{-\frac{4}{3}}, \text{ if } \eta \ll r \ll r_c \tag{38}$$

where $D_{LLL}(r)$ is the third-order structure function, $r$ the distance apart, $\varepsilon$ the energy dissipation rate, $C_1$ the coefficient, $r_c = 1/k_c$, $k_c$ the characteristic wave number of energy input, $C_2$ the coefficient, and $\eta$ the Kolmogorov length scale. In which, the last two terms, i.e. $C_1\left(\frac{r}{r_c}\right)^m$ and $C_2\left(\frac{r}{\eta}\right)^{-\frac{4}{3}}$, correspond to the finite $R_\lambda$ effect. In the author's view, by comparison with Kolmogorov (1941c, page 15, equation (5); page 20, relation (7)) and Novikov (1965, page 1292, equation (3.10)), the above equation (35), i.e. Qian (1999, page 3411, equation (24)), can be called the Kolmogorov–Novikov–Qian equation [being confirmed by Evgeny Alekseevich Novikov, 23 June 2020], which accounts for the finite $R_\lambda$ effect of turbulence, although similar forms are derived in Kaneda, Yoshino and Ishihara (2008, page 3, relation (33)) and Ishihara, Gotoh and Kaneda (2009, page 171, equation (7)). It is also clear from their papers that Novikov (1965, page 1292, equation (3.10)) and Qian (1999, page 341, equation (24)) independently derived the similar equation using the different approaches.

As mentioned in Qian (1998b, page 7328, right column, paragraph 1, lines 6-9), Grossmann, Lohse, L'vov and Procaccia (1994) and Vainshtein and Sreenivasan (1994) have studied the issue of the finite Reynolds number effect on the scaling exponents from different views. Qian's approach [Qian 1998b, 1999] is more convincing since his point of departure is the de Kármán-Howarth equation and he does not have to assume that $\zeta_3$= 1, while this assumption is explicitly made in Grossmann, Lohse, L'vov and Procaccia (1994, page 433, left column, lines 12 and 14). This assumption CANNOT be justifiable if $R_\lambda$ is finite. In any case, Equation (4) in Grossmann, Lohse, L'vov and Procaccia (1994, page 433, the power law (4)), which is purely empirical, unlike Qian's derivation (Qian 1999, page 3409, equation (6)), indicates that the 'anomaly' disappears, regardless of $m$, when $R_\lambda$ goes to infinity.

*Qian's closure approach to high-order structure functions of turbulence*

By using the non-Gaussian statistical model to the closure problem of turbulence in Qian (1998b), Qian (2000) studied the scaling law of high-order structure functions of turbulence. The major procedures include: (a) to determine $D_{LL}(r) = \langle \Delta u_r^2 \rangle$; (b) to obtain $\langle \Delta u_r^3 \rangle$ and $\langle |\Delta u_r|^3 \rangle$ from the Kolmogorov equation and the relation between $\langle \Delta u_r^3 \rangle$ and $\langle |\Delta u_r|^3 \rangle$; (c) to derive the high-order $\langle |\Delta u_r|^p \rangle (p > 3)$ by using the non-Gaussian PDF model. He concluded that the available data of the scaling exponents favour the K41 theory rather than the K62 anomalous scaling predicted by various intermittency models. Clearly, this finding should be of great importance to our understanding of the physics of turbulence since the K41 and K62 theories imply completely different physics of turbulence. Qian's (2000) finding and arguments supported Kraichnan's (1974) view that K62 is not a 'refinement' of K41, but a different picture of turbulence (McComb 2014, page 144).

Qian (2001) wrote a 32 page paper entitled Quasi-closure and scaling of turbulence. Importance of the finite $R_\lambda$ turbulence and the finite $R_\lambda$ effect has been re-emphasized in Qian (2001, page 1115, last paragraph, lines 10-20):

> Finite Reynolds number turbulence has rich and interesting structures, at present we are far from understanding the FRN effect. In most cases, the turbulent flows at finite $R_\lambda$ are turbulent shear flows, and the shear effect, i.e. the effect of large-scale shear motion on small-scale statistics, is an intrinsic part of the FRN effect. In the framework of homogeneous isotropic turbulence, we use an isotropic energy source to model the effect of large-scale shear motion by Batchelor averaging, and can not take into account some nonisotropic aspect of shear effect. A complete understanding of the shear effect as an intrinsic part of the FRN effect is indispensable for resolving issues on small-scale statistics of the shear-flow turbulence at finite $R_\lambda$ (jet flow, atmospheric boundary layer flows, and so on), and is a challenge for turbulencists.

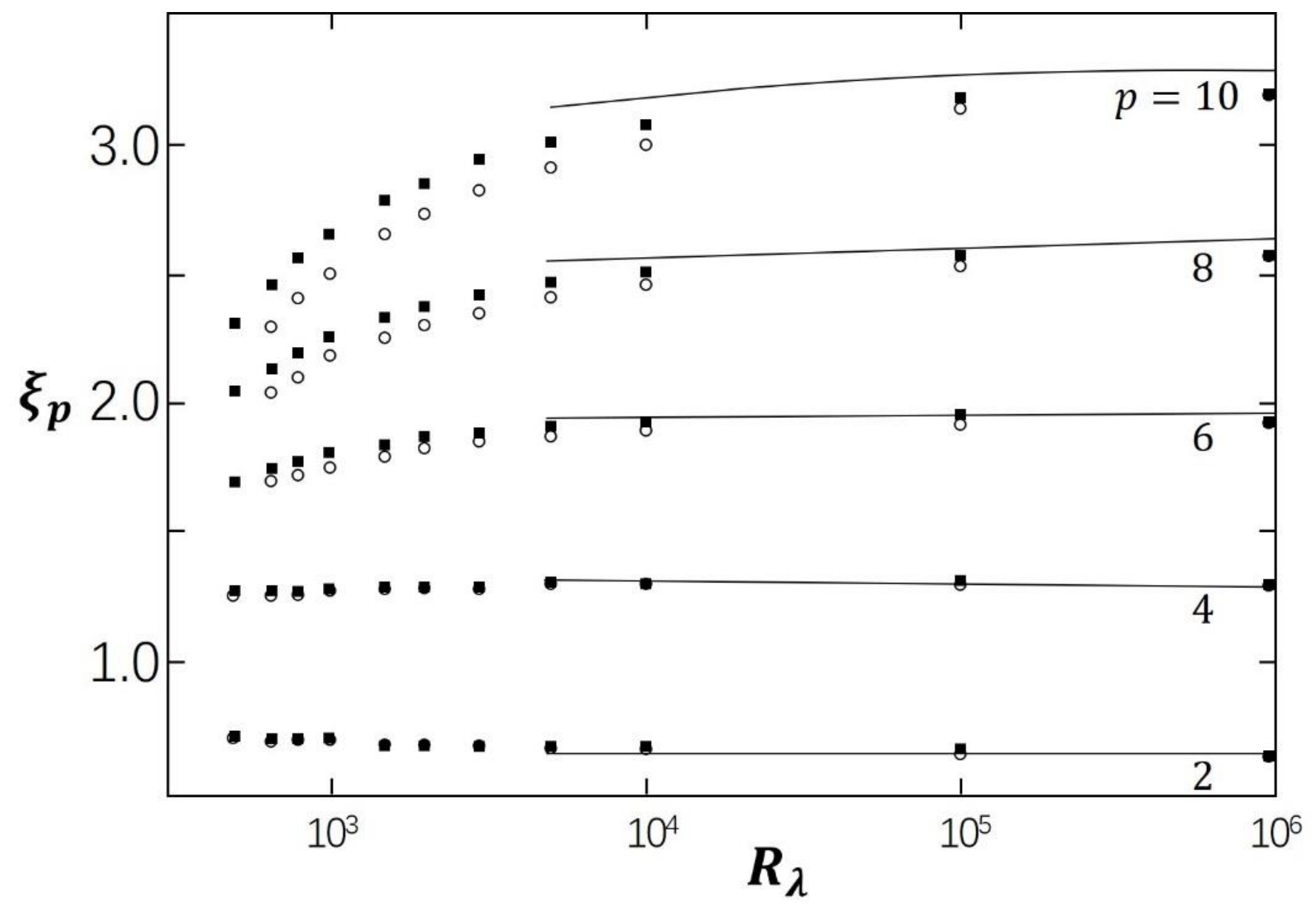

Figure 17 The dependence of the absolute scaling exponent $\xi_p$ upon $C_D$ and $R_\lambda$ for $p$=2, 4, 8, and 10; the line represent $C_D$=0.795; the square $C_D$=0.75; the circles $C_D$=0.7 (Qian 2001, page 1099, Fig. 7).

As shown in Qian (1998, page 7328, FIG. 1) and Qian (2001, page 1099, Fig. 7)(Figure 17), the normal scaling (K41) is approaching as $R_\lambda \to \infty$. By applying spectral dynamics and non-Gaussian statistical model of velocity difference, Qian (2002) studied the scaling of structure functions in homogeneous shear-flow turbulence. It was found that the scaling is proportional to the ratio of the shear length scale ($L_S$) and the viscous length scale ($\eta$). Specifically, when the ratio ($L_S/\eta$) is finite, due to the combined effect of viscosity and mean shear, the scaling deviates from normal one, and its deviation increases with decreasing ratio. When the ratio is small than 100, i.e. the presence of a strong shear, the deviation of the scaling is substantially larger than the predicted one by typical intermittency models. When the ratio ($L_S/\eta$) is infinite, the normal scaling is valid within the inertial range where viscous and shear effects are negligible.

*Qian's work on the equality about the velocity derivative skewness*

Since the $R_\lambda$–dependence of the velocity derivative skewness ($S$) is intimately related to the issue of K41 and K62, Qian (2003) described the two methods of calculating the second-order structure function $D_{LL}(r)$ and applied them to the case of K41 ($\zeta_p$=2/3) as well as K62 ($\zeta_p$=0.7). By using exact relations of isotropic turbulence and various typical models of second-order structure function and energy spectrum, the following equality was obtained (Qian 2003, page 1005, equation (4)):

$$-S = C\left(\frac{k_c}{k_d}\right)^{\beta} \text{ if } R_\lambda \gg 1,\ \beta = 2 \quad (39)$$

where $S$ is the velocity derivative skewness, $C$ a coefficient, $k_c$ the center wavenumber of energy dissipation spectrum, $k_d$ the Kolmogorov wavenumber, and $\beta$ the constant.

In Qian's own words, the main result of Qian (2003) is equality (equation (39), which is applicable in the case of $\zeta_2$>2/3 (intermittency models of K62) as well as in the case of $\zeta_2$=2/3 (K41). The importance of the study of $R_\lambda$–dependence of $\frac{k_c}{k_d}$ is that it can help us to resolve the issue of K41 and K62.

*Qian's work on finite* $R_\lambda$ *turbulence and the two different turbulences* (K41 and K62)

The term 'finite–$R_\lambda$ turbulence' appeared 7 times in Qian (2006). Upto now, the following equation was Qian's last published mathematical and physical relation with regard to the small–scale turbulence (Qian 2006b, page 9, relation (20)):

$$\xi_2 > \zeta_2 \text{ and } \xi_p < \zeta_p \text{ if } p > 3 \quad (40)$$

In his own concluding remarks, the above relation has an interesting implication for the issues of K41 and K62:

> Relation (20) has an interesting implication for the issues of K41 and K62. Based upon the assumption that experimental exponents $\xi_p$ are the same as the inertial-range exponents $\zeta_p$, the experimental fact of $\xi_p$ being anomalous has been interpreted as evidence supporting K62 and against K41. For example, the well-known data ($\xi_p$ being a nonlinear function of $p$) of Anselmet et al. [21][Anselmet, Gagne, Hopfinger, and Antonia 1984] are interpreted as experimental evidence supporting K62, and "have played an important role in the turbulence theory because they have directly motivated the introduction of the multifractal model"(Frisch [6])[Frisch 1995]. However, if the assumption $\xi_p = \zeta_p$ is not valid according to (20) and figure 6, then the data of Anselmet et al. [Anselmet, Gagne, Hopfinger, and Antonia 1984] may be consistent with K41 ($\zeta_p = p/3$ ) and could not be interpreted as an evidence supporting multifractal intermittency models, because the data of Anselmet et al. [Anselmet, Gagne, Hopfinger, and Antonia 1984] are in the range of $\xi_p$ shown in figure 6 (a) [Figure 18].

Clearly, these are not merely coincidentally published scientific ideas of Qian's but highlight the fundamental contribution which he left for our physical understanding of Kolmogorov's theories of small-scale turbulence (K41 and K62).

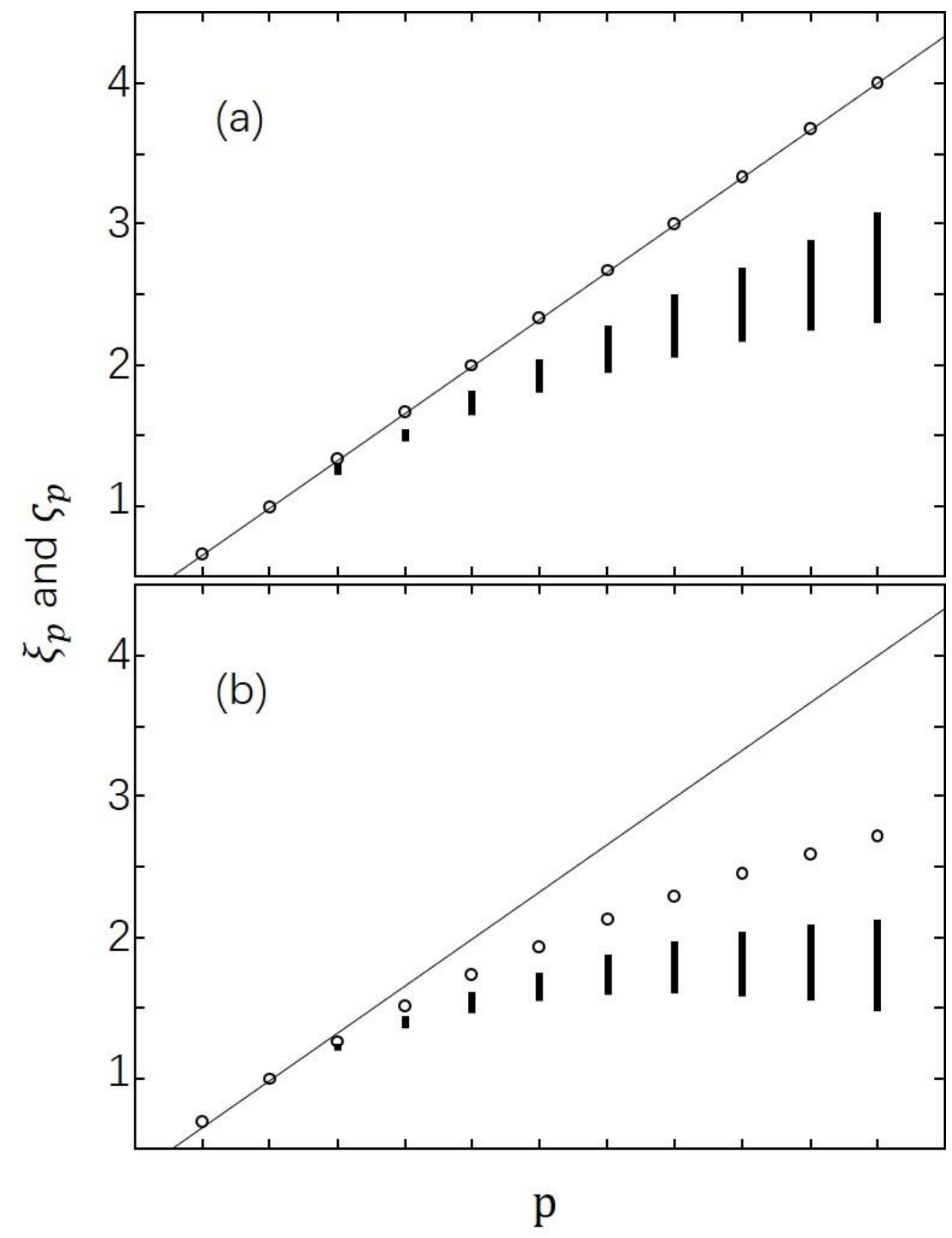

Figure 18 The absolute scaling exponent ($\xi_p$) and the inertial range scaling exponent ($\zeta_p$) against the order ($p$). (a) $D_{LL}(r)$ with $\zeta_p = 2/3$ (K41); (b) $D_{LL}(r)$ with $\zeta_p = 0.7$ (K62). Empty circles refer to the inertial range scaling exponents, the straight line to $p/3$ (K41), the vertical bars to the absolute scaling exponents $\xi_p$ over the range defined by (Qian 2006b, page 8, equations (16a, b)) while modeling flow parameters of Anselmet, Gagne, Hopfinger and Antonia (1984) (Qian 2006b, page 8, Figure 6).

*Others' work on the finite $R_\lambda$ effect of turbulence in parallel with Qian*

George (1989), George and Castillo (1994) and George (1995) proposed two ideas which allow treatment of finite Reynolds number effects in turbulent flows when infinite Reynolds number forms of the equations are available: (a) the Asymptotic Invariance Principle; (b) the Near-Asymptotics analysis. Mydlarski and Warhaft (1996) presented a Near-Asymptotics analysis of the turbulence energy spectrum that accounts for the effects of finite Reynolds number. Gamard and George (1999) developed a theory for the energy spectrum at finite Reynolds numbers. In the author's view, George's studies are independent of Qian (1997, 1998b).

Bowman (1996, page 10, bottom) argued that the asymptotic nature of inertial-range scaling laws must be emphasized. The two terms, 'an infinite inertial range' and 'a finite inertial range', appeared in Bowman (1996, page 10, bottom). He wrote that 'The theoretical scalings are expected only in the limit of an infinite inertial range, i.e. when the dissipation and forcing wavenumbers are widely separated. The wavenumbers at the end of a finite inertial range are influences by the shape of the energy spectrum outside the inertial range and will not exhibit true inertial-range behaviour.'

It was true that Sreenivasan and Antonia (1997) were aware of the finite Reynolds number effect. The notion 'the effects of …finite Reynolds number' appeared in Sreenivasan and Dhruva (1998, page 103, Abstract, line 10). However, a quantitative study of it was not done until Sreenivasan and Bershadskii (2006). The motivation was driven by the fact that experimental or numerical data in turbulence are invariably obtained at finite Reynolds numbers whereas theories of turbulence correspond to infinitely large Reynolds number. Sreenivasan and Bershadskii (2006) used logarithmic expansions to quantify corrections for finite Reynolds numbers. It is unclear why the studies on the finite $R_\lambda$ effect by Qian, especially, the quantitative study of Qian (1999), were not cited in Sreenivasan and Bershadskii (2006). Sreenivasan and Bershadskii (2006) seems to be independent of Qian (1997, 1999).

*Others' work on the finite $R_\lambda$ effect of turbulence after Qian*

The finite $R_\lambda$ effect was implicitly discussed in Lundgren (2002, 2003, 2005). Arenas and Chorin 2006 (page 4352, right column, paragraph 2, lines 9-11) wrote that "However, at finite $R$ [Reynolds number] viscosity reduces the intermittency, and the scaling has to be corrected for "intermittency reductions" that depend on $R$ [Reynolds number]".

The finite Reynolds effect is also implicitly discussed in Kaneda and Ishihara (2006, page 13, Section 4.6 Reynolds number dependence of statistics). The main interest of Section 4.6 of Kaneda and Ishihara (2006, page 13) is on the $E(k)$ in the near dissipation range, in particular the $R_\lambda$-dependence of on the basis of a model of $E(k)$. To study the Kolmogorov's 4/5 law and the Kármán-Howarth-Kolmogorov equation, by high-resolution direct numerical simulations of turbulence, Kaneda, Yoshino and Ishihara (2008, page 1, Abstract, line 5) focused on the influence of the finite Reynolds number, length-scale and the weak but finite anisotropy. In a sense, they provided a quantitative study of the finite Reynolds number effect. Similarly, but they did not explicitly attribute 'the influence of the finite Reynolds number' to Qian, they did cite Qian (1997, 1999). Kaneda, Yoshino and Ishihara (2008, pages 1 and 8) cited many references [4-23]. Qian (1997, 1999) can be placed among the first group (Yukio Kaneda with Personal Communication, 14 April 2020). Ishihara, Gotoh and Kaneda (2009, page 171, Section 3.4) argued in Section 3.4. Influence of Finite Reynolds Number and Scale that "The idea of universality in K41 concerns only the asymptotic statistics of turbulence at large enough $Re$ and small enough scales. However, $Re$ and the scale size are finite in any real turbulence. In general, it is difficult to know a priori how large $Re$ must be for universality to apply or to get a quantitative estimate of the influence of finite $Re$ and scale size. In this regard, Equations 3 and 6 are exceptional because they are exact in the limit of $Re \to \infty$ and $r/L,\ \eta/r,\ k\eta,\ 1/(kL) \to 0$ for stationary HI turbulence. They may give some idea of the influence of finite $Re$ and scale size". Although Ishihara, Gotoh and Kaneda did not explicitly attribute this to Qian, they did cite Qian (1999). It can be inferred that they were inspired by Qian (1999) and 'Influence of Finite Reynolds' by Ishihara, Gotoh and Kaneda is exactly the same as 'finite Reynolds effect' by Qian (1997).

By use of the eddy-damped quasi-normal Markovian model, Bos, Chevillard, Scott and Rubinstein (2012) computed second and third order longitudinal structure functions and wavenumber spectra of isotropic turbulence. They compared their results with those of the multifractal formalism. The following interesting results are found: (i) both models give power-law corrections to the inertial range scaling of the velocity increment skewness; (ii) for the eddy-damped quasi-normal Markovian model, this correction is a finite Reynolds number effect. Qian (1997, 1999) were cited in Bos, Chevillard, Scott and Rubinstein (2012). In their conclusion, Bos, Chevillard, Scott and Rubinstein (2012, page 16, paragraph 2, lines 7-9) highlighted an interesting perspective, i.e. to investigate to what extent intermittency-corrections to higher-order quantities such as the flatness can be distinguished from Reynolds number effects.

The term 'finite-Reynolds effects' appeared in a 122 page review by Biferale and Procaccia (2005, page 53, Section 2.4, lines 3-4 from the bottom; page 55). Since anisotropy decays with small scales, large Reynolds number will bring the two set of exponents close together. This motivated Luca Biferale to think about the finite Reynolds number effect (Personal Communication with Luca Biferale, 26 March 2020). As shown in Abstract, Benzi and Biferale (2015, page 1531) presented a detailed discussion of finite Reynolds effects. Benzi and Biferale (2015, page 1362, lines 12-13) wrote that "Therefore the discrepancy between $\zeta_l(p)$ and $\zeta_{tr}(p)$ is a measure of the finite size $Re$ effects." where $\zeta_l(p)$ and $\zeta_{tr}(p)$ are the corresponding scaling exponents of the third-order structure functions in the Eulerian fame for longitudinal velocity difference and transverse velocity difference, respectively. Clearly, longitudinal and transverse scaling exponents are supposed by theory to be the same for homogeneous isotropic turbulence at very large Reynolds number (Biferale and Procaccia 2005). Anisotropic effects can introduce some sub-leading correction which should vanish when $Re \to \infty$, so the discrepancy between longitudinal and transverse scaling exponents is possibly a consequence of anisotropic finite Reynold numbers (Personal Communication with Luca Biferale, 26 March 2020). Although three different notations, 'finite size effects'(Benzi and Biferale 2015, page 1355, lines 2-3 from bottom; page 1356, line 1), 'finite Reynolds effects' and 'the finite size $Re$ effects', were used by Benzi and Biferale (2015), their physical meanings are identical to 'the finite Reynolds effect' proposed by Qian (1997).

There has been progress on this since Lundgren (2002, 2003). Obligado and Vassilicos (2019) assessed Lundgren (2002, 2003) and compared to data. It was found that the Kolmogorov equilibrium in such turbulence is only reachable as $Re_\lambda \to \infty$ in the vicinity of the Taylor length-scale not somewhere between viscous length scales and integral scales; the rest of the inertial range is increasingly out of equilibrium as the scale increases from the Taylor length to the integral length. This may appear as a small issue as the deviation is small, but it is gradually increasing which means that the famous dissipation scaling of $u^3/L$ cannot be justified in the usual way (and as Kolmogorov himself did). In fact it is not always true, and there are cases where the scaling is different.

## Significance of Qian's work on the finite $R_\lambda$ effect of turbulence

> [George Berkeley's] *The analyst* is a criticism of the calculus, in both its Newtonian and Leibnizian formulations, arguing that the foundations of the calculus are incoherent and the reasoning employed in it is inconsistent. Berkeley's powerful objections provoked numerous responses, and the task of replying to them set the agenda for much of British mathematics in the 1730s and 1740s (Jesseph 2005, page 121, paragraph 1)

Likewise, the controversy surrounding K41, i.e. the intermittency correction and the finite $R_\lambda$ effect, may not be fully clarified, however, it has already provoked numerous interesting studies at these two ends. The $Re$- or $R_\lambda$-dependent studies are mutually related to each other but the finite $R_\lambda$-related study is different from them. Overall, those $Re$- or $R_\lambda$-dependent studies may not be fully aware of the following facts: (i) whether $Re$- or $R_\lambda$ is finite or infinite; (ii) K41 and K62 actually refer to the asymptotic case of infinite Reynolds numbers; (iii) field and laboratory experiments and numerical simulations are made at finite $R_\lambda$; (iv) the finite $R_\lambda$ can be (very) effective in controlling turbulence, i.e. the occurrence of the finite $R_\lambda$ effect ($Q_e$); (v) the finite $R_\lambda$ effect ($Q_e$) can decay either quite fast or slow and can be quantified as the decay law ; (vi) the finite $R_\lambda$ effect ($Q_e$) cannot at all be neglected; and (vii) the finite $R_\lambda$ turbulence should be studied. Clearly, Qian did his work on (iv), (v), (vi) and (vii).

Qian submitted his manuscript entitled Inertial range and the finite Reynolds number effect of turbulence to *Physical Review E* on 31 May 1996 and it was published in January 1997. The term, 'the finite Reynolds number effect', repeatedly appeared four times in Qian (1997, page 337, paragraph 1). Given the fact that Kraichnan (1991) coined the term, 'Finite Reynolds number turbulence', it may be cautiously inferred that Qian had read Kraichnan (1991). Clearly, the term 'finite Reynolds number turbulence' (Qian 1997, page 337, right column, line 9) was identical to 'Finite Reynolds number turbulence' (Kraichnan 1991, page 66, paragraph 4, line 4). The term 'finite Reynolds number turbulence', appeared in Qian (1998b, page 7325, Abstract, line 2) is identical to Kraichnan (1991, page 66, paragraph 4, line 4). However, Kraichnan (1991) was not cited in Qian (1997) but Kraichnan (1991, page 66, paragraph 4) was cited/quoted in Qian (1998b, page 7328, right column, paragraph 2, lines 18-24).

The same term 'the finite Reynolds number effect' also appeared in Sreenivasan and Antonia (1997, page 446, lines 2-3). It was interesting that Publication date of *Annual Review of Fluid Mechanics* Volume containing Sreenivasan and Antonia (1997) was also January 1997. It is clear that Qian, Sreenivasan and Antonia were not able to read other's papers because of the close publication in timing. Apparently, the order of publication does not necessarily reflect the real order in which these two pieces of work were done. The term 'a finite Reynolds number' appeared in Barenblatt and Chorin (1998, page 288, paragraph 2, lines 7-9). They wrote that "A nonzero viscosity, or equivalently, a finite Reynolds number, creates a correction to this ideal intermittency and produces a viscous correction to the asymptotic slope." It is interesting to note that the intermittency can be corrected by a finite Reynolds number. Nevertheless, Qian is fully credited with the finite Reynolds number effect of turbulence and the finite Reynolds number turbulence (Qian 1997, 1998b, 1999, 2000, 2001, 2002, 2003, 2006a, b).

*Qian's indirect response to Batchelor's (1962) concern about K41*

Batchelor (1962, page 93, paragraph 2, lines 9-11) reviewed that "The [K41] theory is an asymptotic one, and its predictions hold with increasing accuracy (if the theory is correct) as $\Re \to \infty [R_\lambda \to \infty]$, but no theoretical estimate has been made of the actual value of $\Re$ $[R_\lambda]$ needed for a given degree of accuracy." By applying the Kolmogorov's 4/5 law, Qian (1997) made the theoretical estimate of the actual value of $R_\lambda$, i.e. 'there is no inertial range when $R_\lambda \leq 2000$ and, .., $R_\lambda$ should be higher than $10^4$ in order to have an inertial range wider than one decade.'

*Robert A. Antonia's early relation with Qian*

Robert A. Antonia, Emeritus Professor at the University of Newcastle, Australia, has been working on small-scale turbulence since the 1970s (e.g. Antonia 1973). Antonia was ever an e-mail friend of Qian's. In his first e-mail to Antonia dated 2 January 2000 (Figure 19), Qian e-mailed that "Most scholars believe in the anomalous scaling of turbulence due to intermittency. In contrast to this prevailing point of view, we argue that the available data actually favors the K41 normal scaling rather than the anomalous scaling predicted by various intermittency models, if the data are properly interpreted by taking into account the finite Reynolds number effect which decays slowly. This result will change our understanding of

turbulence….It is interesting to make further experiments to prove or disprove the results of my papers.” Qian further wrote that “…… The important issue is how $F$ changes as the Reynolds number $R_\lambda$ increases. Is there new experimental data to help solve this issue?......The large values of $F$ around $R_\lambda = 10^4$ is obtained in a shear flow such as atmosphere boundary layer. For the case of isotropic turbulence, is it possible that $F$ approaches a constant as $R_\lambda \to \infty$? …” However, Qian did not explain how ‘This result will change our understanding of turbulence.’ until Qian (2006, page 9, relation (20)).

To: Prof. R. A. Antonia
Department of Mechanical Engineering
University of Newcastle, Newcastle, Australia 2038

From: J. Qian
Department of Physics
Graduate School of Academia Sinica
P. O. Box 3908, Beijing 100039, China

Date: 2 January 2000

Dear Prof. R. A. Antonia ,

Most scholars believe in the anomalous scaling of turbulence due to intermittency. In contrast to this prevailing point of view, we argue that the available data actually favors the K41 normal scaling rather than the anomalous scaling predicted by various intermittency models, if the data are properly interpreted by taking into account the finite Reynolds number effect which decays slowly. This result will change our understanding of turbulence. Enclosed please find my recent papers on this topic. It is interesting to make further experiments to prove or disprove the results of my papers.

I have read your interesting papers on the flatness factor F of the longitudinal velocity derivative $\partial u/\partial x$, for example your Phys. Fluids 1980 paper and Ann. Rev. Fluid Mech. 1997 paper. The important issue is how F changes as the Reynolds number $R_\lambda$ increases. Is there new experimental data to help solve this issue? Could you please recommend relevant papers? The large values of F around $R_\lambda = 10^4$ is obtained in a shear flow such as atmosphere boundary layer. For the case of isotropic turbulence, is it possible that F approaches a constant as $R_\lambda \to \infty$? I would appreciate it very much if you would let me know your opinion.

Best Wishes for A Happy New Year !

Sincerely,

J. Qian
email address: jianqian@public3.bta.net.cn

Figure 19 Qian’s e-mail to Antonia dated 2 January 2000.

Note that both $R_\lambda$ and $Re_\lambda$ refer to the Reynolds number of the turbulence or the Taylor microscale Reynolds number. There are a few interesting points: (a) none of Qian’s papers were cited in Sreenivasan and Antonia (1997); (b) Antonia was a strong supporter of the K62 phenomenology, though always in the context of trying to explain the Reynolds number dependence of various turbulence statistics, prior to Qian’s e-mail dated 2 January 2000. This email triggered many subsequent email exchanges between Antonia and Qian during 2000 and 2001.

*Re-examination of Batchelor and Townsend (1949, page 249, Figure 5)*

To answer Qian’s question ‘For the case of isotropic turbulence, is it possible that $F$

approaches a constant as $R_\lambda \to \infty$?', based on the grid-generated isotropic turbulence data of Batchelor and Townsend (1949, page 249, Figure 5), i.e. Flattening factors of velocity derivatives, we replot a new graph showing the flatness factor ($F$) against the mesh Reynolds number ($R_M$) (Figure 20). It can been seen that the flatness factor ($F$), for the case of isotropic turbulence, seems to asymptotically approach a constant as $R_M$ increases for the different order of the velocity derivative, $n$ =0, 1, 2 and 3, respectively, i.e. $R_\lambda \to \infty$. This is only true for a fixed downstream location (constant $x/M$). Note that we use 'flatness'.

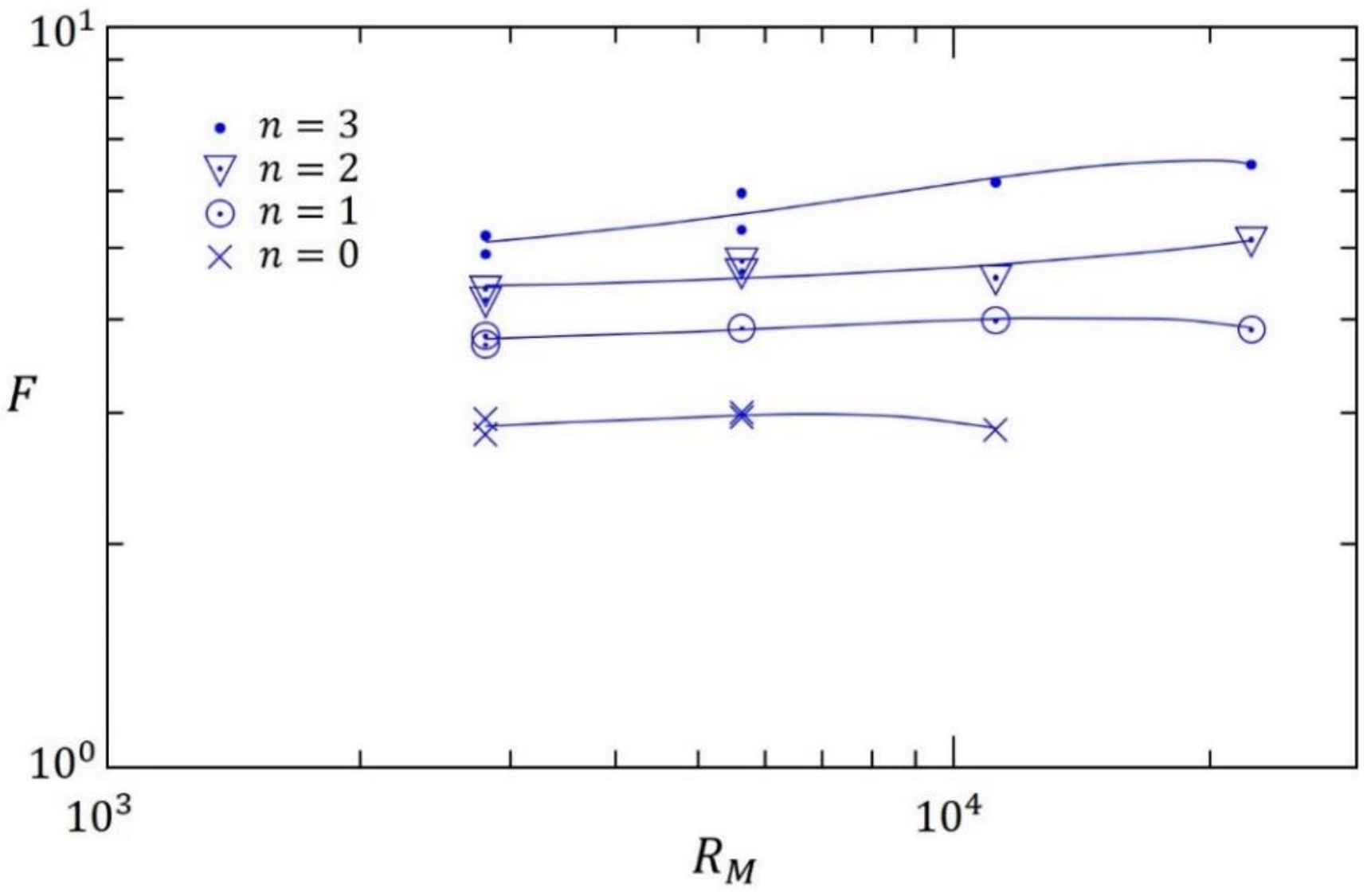

Figure 20 The flatness factor ($F$) against the mesh Reynolds number ($R_M$) for the grid-generated isotropic turbulence.

Clearly, our reanalysis of Batchelor and Townsend (1949, page 249, Figure 5) confirms the possible existence of the boundedness of the flatness factor of the velocity derivation when $R_{M\lambda}$ as found in Qian (1986a). It may suggest that the intermittency of turbulence discovered by Batchelor and Townsend (1949) was due to the finite Reynolds number effect. More complete set of measurements have since been made (e.g. Meneveau and Sreenivasan 1998).

*Antonia's earliest recognition of the significance of Qian's work on* $Q_e$

To the best knowledge of the author, it was Antonia who first fully recognized the merit and ramifications of Qian's work on 'the Finite Reynolds number (FRN) effect'. Although Antonia was well aware of the finite $R_\lambda$ effect of turbulence, e.g. Van Atta and Antonia (1980), Antonia, Chambers and Satyaprakash (1981), Antonia, Satyaprakash and Chambers (1982), and Sreenivasan and Antonia (1997, page 446, line 6), it was Qian's e-mail that started to steer Antonia and his associates along the right path to an improved understanding of homogeneous, isotropic turbulence. Antonia had not, in the context of the 4/5 law and the evidence for the so-called 'anomalous scaling', paid sufficient attention to 'the effect of a finite Reynolds number' which was first proposed by Qian (1997) until his joint work (Danaila, Anselmet, Zhou and Antonia 1999) and his paper entitled Approach to the 4/5 law in homogeneous, isotropic turbulence (Antonia and Burattini, 2006, page 175, Abstract, lines 2-3). However, Qian (1997) was not cited, instead, Qian (1999, 2000) were cited in Antonia and Burattini (2006). Antonia and his associates have published a series of papers on this topic (e.g. Antonia and Burattini 2006; Antonia, Tang, Djenidi and Danaila 2015; Tang, Antonia,

Djenidi, Danaila and Zhou 2017; Tang, Antonia, Djenidi, Danaila and Zhou 2018; Antonia, Tang, Djenidi and Zhou 2019).

*Antonia and Burattini's (2006) support for Qian's work on* $Q_e$

In particular, it led to the realization (Antonia and Burattini 2006) that Kolmogorov's famous 4/5 law can only be validated at extremely large Reynolds numbers for decaying homogeneous, isotropic turbulence. Antonia's more recent work, in collaboration with Shunlin Tang, Lyazid Djenidi, Luminita Danaila and Yu Zhou (Antonia, Tang, Djenidi and Danaila 2015; Antonia, Djenidi, Danaila, and Tang 2017; Tang , Antonia, Djenidi, Danaila and Zhou 2017, 2018), provides strong support for Qian's prediction that both the skewness and flatness factor of the velocity derivative approach constant values at large Reynolds numbers, which is in accordance with K41, as well as Qian's contention that the anomalous scaling most likely represents a Reynolds number effect. Antonia, Tang, Djenidi and Danaila (2015, page 727, Abstract/line sentence) found that "The constancy of [the velocity derivative skewness] $S$ at large $Re_\lambda$ has obvious ramifications for small-scale turbulence research since it violates the modified similarity hypothesis introduced by Kolmogorov (1962) but is consistent with the original similarity hypothesis (Kolmogorov 1941)". Qian (1986, 1994, 1999, 2003) were cited in Antonia, Tang, Djenidi and Danaila (2015).

*Tchoufag, Sagaut and Cambon's (2012) support for Qian's work on* $Q_e$

Tchoufag, Sagaut and Cambon (2012) also recognized the merits of the finite $R_\lambda$ effect of turbulence proposed by Qian (1997, 1999) in which a spectral approach was used. At page 015107-1/Abstract/lines 12-13 of Tchoufag, Sagaut, and Cambon (2012) wrote that " in the line of Qian [Phys. Rev. E **55**, 337 (1997), Phys. Rev. E **60**, 3409 (1999)] ". Qian's (1999) decaying law was listed in Tchoufag, Sagaut and Cambon (2012, page 3, Table 1, page 4, Table 2). Furthermore, Qian's (1984) shape function of the kinetic energy spectrum was listed in Sagaut and Cambon (2018, page 118, Table 4.7). Qian's (1999) decaying law was listed in Sagaut and Cambon (2018, page 115, Table 4.4; page 116, Table 4.5).

*McComb's (2014, pages 143-187) support for Qian's work on* $Q_e$

In his Chapter 6 entitled Kolmogorov's (1941) theory revisited (McComb 2014, pages 143-187), McComb also recognized the merit and significance of Qian's work on the finite Reynolds number effect. He presented an overview of "Qian's method" which appears in his book which devotes three pages to Qian's work, published in *Physical Review Letters* in 2000, on high-order structure functions of turbulence. Notably, Qian investigates whether the second-order exponent corresponds to normal Kolmogorov scaling or anomalous scaling. Qian's method makes use of exact relationships incorporated with well-established data correlations available in the literature, so as to extract as much physics as possible from the experimental results.

*McComb, Yoffe, Linkmann, and Berera's (2014) support for Qian's work on* $Q_e$

McComb, Yoffe, Linkmann, and Berera (2014, page 9, VII. Conclusions/paragraph 2, lines 1-3) explicitly recognize the importance of the finite $R_\lambda$ effect in their anomalous value of $\zeta_2$ that "Our result that $\zeta_2 = \Gamma_1 + 1 \to 2/3$ is an indication that anomalous values of $\zeta_2$ are due to finite Reynolds number effects, consistent with the experimental results of Mydlarski et al. [Mydlarski and Warhaft 1996] which point in the same direction".

*Boschung, Gauding, Hennig, Denker and Pitsch's (2016) support for Qian's work on* $Q_e$

In their literature review of I. INTRODUCTION, Boschung, Gauding, Hennig, Denker and Pitsch's (2016, page 2, paragraph 3, lines 2–3) argue that "For that reason, modifications of the asymptotic results which include the finite Reynolds number effects were proposed for different kinds of flows." After that, they reviewed a number of studies within this category. Clearly, based on the publication date, Qian (1997, 1999), which were cited/reviewed by them, 'the finite Reynolds number effects' is firstly due to Qian's work on $Q_e$.

*Antonia, Djenidi, Danaila, and Tang's (2017) support for Qian's work on* $Q_e$

Antonia, Djenidi, Danaila and Tang (2017, page 1, Abstract) explicitly stated that "Failure to recognize the importance of the finite Reynolds number effect on small scale turbulence has, by and large, resulted in misguided assessment of the first two hypotheses of Kolmogorov…or K41 as well his third hypothesis…or K62". Antonia, Djenidi, Danaila, and Tang (2017, page 2, paragraph 1, end of I. Introduction) further wrote that "The main purpose of this paper is to draw attention to the influence that the FRN effect can have on small scale statistics. Failure to properly account for this effect has led to misinterpretations of the data and, more seriously, has hampered our ability to distinguish between K41 and K62 unambiguously." This paper was written for the special issue in memoriam of John Lumley but could well have been dedicated to Qian Jian also.

*Tang, Antonia, Djenidi, Danaila and Zhou's (2017) support for Qian's work on* $Q_e$

Along the line shown in Qian (2006, page 8, Figure 6), Tang, Antonia, Djenidi, Danaila and Zhou (2017, page 359, Figure 14) re-examined those previous published data including Anselmet, Gagne, Hopfinger and Antonia (1984, page 86, Figure 14). As shown in Figure 21, Figure 14 in Anselmet, Gagne, Hopfinger and Antonia (1984, page 86) has been regarded as the classic result. According to Google, it has been cited more than 2000 times in the literature. It was also cited in Moffatt and Dormy (2019, page 420, Figure 15.2). However, after taking the Finite Reynolds number effect into account, the data presented in Anselmet, Gagne, Hopfinger and Antonia (1984, page 86, Figure 14) becomes a single red "+" ($R_\lambda$=852, $\alpha_6 = 1.64$) in Figure 14 in Tang, Antonia, Djenidi, Danaila and Zhou (2017, page 359, Figure 14).

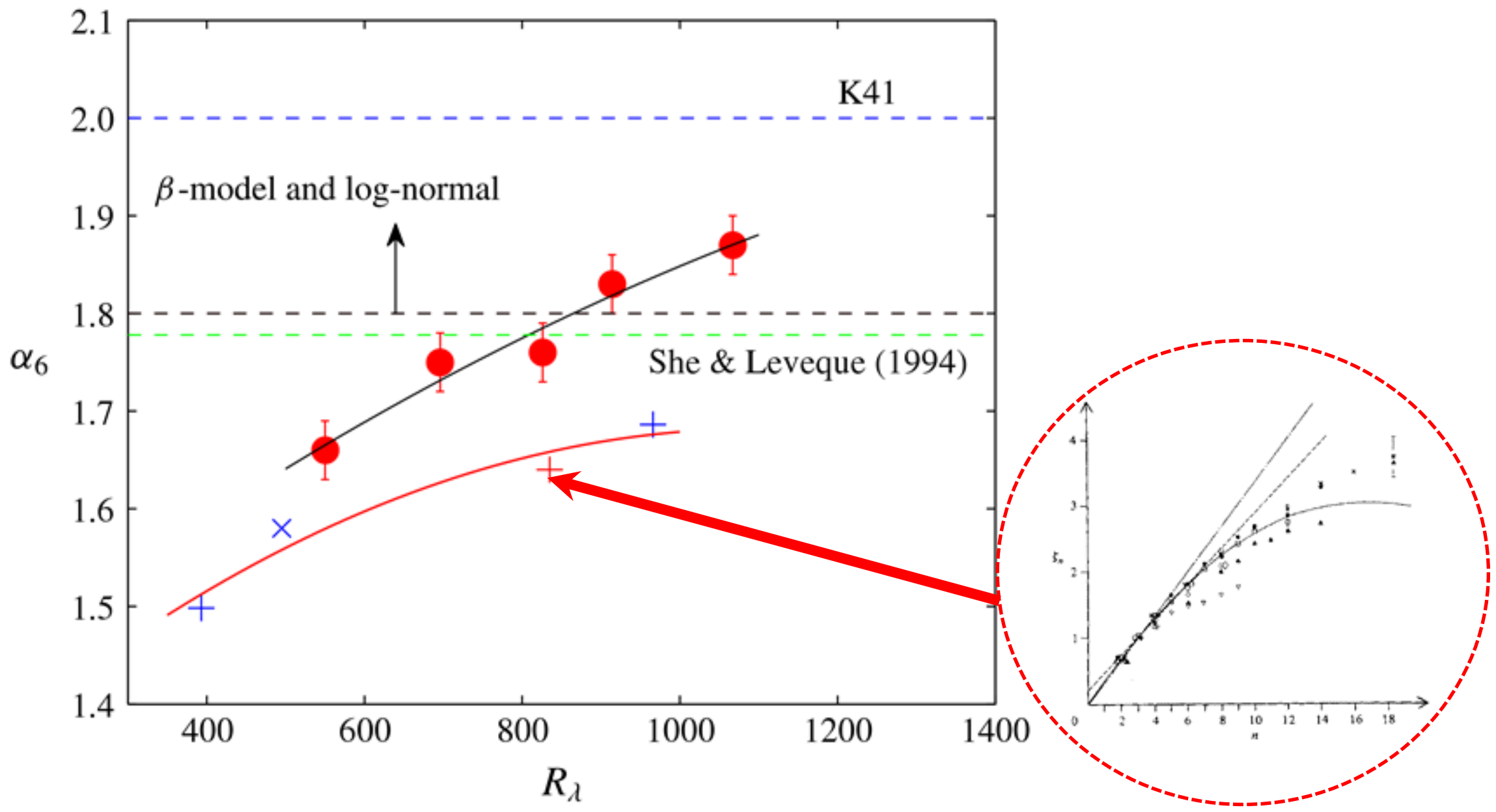

Figure 21 The data presented in Anselmet, Gagne, Hopfinger and Antonia (1984, page 86, Figure 14) becomes a single red "+" ($R_\lambda$=852) in Figure 14 in Tang, Antonia, Djenidi, Danaila and Zhou (2017, page 359, Figure 14). Right (red-dashed circle): Variation of exponent $\zeta_n$ as a function of the order $n$ (Anselmet, Gagne, Hopfinger and Antonia 1984, page 86, Figure 14). Note that $\alpha_6$ in Anselmet, Gagne, Hopfinger and Antonia (1984) was 1.8 while Tang, Antonia, Djenida, Danaila and Zhou (2017, page 357, circles in Figure 11) re-estimated it as 1.64.

Clearly, Tang, Antonia, Djenidi, Danaila and Zhou (2017, page 359, Figure 14) provide one of the strongest supports for the significance of Qian's work on the finite Reynolds number.

Qian (1999) predicted that different flows have different max [-$< du^3 >/r$] while their $R_\lambda$ are the same (Qian's e-mail to Antonia, 21 March 2000). He emphasized the slow nature of the decay of the finite-Reynolds-number effect. There is a flow dependence besides the $R_\lambda$ dependence. Zhou and Antonia (2000) showed that the magnitude of $\langle(\delta u_*)^3\rangle$ over the inertial range is smaller than $\frac{4}{5}$, the difference tending to become smaller as $R_\lambda$ increases. Based on those published extensive experimental data, Pearson and Antonia (2001) provided a fairly comprehensive description of the finite Reynolds number effect in different flows. It can be inferred from their studies that there is a flow dependence at finite Reynolds number. Qian agreed with Antonia that one needs to account for a 'flow dependence' as well as $R_\lambda$ dependence (Qian's e-mail to Antonia, 21 March 2000). Notably, the flow dependence simply reflects the different contributions of the large scale (or non-stationary) term in the von Kármán–Howarth equation, which is neglected in K41, when different flows are taken into account. This was nicely illustrated in Tang, Antonia, Djenidi, Danaila and Zhou (2017).

As shown in Figure 21, one may argue that one can never prove this by measurement alone. Nevertheless, the $R_\lambda$ dependence of $\alpha 6$ is naturally interesting. There are many other arguments in favor of anomaly, not a finite-Re effect. As an interesting exception, the same anomaly is found in atmospheric turbulence at both a much higher and lower $R_\lambda$ (Sreenivasan and Dhruva 1998). They found that anomaly prevails even for very low orders. Is there scaling in high $R_\lambda$ turbulence? Sreenivasan and Dhruva (1998, page 117) made a case that, despite some problems due to finite Reynolds numbers and finite shear, inertial range scaling does exist in high-Reynolds-number turbulence. Sreenivasan and Dhruva (1998) qualitatively concluded that it is difficult to discuss the scaling effectively without first understanding quantitatively the effects of finite shear and finite Reynolds numbers. By using the third-order structure function, Qian (1999) quantitatively studied the effect of finite Reynolds number on the scaling.

*Tang, Antonia, Djenidi, Danaila and Zhou's (2018) support for Qian's work on* $Q_e$

Let us now re-look at the three questions in Qian's e-mail to Antonia in 2000: (a) how $F$ changes as the [Taylor-]Reynolds number $R_\lambda$ increases? (b) is there new experimental data to help solve this issue? (c) is it possible that $F$ approaches a constant as $R_\lambda \rightarrow \infty$? Those questions were answered by Tang, Antonia, Djenidi, Danaila and Zhou (2018, page 244, Abstract) as follows:

> (a) $F$ can differ from flow to flow at moderate $Re_\lambda[R_\lambda]$; (b) $F$ may also depend on the initial conditions; (c) the data for $F$ in various flows, e.g. along the axis in the far field of plane and circular jets, and grid turbulence, show that it approaches a constant, with a value slightly larger than 10, when is sufficiently large; (d) this behavior for $F$ is qualitatively supported by our analytical considerations; (e) the constancy of $F$ at large $Re_\lambda$ violates the refined similarity hypothesis introduced by Kolmogorov (1962); (f) it is not, however, inconsistent with Kolmogorov's (1941a) original similarity hypothesis; (g) the power-law relation $F{\sim}Re_\lambda^{\alpha_4}$ (Kolmogorov 1962), which is widely accepted in the literature, has in reality almost invariably used to 'model' the finite Reynolds number effect for the laboratory data and has been strongly influenced by the weighting given to atmospheric surface layer data. (h) This inclusion of the latter data [atmospheric surface layer data] has misled previous investigations of how $F$ varies with $Re_\lambda$ $[R_\lambda]$.

Those findings further substantiate Qian's following foreseeing view in 2000 in detail, i.e. "This result should have important consequences in the field of turbulence, since the K41 and K62 theories imply completely different understanding of the physics of turbulence." (Qian 2000, page 646, left column, lines 38-41).

*Antonia, Tang, Djenidi, and Zhou's (2019a, b) support for Qian's work on* $Q_e$

In the author's view, after Antonia's joint review paper (Sreenivasan and Antonia 1997), his own review paper entitled Finite Reynolds number effect and the 4/5 law is an important one in the field of small-scale turbulence. In Antonia, Tang, Djenidi, and Zhou (2019b, page 14, last paragraph, lines 12-14), Antonia and his associates highlight the merit and significance of Qian (1997) by saying that "The present results, like those of McComb et al. [36][McComb, Yoffe, Linkmann, and Berera 2014], indicate that the anomalous values of $\zeta_2$ are very likely associated with the FRN effect; Qian (1997) was in fact first to draw attention to this."

Antonia, Tang, Djenidi, and Zhou (2019b, page 14, last paragraph, lines 14-16) further conclude that "More importantly, the present results also emphasize that $\zeta_3$ should never have been assumed to be equal 1; the latter value can only be reached at extremely large Reynolds numbers." This important finding is the result of Qian's work on $Q_e$. As discussed before, Grossmann, Lohse, L'vov and Procacia (1994) assumed $\zeta_3$=1 in their Finite size corrections to scaling high Reynolds number turbulence.

*Djenidi and Antonia's (2020) support for Qian's work on* $Q_e$

By using a numerical model, Djenidi and Antonia (2020) attempt to understand the effect of large-scale forcing on the statistical behavior of small scales in isotropic turbulence. In their review of the effect of the large-scale motion, or equivalently the finite Reynolds number (FRN) effect on the third-order longitudinal velocity structure functions, Djenidi and Antonia (2020, page 1, right column, line 11) cited Qian (1997, 1999) which are among the top of those cited references. By taking 'the finite Reynolds number effect' into account, Djenidi and Antonia (2020) were able to compare Oboukhov's (1962) and Kolmogorov's (1962) corrections to the 2/3 law.

*Tang, Antonia, Djenidi and Zhou's (2020) support for Qian's work on* $Q_e$

Based on hot-wire measurements in grid turbulence, Tang, Antonia, Djenidi and Zhou (2020) examine the scaling of the two-point correlation for the turbulent energy dissipation rate ($\epsilon$), over a range of $R_\lambda$. As $R_\lambda$ >300, i.e. a condition achieved for both plane and circular jets, the normalized dissipation correlation function ($\Psi(r)$) collapses over nearly all values of $r$ when Kolmogorov local scale of turbulence ($\eta$) is used for the normalization. On the contrary, there is no collapse in either the power-law range or dissipative range when the integral (or external) length scale ($L$) (Figure 3) is used for the normalization, suggesting that there is no self-similarity based on external scales. It is found that the behavior of turbulent energy dissipation correlation function on the axes of plane and circular jets seems consistent with the first similarity hypothesis of K41 but not with K62 as $R_\lambda \approx 10^3$. Tang, Antonia, Djenidi and Zhou's (2020) finding provides strong support for the significance of Qian's work, particularly Qian (1997, 1998).

*Other possible implications of Qian's work*

To testify K62 an O62, Siggia (1977) found that scaling exponents and probability distributions are in crude agreement with experiment and the log-normality hypothesis. Clearly, 'crude agreement' is mainly due to the fact that experiment was made in the finite Reynolds number turbulent flow.

Hunt and Vassilicos (1991b) added a note as errata to Hunt and Vassilicos (1990a) that "there have been several sets of experimental measurement of turbulent flows at high Reynolds number which are *not* consistent with some of the key assumptions and results of K41a (see, for example, Mestayer, P. *J. Fluid Mech* 117, 27-43 (1982)). There is no generally agreed explanation for the difference between these findings and those that are consistent with Kolmogorov's theory." Mestayer (1982) found that the local-isotropy assumption is not satisfied by their measured velocity field in the expected inertial subrange. In the author's view, this can be explained by the finite $R_\lambda$ effect of turbulence, because their measurements were made in the I.M.S.T. Air-Sea Interaction Simulation Tunnel which was the finite $R_\lambda$ turbulent flow.

As another possible implication of Qian's work, Saddoughi and Veeravalli (1994, page 350, Figure 9) ever compiled those published data (taken from Chapman 1979) together with their own data to show Kolmogorov's universal scaling for one-dimensional longitudinal power spectra. In their Figure 9, a $k^{-5/3}$ range is obtained by making a log-log plot of the one-dimensional longitudinal spectra against the wave number $k_1\eta$. According to Qian (1997), the so-called 'inertial range' in Saddoughi and Veeravalli (1994, page 350, Fig. 9) is just a 'scaling range' and not the same as Kolmogorov's real inertial range.

As an implication of Qian's work, tidal-channel turbulence measurements (Grant,

Moilliet and Stewart 1959; Grant, Stewart and Moilliet 1962) ever strongly supported K41. However, Long (2003) considered their measurements questionable. This controversy may be due to the finite $R_\lambda$ turbulence was measured in the tidal channel.

*Summary*

As a first note, in an interview, Sreenivasan (2018) shared his view "It shows that if an idea is powerful, it will occur to a number of people, and more or less at the same time." As a second note, William K. George shared the following view with the author "It always amazes the author and others how/why different people in different places come up with similar ideas at the same time. It is almost like the ideas themselves are controlling when they reveal themselves to us. Those important ideas are revealed to us when the time is right. When it is 'their time'. And often simultaneously to many in far-away places" (William K. George with Personal Communication, 8 April 2020)".

There are four approaches to quantify corrections for the finite $R_\lambda$ effect of turbulence, i.e. (a) the Asymptotic Invariance Principle (e.g. George 1989; George and Castillo 1994; George 1995); (b) the Near-Asymptotics (e.g. George 1989; George and Castillo 1994; George 1995); (c) the Kolmogorov equation (e.g. Qian 1997, 1998a, b, 1999, 2000, 2001, 2002, 2003, 2006), or Kármán-Howarth-Kolmogorov equation (e.g. Gotoh, Fukayama and Nakano 2002, page 1171, VI; Kaneda, Yoshino and Ishihara 2008, page 2); and (d) logarithmic expansions (e.g. Sreenivasan and Bershadskii 2006). Plus George (2013) (from Marseille 2011) has laid out an additional case that flows in near statistical equilibrium (like most shear flows and forced turbulence) are different from those that are not (like decaying turbulence or homogeneous shear flow turbulence). These can be easily confused with finite Reynolds number effect.

Iyer, Sreenivasan and Yeung (2020) mentioned that there do exist occasional claims that departures from self-similarity are artifacts of finite Reynolds numbers, and so will vanish in the limit of very large Reynolds numbers under ideal circumstances. The latter can be best illustrated by Tang, Antonia, Djenidi, Danaila and Zhou (2017, page 359, Figure 14). Qian (1998) was among the top list of their cited five papers in Iyer, Sreenivasan and Yeung (2020). It can be inferred from Iyer, Sreenivasan and Yeung (2020) that they strongly believe that intermittency is of paramount importance and that scaling exponents do saturate. As W. David McComb has pointed out, there has been a strong polarization of views in the community (mainly 'intermittency' versus 'FRN' effects). It is still believed that Qian is correct in the sense that the available Reynolds numbers are not sufficient to allow us to dismiss K41, i.e. the so-called 'anomaly' needs to be treated with extreme caution, if not with a grain of salt. Iyer, Sreenivasan and Yeung (2020) claimed that there is a modest 4/5 law. This throws a major doubt on whether we can really talk about rigorous power laws since the inertial range has yet to be established. One also needs to keep in mind that the conclusions in Iyer, Sreenivasan and Yeung (2020) are based on 'forced' box turbulence.

Does "a strong polarization of views on 'intermittency correction' versus 'finite $R_\lambda$ effect' in the community of small-scale turbulence" mean that they are totally different, i.e. contradicted to each other? Does it mean that one is correct and the other wrong? Which is correct? Which is wrong? Can they be 'unified'? As presented before, Qian (1986a) already concluded that both the Kolmogorov's $k^{-5/3}$ law and the high degree of intermittency can coexist. Based on those limited historical overview and the state-of-the art review above, in the author's view, they may be 'unified' in a sense. As we all know, there is no doubt about the existence of the intermittency of turbulence. However, its effect will be $R_\lambda$-dependent. All kinds of turbulent flows in nature, the laboratory, or numerical modeling are actually at finite $R_\lambda$, i.e. the finite $R_\lambda$ turbulence. Clearly, a full understanding of the dynamics of the finite $R_\lambda$ turbulence is likely to be able to resolve the so-called controversial question of intermittency corrections to K41. Among various types of $R_\lambda$-related studies above (Section V), (xiii) (the anomalous values of ) the scaling exponents $\xi_p$ $[\xi_n]$ of $p$th $[n$th] order structure function versus the finite $R_\lambda$, in particular, the absolute and inertial range scaling exponents of velocity structure functions versus $R_\lambda$, is the right path to this end. It was Qian (1997, 1998, 1999) who among the first draw our attention to this right path. In this respect, Qian's contribution to small-scale turbulence is of fundamental importance.

## The continuing legacy of Qian

To remain unsoured even though one's merits are unrecognized by others, is that not after all what is expected of a gentleman?
The Master Confucius (551-479 BC) [THE ANALECTS BOOK I] (Waley 2000)

In the field of small-scale turbulence, there is no doubt that Kolmogorov, Batchelor and Kraichnan have played leading roles. From his first paper (Qian 1983) to his last one (Qian 2006), it can be seen that Qian, following up Kolmogorov, and then Batchelor and finally Kraichnan, gradually tried to fully understand Kolmogorov's work on small-scale turbulence. The name 'Kolmogorov' appeared more than 294 times in the texts of Qian's 26 single-authored international peer-reviewed papers. In the author's view, Kolmogorov was no doubt Qian's hero in small-scale turbulence studies while Qian was one of the best indirect disciples of Kolmogorov's. In Qian's study of small-scale turbulence, the following Batchelor-related studies were made: (i) the Batchelor constant $Ba$ (Qian 1987, 1988); (ii) Batchelor's $k^{-1}$ spectrum (Qian 1990); (iii) the Batchelor wave number (Qian 1995, Abstract, line 7); (iv) the Batchelor interpolation formula or Batchelor's model of the structure function (Qian 1998a, page 3201, Section 4); (v) the Batchelor averaging (Qian 1999; Qian 2001, page 1115, line 16); and (vi) Batchelor's fit (Qian 2000, page 646; Qian 2001, page 1096, Section 3.2; Qian 2003, page 1006, Section III.). The name 'Batchelor' appeared 6 times in Qian (1990a), 10 times in Qian (1990b), 11 times in Qian (1998b), 4 times in Qian (1999), and 11 times in Qian (2003). The name 'Kraichnan' appeared 13 times in his first paper (Qian 1983). Kraichnan's 7 papers were cited in Qian (1983). Qian's readings were intensive, including most of the published papers on small-scale turbulence available in the literature. Kraichnan's (1991, page 66, paragraph 4, lines 5-7) view, i.e. "it is likely that the question of intermittency corrections to K41 can be resolved only when a detailed understanding of the dynamics at finite Reynolds numbers has been achieved", was indeed the scientific 'voice of one crying in the wilderness'. Qian luckily heard about this voice and became a follower of Kraichnan's.

Most importantly, Qian has significantly stimulated our physical understandings of K41 and K62. More specifically, through (i) closure approach to higher-order structure functions of turbulence, (ii) the finite Reynolds number effect of turbulence, and (iii) the finite Reynolds number turbulence, Qian had sought to make his own contributions to the theory of homogeneous turbulence, to which Batchelor had already contributed greatly. It is unclear whether Batchelor ever read Qian's work, in particular, Qian (1997, 1998, 1999). The author feels reasonably sure that Batchelor would be interested to see how the problem of intermittency, which was first inferred by Corrsin (1943), Townsend (1948b) and Batchelor and Townsend (1949), was studied and understood by Qian from the finite Reynolds number perspective.

To anyone familiar with the scientific and academic worlds, it is not surprising that some new approach or idea might at first be too novel for those who reviewed the submitted manuscripts. Qian may have experienced this kind of obstacle. Likewise, simply because of Qian's deep thinking of small-scale turbulence, which was new for the reviewers who reviewed his papers, in Qian's own words, it was entirely due to the Editor who decided whether his submission was acceptable or not.

Because of Qian's outstanding contribution to turbulence, he was invited to present a talk at the International Conference on Fluid Mechanics and Theoretical Physics in Honour of Peiyuan Chou's 90th Anniversary, June 1-3, 1992, Peking University, China. Despite Qian's strong academic background, and his scientific contribution, he was never elected an Academician of the Chinese Academy of Sciences. In the author's review, this may have been partially due to the fact that Qian was socially unsophisticated. This did not prevent him from working hard. He enjoyed a life in search of scientific truth about turbulence. Many Chinese physicists were aggrieved at that Qian was not an Academician yet. For example, Ru-Zeng Zhu, a senior Chinese physicist, once aggrievedly said "Qian Jian is more Academician than Academician".

After Qian's English obituary was published online in *Physics Today* (Shi 2018), Antonia encouraged the present author by saying that 'Your obituary was needed and represents an important first step towards making young physicists aware of Qian's work.' Qian's Chinese obituary was published in *Chinese Journal of Theoretical and Applied Mechanics* (Shi 2019). This should assist in making Chinese physicists aware of Qian's work. After the publication of Qian's obituary in English and Chinese, the author received the warmest compliments from the Chinese turbulence community. One of them, which the author appreciates, was that "You made one of the greatest contributions to the Chinese turbulence community". Batchelor had inspirational influence on Chinese fluid dynamicists as presented in detail in Shi (2020). Among them, Qian was one of the best Chinese disciples of Batchelor's. Qian's work with Antonia's compliment was briefly highlighted in Shi (2020). This can be regarded as a third important step towards making more fluid dynamicists aware of Qian's work.

Undoubtedly, Antonia and his associates (S.L. Tang, L. Djenidi and Y. Zhou) made a fourth important step towards awareness of Qian's work. There was a Workshop on Reynolds Number Effects: Implications for Understanding and Controlling Turbulence which celebrated Robert A. Antonia's 75th birthday in March 2019 in Shenzhen, China. At this Workshop, Antonia kindly paid his tribute to Qian Jian (Figure 22). His talk was entitled The finite Reynolds number effect on small scale turbulence: a tribute to QIAN JIAN. At the end of his talk, Antonia cited two paragraphs in Obituary for Qian Jian (Shi 2018). Antonia, at his own birthday workshop, paid his tribute to a Chinese turbulence man, Qian Jian. Many people including the author were moved by what/how Antonia did. Unfortunately, the author was not present at the Workshop.

Qian fully deserves to be 'better' recognized/appreciated by a larger community of small-scale turbulence researchers. Through his study of the finite $R_\lambda$ effect, Qian has raised a significant doubt on the so-called 'anomalous' scaling of turbulence - which undermines Kolmogorov (1962) and the need to make intermittency corrections. In fact, Qian never, of course, queried the fact that intermittency is real and departures from Gaussianity are important. Qian's contribution to the finite $R_\lambda$ effect (and its importance) is much more fundamental.

In summary, like many others before him, Qian used the von Kármán–Howarth (as well as Lin's) equation together with a variety of methodologies (he developed his own closure but also used, for example, a form of $du^2$/spectrum) to mainly estimate the dependence of $du^3$ (or $T(k)$) on the Reynolds number. He quickly reached the conclusion that the inertial range, as defined by Kolmogorov, is out of reach. In particular, the Reynolds number has to be incredibly large for the viscous effect as well as the large scale term effect to disappear. He pointed out, almost ad nauseam, that the scaling range that intermittency modellers refer to is really artificial , thus correctly querying the 'anomalous scaling' that these modellers claim and hence casting a major doubt on K62. Qian's view is really his major, invaluable, legacy to turbulence research.

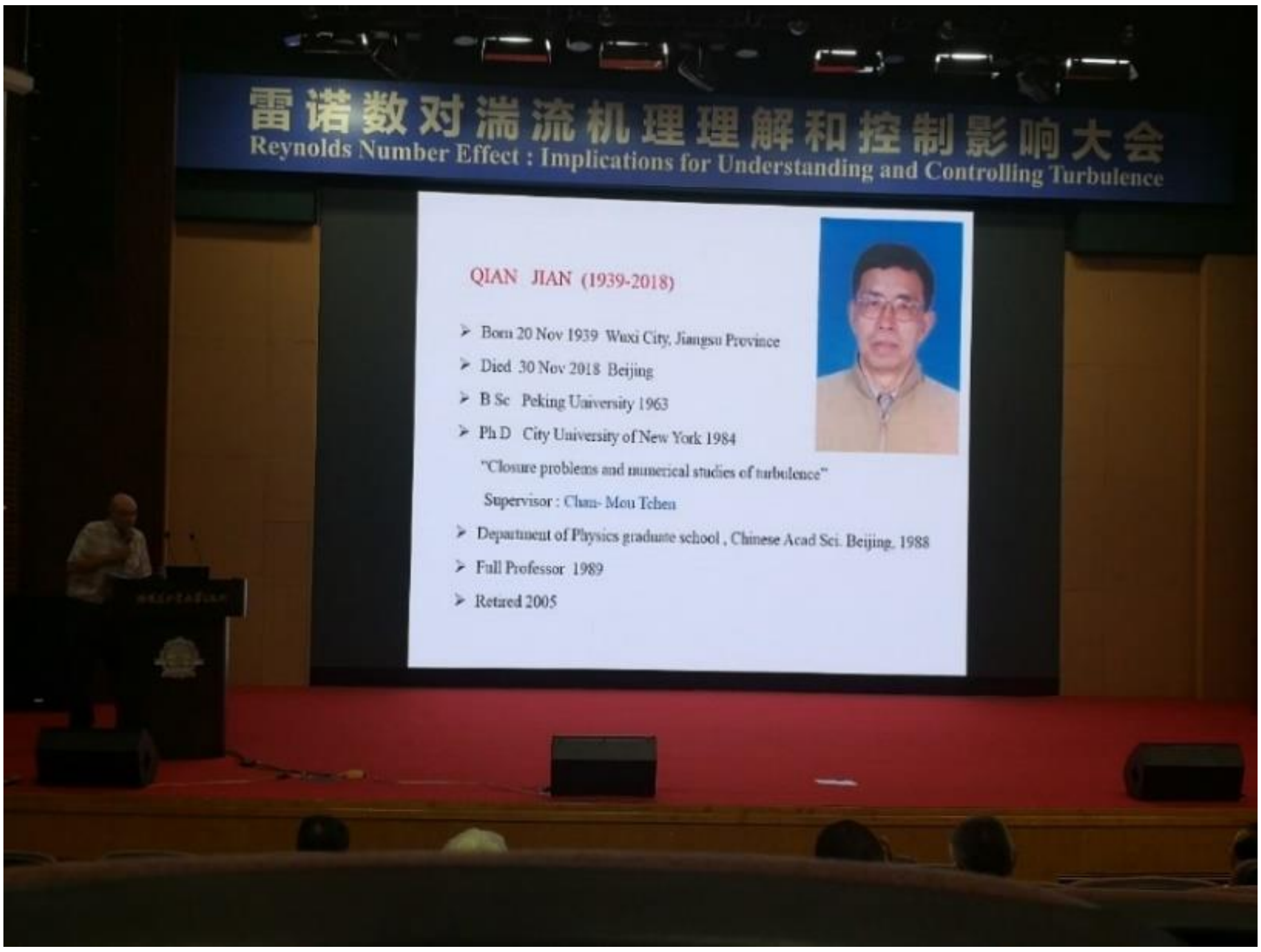

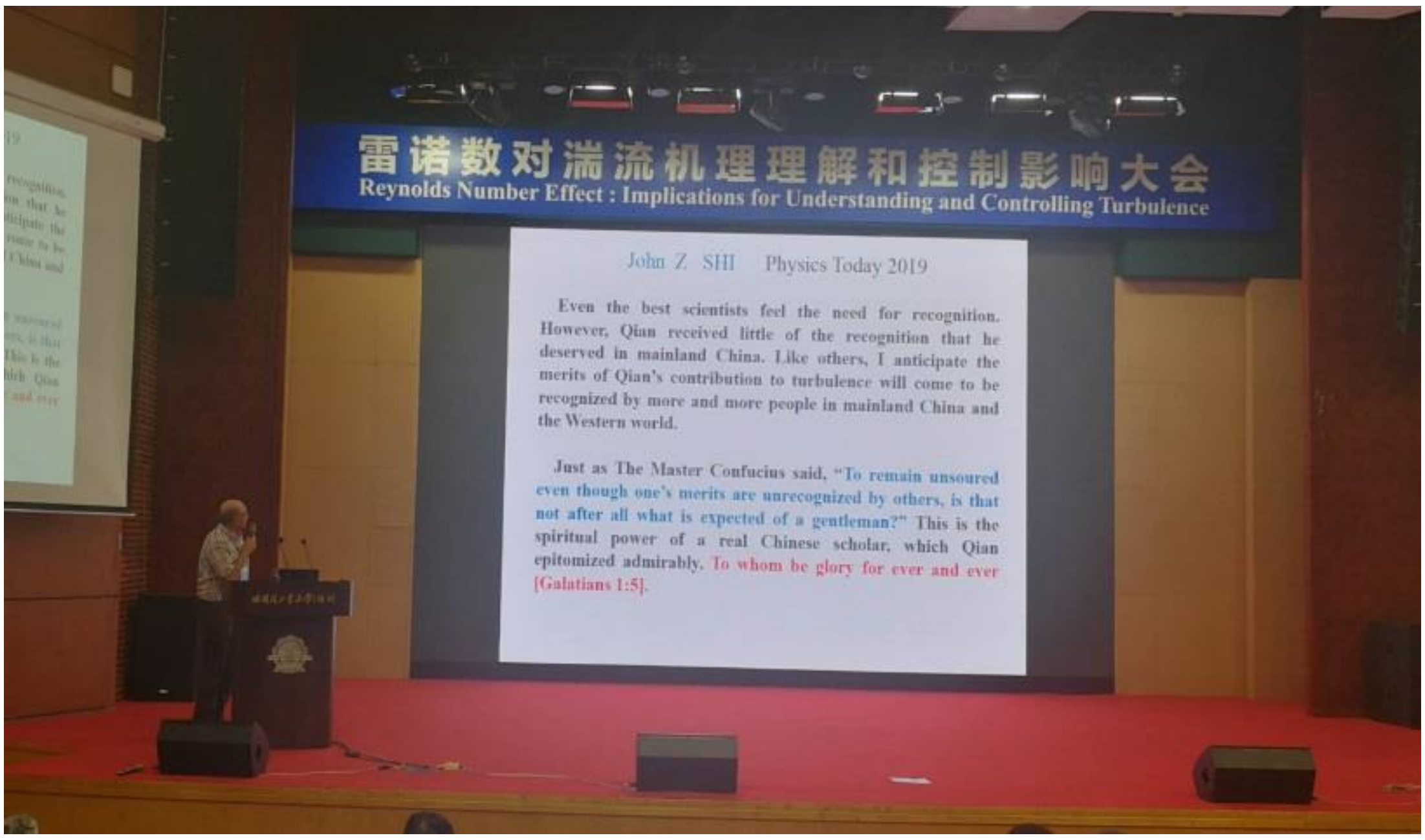

Figure 22 A workshop was held in honour of Robert A. Antonia's 75th birthday in March 2019 in Shenzhen, China. Top: Antonia presenting Qian's biography; bottom: Antonia quoting the last paragraph of Qian's obituary in *Physics Today* by Shi (2018). Photo Credit: Nan Gao.

## Concluding remarks

Of all physicists, Dirac has the purest soul.
the remark by Niels Bohr (1885-1962)

The author would say 'Of all Chinese physicists in mainland China, Qian Jian has the purest soul.' He was a disciple of small-scale turbulence, and he focused on this topic to the exclusion of all else. He was clearly a major investigator and writer in the field of small-scale turbulence, with many substantial papers to his own credit.

According to Kolmogorov (1941a, c; 1962), Monin and Yaglom (1975), Nelkin (1994), Frisch (1995), and Qian (2000, page 627, right column, paragraph 2) and Qian (2003, page 1005, Abstract, lines 9-10), the Kolmogorov theories of small-scale turbulence (K41 and K62) are different ones, i.e. in the real inertial range corresponding to the asymptotic case of $R_\lambda \to \infty$, $\langle |\Delta u_r|^p \rangle \sim r^{\zeta_p}$ or $\langle \Delta u_r{}^p \rangle \sim r^{\zeta_p}$, and $\zeta_p$ is the scaling exponent of order $p$; furthermore, the K41 theory predicts that $\zeta_p = p/3$ (normal scaling), while the K62 that $\zeta_2 > 2/3$ and $\zeta_p < p/3$ if $p > 3$ (anomalous scaling). The closure scheme developed by Qian (1986a, 2000) can be applied to predict high-order universal constants of inertial range scaling. The available data of the scaling exponents favour the K41 normal scaling theory rather than the K62 anomalous scaling theory predicted by numerous intermittency models. The finite $R_\lambda$ number effect $Q_e \to 0$ as $R_\lambda \to \infty$.

It took about 8 years from Kolmogorov (1941a, c) to the discovery of the small-scale intermittency of turbulence by Corrsin (1943), Townsend (1948b) and Batchelor and Townsend (1949). However, 'a veritable industry', the term used by McComb (2014, Chapter 6, page 143, line 1), regarding the search for so-called intermittency corrections, had a time span of more than 40 years since Corrsin (1943), Townsend (1948b) and Batchelor and Townsend (1949) before the concept of 'the finite Reynolds number effect' (Kraichnan 1991), Finite-size corrections (Grossmann, Lohse, L'vov and Procacia 1994), and the detailed study of the finite Reynolds number effect of turbulence (Qian 1997). It was truly a slow march to the path to the great ocean of truth about the physics of small-scale turbulence.

Qian, together with others, including Novikov, Sato, Kraichnan, and Antonia, steered all of us to initially the alternative path to the possible solution of the controversy surrounding K41 and eventually the right one to it through a detailed understanding of the dynamics of turbulence at the finite $R_\lambda$. In a sense, Qian identified the importance of the finite $R_\lambda$ effect of turbulence and steered all of us along the right path to an improved understanding of turbulent fluid dynamics and solutions to its problems. However, we are still far away from our full understanding of the finite $R_\lambda$ effect of turbulence and the finite $R_\lambda$ turbulence.

Man's life is two-fold, i.e. Procreate and Create. Qian's Procreate life is not so successful as his Create life. Qian has the sympathy for his Procreate life from the author and the admiration for his Create life from the author. Thanks be to the great Creator for the beauty, complexity and mystery of turbulence, and for his gift to human beings of the capacity for inquiry, analysis and wonder; for the understanding and insight that he has given to all mathematicians and scientists, among them Qian Jian should be remembered.

## Acknowledgements

The author first learned about the Kolmogorov–Oboukhov energy spectrum, i.e. E(k){proportional_to}k{sup -5/3}, when he was attending the IUTAM Summer School on Statistical and Hierarchical Structures in Turbulence, which Zhen-Su She organized, at State Key Laboratory for Turbulence and Complex Systems, Peking University, China, on 3-9

August 2001. The author's interest in small-scale turbulence was strongly re-motived and stimulated by his two sabbaticals during the Michaelmas Terms 2012 and 2019 at the G.K. Batchelor Laboratory, Department of Applied Mathematics and Theoretical Physics, University of Cambridge, U.K. Julian C.R. Hunt, Paul F. Linden and Stuart B. Dalziel are thanked for their full support. Zhen Ran is thanked for having drawn the author's attention to Qian's work on the finite Reynolds number effect of turbulence on 26 July 2018 at the Forum on New Thoughts on Turbulence held at Nanchang University, China. The writing up of this article was initially motivated by McComb's (2014, pages 168-170) book entitled Homogeneous, Isotropic Turbulence, in which Qian's method was introduced in detail, and subsequently motivated by Robert A. Antonia's recognition of the merit of Qian's work on the finite Reynolds number effect of turbulence. This article is dedicated to Xing-Chun Shi (1938-2003). The Wren Library, Trinity College Cambridge, is thanked for having permitted the author to access the Batchelor archive. Yiannis Andreopoulos is thanked for having found the author the pdf copy of Qian's (1984a) Ph.D. thesis and the portrait of Chan-Mou Tchen. Many people are thanked for their clarifications and discussions that contributed to this article, in particular, Robert A. Antonia, William K. George, Detlef Lohse, Yukio Kaneda, W. David McComb, Katepalli R. Sreenivasan, Shunlin Tang, and John-Christos Vassilicos. Kaneda found the author Sato, Yamamoto and Mizushina (1983, 1984) from Japan. Detlef Lohse found the author Lohse and Grossmann (1993) and Lohse (1994). Les J. Hamilton found the author Townsend (1948b) in Australia. Les J. Hamilton and Robert A. Antonia are thanked for carefully correcting the two draft versions of the Abstract of this article. William K. George and Les J. Hamilton have carefully read and corrected the two different draft versions. Other people who read the different draft versions include: Luca Biferale, Wouter Bos, Le Fang, Harindra Joseph S. Fernando, Mohamed Gad-el-Hak, and Guowei He. Yu-Hang Li is thanked for plotting Figure 7 and Dan Zhong for replotting the rest of Figures. Wen-Xuan Niu is thanked for kindly providing the author with Qian's colour portrait photo shown in the frontispiece. A good part of this article was written in COVID-19 lockdown.